\documentclass[aps,prc,twocolumn ,superscriptaddress]{revtex4-1}

\usepackage{graphicx}
\usepackage{amsmath}
\usepackage{braket}
\usepackage{url}

\usepackage[colorlinks,
linkcolor=blue,
anchorcolor=red,
urlcolor=red,
citecolor=red]{hyperref}

\begin{document}


\title{Proton decays in $^{16}$Ne and $^{18}$Mg and isospin-symmetry breaking in carbon isotopes and isotones}



\author{N. Michel}\email[]{nicolas.michel@impcas.ac.cn}
\affiliation{Institute of Modern Physics, Chinese Academy of Sciences, Lanzhou 730000, China}
\affiliation{School of Nuclear Science and Technology, University of Chinese Academy of Sciences, Beijing 100049, China}
\author{J. G. Li}
\affiliation{School of Physics,  and   State Key  Laboratory  of  Nuclear  Physics   and  Technology, Peking University, Beijing  100871, China}
\author{F. R. Xu}
\affiliation{School of Physics,  and   State Key  Laboratory  of  Nuclear  Physics   and  Technology, Peking University, Beijing  100871, China}
\author{W. Zuo}\email[]{zuowei@impcas.ac.cn}
\affiliation{Institute of Modern Physics, Chinese Academy of Sciences, Lanzhou 730000, China}
\affiliation{School of Nuclear Science and Technology, University of Chinese Academy of Sciences, Beijing 100049, China}



\date{\today}

\begin{abstract}
Proton-rich nuclei possess unique properties in the nuclear chart. Due to the presence of both continuum coupling and Coulomb interaction, phenomena such as halos, Thomas-Ehrman shift, and proton emissions can occur. Experimental data are difficult to be obtained therein, so that theoretical calculations are needed to understand nuclei at drip-lines and guide experimentalists for that matter. In particular, the $^{16}$Ne and $^{18}$Mg isotopes are supposed to be one-proton and/or two-proton emitting nuclei, but associated experimental data are either incomplete or even unavailable. Consequently, we performed Gamow shell model calculations of carbon isotones bearing $A=15\text{-}18$. Isospin-symmetry breaking occurring in carbon isotones and isotopes is also discussed. It is hereby shown that the mixed effects of continuum coupling and Coulomb interaction at drip-lines generate complex patterns in isospin multiplets. Added to that, it is possible to determine the one-proton and two-proton widths of $^{16}$Ne and $^{18}$Mg. Obtained decay patterns are in agreement with those obtained in previous experimental and theoretical works. Moreover, up to the knowledge of authors, this is the first theoretical calculation of binding energy and partial decay widths of $^{18}$Mg in a configuration interaction picture.

\end{abstract}


\maketitle

\section{Introduction}  
The proper description of nuclei at drip-lines is one of the main challenges of nuclear theory.
Phenomena absent in the valley of stability appear at the limits of the nuclear chart because of the decreasing separation energy of nuclear ground state. The most striking among them is the appearance of halos in the asymptotic region, of one-nucleon and two-nucleon types \cite{Al-Khalili2004,PhysRevLett.55.2676,PhysRevLett.99.252501,PhysRevLett.112.142501,PhysRevLett.112.242501,PhysRevLett.104.062701,Shi97,TANIHATA2013215,PhysRevLett.112.242501,RevModPhys.66.1105,Hansen95,RIISAGER1992393,PhysRevLett.79.3837,PhysRevLett.126.082501}, and the fact that ground states can become unbound \cite{PhysRevC.78.041302,PhysRevC.85.034327,PhysRevC.89.044610,PhysRevC.102.044614,PhysRevLett.108.142503,Baumann20071022}. One can also mention the formation of cluster structures in drip-line nuclei \cite{RevModPhys.90.035004,PhysRevLett.112.162501,PTP_clusters}. The fundamental reason leading to such phenomena is the strong coupling to continuum occurring in drip-line nuclear states. This occurs because valence nucleons mainly occupy shells lying close to particle-emission threshold. In fact, drip-line nuclei exhibit a large enhancement of inter-nucleon correlations compared to those of the valley of stability \cite{DOBACZEWSKI2007432}. 

The largest number of halo nuclei is found at neutron drip-line \cite{Al-Khalili2004}. This arises because neutrons are only subject to the centrifugal barrier in the asymptotic region. Indeed, neutrons occupy $sp$ partial waves in light nuclei for a large part, so that the small or even nonexistent centrifugal barrier therein allows to generate halos \cite{Al-Khalili2004,RIISAGER1992393}. One can cite for that matter the one-neutron halos $^{11}$Be, $^{19}$C, $^{31}$Ne, $^{37}$Mg \cite{Al-Khalili2004,PhysRevLett.112.142501,PhysRevLett.112.242501,Hansen95,TANIHATA2013215,PhysRevLett.112.242501,RevModPhys.66.1105} and the two-neutron halos $^{6,8}$He,$^{11}$Li, $^{14}$Be, $^{17}$B and $^{22}$C \cite{Al-Khalili2004,PhysRevLett.55.2676,PhysRevLett.99.252501,Hansen95,PhysRevLett.104.062701,Shi97,PhysRevLett.126.082501,TANIHATA2013215,RevModPhys.66.1105} and maybe a four-neutron halo in $^8$He \cite{PhysRevLett.99.252501}. 

$^{31}$F is suspected to be a halo nucleus \cite{PhysRevC.64.011301}, which is supported from Gamow shell model (GSM) calculations done by the authors \cite{PhysRevC.101.031301}. In fact, $^{31}$F might well be the heaviest two-neutron halo nucleus synthesized experimentally \cite{PhysRevC.101.031301}.

A clear consequence of the strong intertwining of the degrees of freedom of continuum and nucleon correlations at neutron drip-line is the odd-even staggering in the helium chain \cite{PhysRevC.67.054311,ensdf}. Indeed, the odd isotopes of the helium chain except $^3$He are all unbound and bear widths between 100 keV to 700 keV, whereas all even isotopes of the helium chain are bound except $^2$He and $^{10}$He \cite{ensdf}. Another important effect arising at neutron drip-line is the redistribution of magic numbers induced by continuum coupling \cite{SORLIN2008602}. As a consequence, nuclei of the neutron drip-line are widely studied from both theoretical and experimental point of views (see Refs.\cite{1402-4896-2013-T152-014022,DOBACZEWSKI2007432,0954-3899-36-1-013101,Navratil_2016} and Refs.\cite{RevModPhys.66.1105,RevModPhys.84.567} for reviews on these two respective subjects).

Apparently, nuclei at the proton drip-line would only present mild differences compared to well-bound nuclei, contrary to neutron-rich nuclei. This is due to the presence of the Coulomb barrier in proton-rich nuclei, which confines protons in the nuclear region and thus prevents from halos and large widths to develop. However, proton-rich nuclei are far from possessing a nuclear structure similar to nuclei of the valley of stability \cite{RevModPhys.84.567}. As the Coulomb barrier is weak in the lightest nuclei, halos and unbound nuclear states of sizable width can develop at proton drip-line, proton halo states being in smaller number than in the neutron drip-line, however \cite{Al-Khalili2004}. Indeed, $^8$B \cite{RevModPhys.66.1105} and the first excited state of $^{17}$F \cite{PhysRevLett.79.3837} are one-proton halo states, while $^5$Li and $^7$B are unbound and bear a proton-emission width close to 1 MeV \cite{ensdf}. In fact, the repulsive character of the Coulomb Hamiltonian hereby increases proton-emission width in the two latter nuclei compared to their unbound helium mirrors. Note that $^{17}$Ne is suspected to be a two-proton halo from recent experiments \cite{PhysRevC.82.044309,lehr2021unveiling}, supported by the GSM calculations done in this context \cite{MA2020135673}.  Added to that, continuum coupling has been shown to act differently on ground and excited states of proton-rich nuclei \cite{PhysRevC.100.064303}. While Coulomb energies in resonance ground states usually follow the same trend as that of well bound nuclei, namely the isobaric multiplet mass equation (IMME) \cite{PhysRevC.55.2407,IMME,PhysRevC.74.034315}, the Coulomb energies of excited states depart from the IMME due to their more extended wave functions in the pre-asymptotic region \cite{PhysRevC.100.064303}. Another interesting phenomenon has been discovered at proton drip-line, which consists in the formation of a very narrow $1/2^- $ resonance of the $^{15}$F spectrum above the Coulomb barrier, whose width is only about 40 keV \cite{DeGrancey:2016bez}. This many-body resonance was analyzed theoretically in Ref.\cite{DeGrancey:2016bez}, where it was shown that its narrow width could develop only because of the smallness of available phase space for particle decay \cite{DeGrancey:2016bez}. 

In fact, one of the most important drip-line phenomenon, namely two-proton decay, occurs at the proton-drip line \cite{RevModPhys.84.567}. Goldanskii predicted decades ago that two-proton decay is likely to occur in proton-rich nuclei verifying a simple condition function of their energy and width \cite{goldanskii60}. Two-proton radioactivity was then experimentally discovered a few years ago with the examples of $^{48}$Ni \cite{PhysRevC.83.061303}, $^{45}$Fe \cite{PhysRevLett.99.192501} and $^{54}$Zn \cite{PhysRevLett.94.232501}. More recently, the $^{67}$Kr isotope has been noticed to decay by two-proton emission \cite{PhysRevLett.117.162501} (see also Ref.\cite{PhysRevLett.120.212502} for its theoretical study with a three-body model). Other nuclei decaying partially or totally by two-proton decay are $^{12}$O \cite{PhysRevLett.74.860}, $^{16}$Ne and $^{19}$Mg \cite{PhysRevC.82.054315,PhysRevC.77.061303}. While $^{18}$Mg has not been observed, it can decay in principle by proton and/or two-proton emission as both its one-proton and two-proton separation energies are negative \cite{PhysRevC.87.014313}.
A microscopic theory of two-proton radioactivity has been developed in the frame of the real-energy continuum shell model \cite{PhysRevLett.95.042503,ROTUREAU200613} (see also Ref.\cite{Blank_2008} for a review on this topic and on associated experiments). The two-proton emitters $^{48}$Ni, $^{45}$Fe, $^{54}$Zn have been considered in that approach, albeit with sequential or cluster approximations \cite{ROTUREAU200613}.

Clearly, it is needed to study theoretically the one-proton and two-proton decays of $^{16}$Ne and $^{18}$Mg in order to understand and guide experiments on this topic (see for example Refs.\cite{PhysRevC.82.054315,PhysRevC.92.034329}). For this, we will employ the GSM \cite{PhysRevLett.89.042502,PhysRevC.96.024308,0954-3899-36-1-013101,PhysRevC.103.034305}, as it allows to calculate particle-emission widths of many-body nuclear states. One can note that $^{16}$Ne and $^{18}$Mg are mirror nuclei of carbon isotopes, so that the isospin mixing induced by the Coulomb Hamiltonian will appear explicitly. The situation is all the more interesting as partial dynamical symmetry occurs therein \cite{LEVIATAN201193,PhysRevLett.89.222501}, i.e.~the isospin operator does not commute with the used Hamiltonian whereas isospin is conserved in the considered many-body wave functions. We will not consider the two-proton decay of $^{19}$Mg with GSM, however. Indeed, its width is about $10^{-4}$ eV experimentally \cite{ensdf}. Assuming that the $^{19}$Mg ground state is accurately modelled in GSM, its calculation would then be well beyond the current capabilities of GSM, whose numerical precision is about 0.1 keV \cite{PhysRevC.101.031301}.

The paper is then structured as follows. We will firstly present the basic features of GSM. Then, we will describe the Hamiltonian used to calculate many-body wave functions, which is rooted in effective field theory (EFT). After this, we will depict the results obtained with the devised model in carbon isotopes and isotones of $A=$15-18, in which the $^{16}$Ne and $^{18}$Mg nuclei can be found. Energies, particle-emission widths and Coulomb energies will be dealt with. As mentioned above, the isospin partial dynamical symmetry will be discussed in relation to the obtained results. The method used to derive one-proton and two-proton decay widths from GSM many-body wave functions will also be detailed. Conclusion will be made afterwards.

\section{Method}

\subsection{Theoretical background of GSM} 
GSM is a configuration interaction framework based on the Berggren basis \cite{BERGGREN1968265}. The Berggren basis consists of one-body states generated by a finite-range potential, such as a Woods-Saxon potential. The Berggren basis is obtained from the real-energy Newton completeness relation \cite{newton1966scattering},  by deforming the contour of real energy states in the complex plane \cite{BERGGREN1968265}. Narrow resonance states sufficiently close to the real axis must be included with the bound states and complex scattering states \cite{BERGGREN1968265}. The bound, resonance, and complex-energy scattering states of the Berggren basis form a complete set of states:
\begin{equation}
\sum_n \ket{u_n} \bra{u_n} + \int_{L^+} \ket{u(k)} \bra{u(k)}~dk = \mathbf{\hat{1}}, \label{Berggren}
\end{equation}
where $\ket{u_n}$ is a bound or resonance one-body state and $\ket{u(k)}$ is a scattering state belonging to the $L^+$ contour of complex momenta (see Ref.\cite{0954-3899-36-1-013101} for details). In order to use the Berggren basis in numerical applications, one discretizes the $L^+$ contour of Eq.(\ref{Berggren}) with the Gauss-Legendre quadrature \cite{0954-3899-36-1-013101}. One typically needs 30-50 discretized states in order to have converged results \cite{PhysRevC.83.034325}. 

The discretized Berggren completeness relation can be formally identified to that generated by a set of harmonic oscillator (HO) states. Consequently, one can build configurations in GSM from the Berggren basis states of all partial waves, similar to the standard shell model (SM) \cite{0954-3899-36-1-013101}. The nuclear Hamiltonian is then represented by a complex symmetric matrix in GSM. 

One of the fundamental difference from SM is the appearance of many-body scattering eigenstates in the Hamiltonian spectrum. Indeed, many-body resonance states are hidden among the many-body scattering eigenstates of the Hamiltonian. The scattering eigenstates apparently form the vast majority of the eigenspectrum, but they are not part of the eigenstates of the lowest energies, contrary to the eigenstates of interest in SM. Therefore, one had to develop the overlap method to solve the so-called identification problem \cite{PhysRevC.67.054311,0954-3899-36-1-013101}. For that matter, the Hamiltonian is firstly diagonalized using pole approximation, i.e.~by suppressing all scattering configurations of the many-body Berggren basis, so that only resonant configurations, built from $S$-matrix poles, remain. The obtained eigenstate are zeroth-order approximations of the exact resonant eigenstates of the Hamiltonian.
One diagonalizes the Hamiltonian in full space afterwards. The resonant eigenstate of interest is then that which bears the largest overlap with the eigenstate obtained at pole approximation level. This allows to uniquely determine the resonant eigenstates of the Hamiltonian, as the pole configurations are always dominant in practice compared to those associated to the non-resonant continuum. 

The GSM Hamiltonian matrix is diagonalized using the Jacobi-Davidson method \cite{Jacobi_Davidson}. The Jacobi-Davidson method targets the eigenstate closest to the used zeroth-order approximation, so that it converges quickly to the sought resonant state. In fact, the Jacobi-Davidson method in GSM replaces the Lanczos method of SM where only the eigenstates of lowest energies can converge quickly.
The GSM code has also been recently parallelized using the efficient two-dimensional partitioning algorithm \cite{MICHEL2020106978}, so that it can be efficiently used on powerful parallel machines.

\subsection{Hamiltonians and model spaces} \label{GSM_Hamiltonians}
The Hamiltonian and model space used in the calculation of the many-body nuclear states of carbon isotopes and isotones are very close to those used in Ref.\cite{PhysRevC.100.064303} in the context of proton-rich oxygen isotones, so that we will only describe its overall features.

We work in the frame of the core + valence nucleon picture. The considered core is $^{14}$C for carbon isotopes, whereas it is the mirror $^{14}$O for carbon isotones. 

A potential of Woods-Saxon (WS) type is used to mimic the core, which is fitted to the single-particle spectrum of $^{15}$C. The same WS potential is used for the $^{14}$O core, to which a Coulomb potential generated by the core charge density is added, which is taken of a Gaussian form for simplicity \cite{PhysRevC.83.034325}. The parameters of the WS potential are: $d=0.65$ fm for diffuseness, $R_0=2.98$ fm for radius, $V_0=$ 51.5 MeV ($\ell = 0$) or 49.75 MeV ($\ell > 0$) for central potential depth, and $V_{so}=$ 6.5 MeV for spin-orbit potential depth.

We take into account $spdf$ partial waves in the valence space, which are of proton (neutron) type for carbon isotones (isotopes). Partial waves of the $spd$ type are represented using the Berggren basis, whereas the HO states are used for the $f$ partial waves. This is justified by the large centrifugal barrier of the $f$ partial waves, whose effect on asymptotic many-body wave functions is negligible. The residual nucleon-nucleon interaction used is generated from EFT \cite{MACHLEIDT20111}, to which A-dependence is added to simulate missing three-body interactions \cite{PhysRevC.100.064303}. The Coulomb interaction is added when considering valence protons. 

The parameters of the used nuclear interaction and the method used to deal with the infinite-range of the Coulomb Hamiltonian are the same as those used in Ref.\cite{PhysRevC.100.064303}. Therefore, we refer the reader to that paper for details.

\section{Results}

\subsection{Energy spectrum and widths of carbon isotones}
\begin{figure*}[!htb]
\includegraphics[width=1.60\columnwidth]{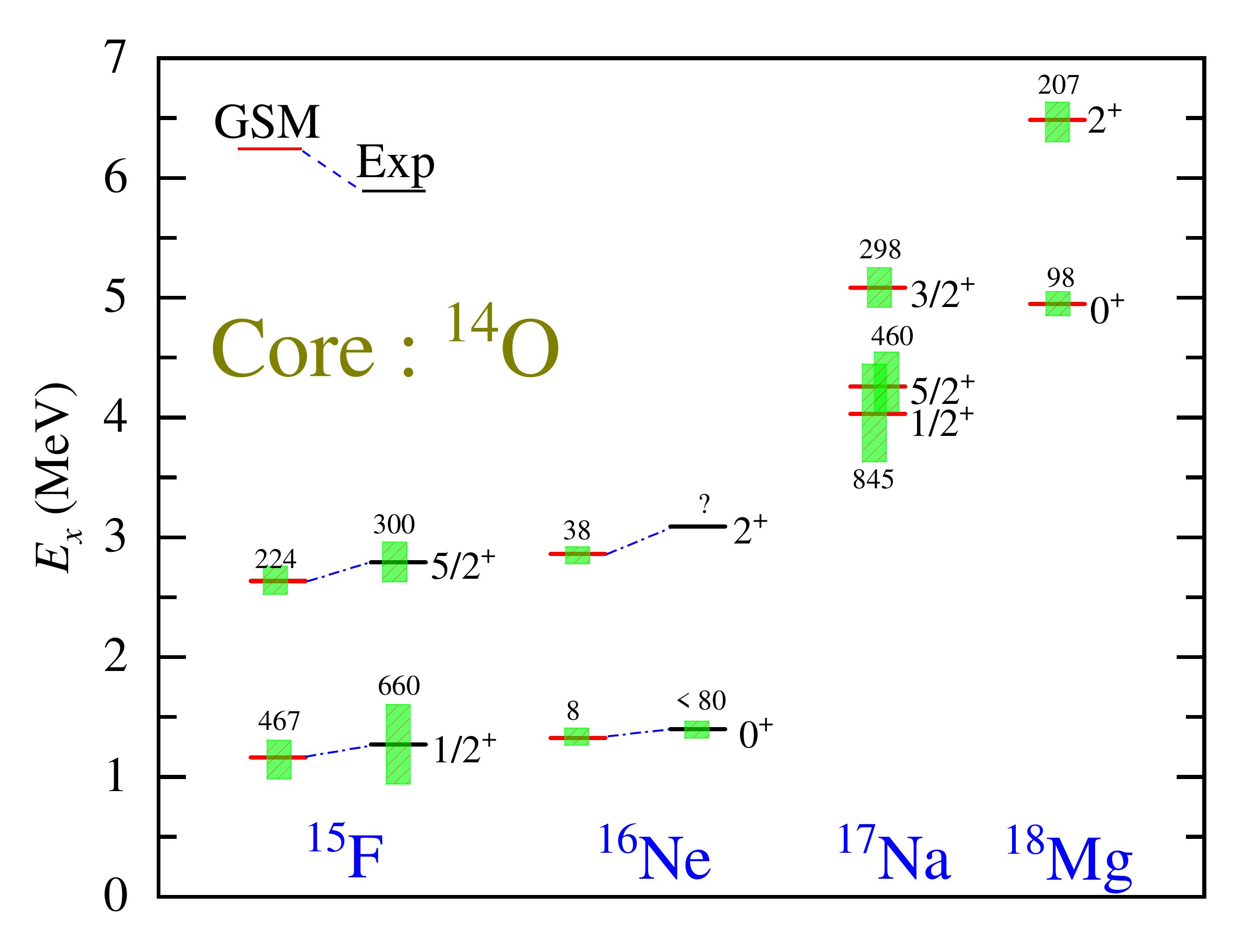}
\caption{Excitation energies ($E_x$, in MeV) and widths (in keV) of ground and excited states of carbon isotones. The GSM calculations are compared to available experimental data. Energies are given with respect to the $^{14}$O core. Widths are represented by green striped squares, and their explicit values are written above.  Experimental data of $^{15}$F are taken from Ref.\cite{ensdf} and those of $^{16}$Ne from Ref. \cite{PhysRevLett.113.232501}.}{\label{C_isotones_spectra}}
\end{figure*}

The obtained energy spectrum of carbon isotones is depicted in Fig. \ref{C_isotones_spectra}. One can see that both energies and widths of experimentally known eigenstates are well reproduced. 

The two first states of $^{15}$F have a single-particle character, i.e.,~the $1/2^+$ ground state and $5/2^+$ first excited state correspond to the proton $1s_{1/2}$ and $0d_{5/2}$ one-body states, respectively. Hence, as the $^{15}$C single-particle states have been fitted to their experimental data, one could expect their isobaric analog states to be close to experimental data as well. 

The experimental energies of the $0^+$ and $2^+$ states of $^{16}$Ne are well reproduced, even though these states were not fitted. Consequently, theoretical widths are fully predictive. One can see that the calculated width of the ground state of $^{16}$Ne is about 10 keV, which is smaller than the experimental value of 122 keV of Refs.\cite{PhysRevC.27.27,PhysRevC.17.1929}. However, our value is close to that obtained in the three-body model by Grigorenko \textit{et al.} \cite{PhysRevLett.88.042502}, where it is estimated to be in the interval of 0.15-3.1 keV. A width of a few keV at most is also supported by other experimental works where 80 keV is an upper bound \cite{PhysRevLett.113.232501}. Consequently, we also support a small value for the particle-emission width of $^{16}$Ne, of the order of 10 keV. Similarly, we obtain a small width for the $2^+$ first excited state of $^{16}$Ne, of about 40 keV, which is close to the values measured in the experimental studies of Refs.\cite{PhysRevLett.112.132502,PhysRevC.92.034329} and calculated in the theoretical analyses done in Refs.\cite{PhysRevC.96.064313,PhysRevLett.88.042502}.

One has no definitive experimental data for the ground and excited states of $^{17}$Na \cite{ensdf}. Indeed, even though $^{17}$Na has been synthesized, it might well be a combination of three different eigenstates \cite{ensdf}. Consequently, we can only compare our results to those arising from other models.
The energies and widths of the three first eigenstates of $^{17}$Na have been considered in an 
empirical potential model in Refs. \cite{PhysRevC.87.044315,PhysRevC.82.027310}. 
The energies and widths of the $1/2^+$ ground state and $5/2^+$ excited state of $^{17}$Na agree well with our results. In particular, the $1/2^+$ ground state width differs by only about a factor 2, as it is 845 kev in our model and 1.6 MeV in that of Ref.\cite{PhysRevC.82.027310}. However, the $3/2^+$ state is about 1 MeV higher in our calculations compared to those of Ref.\cite{PhysRevC.82.027310}, where it is only 170 keV above the $1/2^+$ ground state. Consequently, we also have different widths for the $3/2^+$ state, as it is about 300 keV in our model and 20 keV in that of Ref.\cite{PhysRevC.82.027310}.

As $^{18}$Mg has not been observed experimentally, only a comparison between our calculations and other theoretical estimates can be done. The binding energy of $^{18}$Mg has been calculated with an empirical potential model in Refs.\cite{PhysRevC.87.044315,PhysRevC.94.044305}. The obtained value is very close to ours as it differs by about 100 keV. However, those authors did not provide information of the width of $^{18}$Mg, except for a rough estimate of 9 keV. Added to that, the question of the identification of one-proton and two-proton widths has not yet been effected, as we concentrated on total widths for the moment. We will consider the partial emission widths associated to the one-proton and two-proton channels of $^{16}$Ne and $^{18}$Mg in Sec.(\ref{1p_2p_widths}).

\subsection{Isospin symmetry breaking in the mirroring states of carbon isotopes and isotones}
The Hamiltonians described in Sec.(\ref{GSM_Hamiltonians}) allow to describe both carbon isotopes or isotones. As they are mirroring nuclei, it is interesting to compare their energy spectra. Moreover, carbon isotopes of $A=$ 15-18 are well known experimentally \cite{ensdf}, so that they can give insight to the structure of the proton-rich carbon isotonic states which cannot be measured experimentally (see also Ref.\cite{PhysRevLett.113.142502} for an \textit{ab initio} calculation of carbon isotopes).

\begin{figure*}[!htb]
\includegraphics[width=1.7\columnwidth]{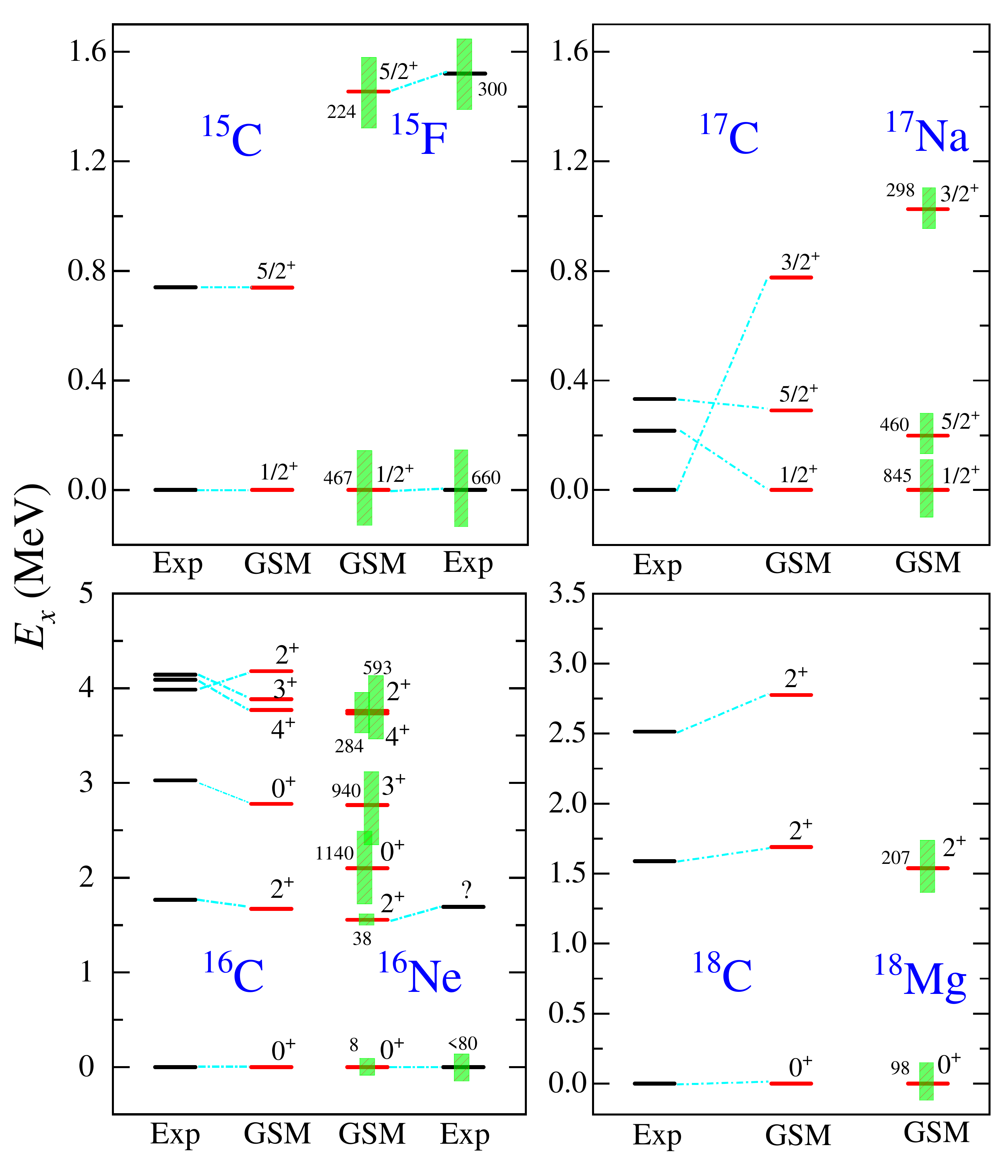}
\caption{Comparison of the excitation energies and widths of mirroring nuclear states of carbon isotones and isotopes. Excitation energies of a nucleus are given with respect to its ground state energy.  Experimental data of $^{16}$Ne are taken from Ref.\cite{PhysRevLett.113.232501}, whereas all other data are taken from Ref.\cite{ensdf}. See Fig. \ref{C_isotones_spectra} for details and notations.}{\label{C_isotopes_isotones_spectra}}
\end{figure*}

The GSM calculations of carbon isotopes and isotones of $A=$ 15-18 are illustrated in Fig. \ref{C_isotopes_isotones_spectra}. 
One can see that the spectra of carbon isotopes are well described, except for the $3/2^+$ excited state of $^{17}$C, which is too high by 800 keV. Consequently, its mirroring state in $^{17}$Na is presumably too high and too wide in energy and width as well. Conversely, all considered states
of the spectra of $^{16}$C and $^{18}$C reproduce experimental data properly, as theoretical and experimental energies differ by a few hundreds of keV at most. Therefore, the calculated excited states of $^{16}$Ne and $^{18}$Mg should be close to experiment.  

All the many-body wave functions of carbon isotopes and isotones have an exact isospin quantum number. This is the case because they only have valence neutrons or protons, respectively, i.e.~they are isospin aligned (see Ref.\cite{PhysRevC.82.044315} where a similar situation had been encountered in $A=6$ systems). Moreover, one can consider that they are generated by the same Hamiltonian, even though one uses different cores for carbon isotopes and isotones. Indeed, it is equivalent to consider either two different Hamiltonians for carbon isotopes and isotones, which differ only by way of the Coulomb Hamiltonian, or a single Hamiltonian built from the same nuclear and Coulomb interactions as in Sec.(\ref{GSM_Hamiltonians}), but defined with a $^{12}$C core, where nucleons in the $0p_{1/2}$ shells are only subject to the WS potential of Sec.(\ref{GSM_Hamiltonians}). Due to the presence or absence of Coulomb Hamiltonian in carbon isotones and isotopes, respectively, energies of mirror states are different. Thus, this shows the presence of partial dynamical symmetry of isospin \cite{LEVIATAN201193,PhysRevLett.89.222501}. Indeed, one has $[\hat{H},\hat{T}^2] \neq 0$ because of the presence of the Coulomb Hamiltonian, whereas $[\hat{H},\hat{T}^2]\ket{\Psi} = 0$ for a many-body wave function $\ket{\Psi}$ of the considered carbon isotopes and isotones, because it is isospin aligned. Consequently, it is necessary to consider other operators instead of $\hat{T}^2$ to assess isospin-symmetry breaking. Therefore, we will concentrate in the following on the energy shifts between mirror nuclear states, and also on the Coulomb contribution, as the latter are maximal in carbon isotones and nonexistent in carbon isotopes.

The energy shift associated to isospin-symmetry breaking is the Thomas-Ehrman shift \cite{PhysRev.81.412,PhysRev.88.1109}. It is particularly strong in the presence of many-body nuclear resonances. One can clearly see its effect in Fig. \ref{C_isotopes_isotones_spectra}. Indeed, in the absence of the Coulomb interaction, spectra would be identical. The Thomas-Ehrman shift is generated by the different asymptotes of the many-body wave functions of the carbon isotopes and isotones. Consequently, it can be expected to be small when the considered states are bound or narrow, and, conversely, to be the largest in the presence of broad resonance states. Isotonic resonance states of smallest widths are the $0^+$ and $2^+$ states of $^{16}$Ne and $^{18}$Mg. As a consequence, the Thomas-Ehrman shift is very mild for these states, as it is around 150 keV. Conversely, the Thomas-Ehrman shift is the strongest in $A=15$ nuclei and in the highest excited states of $A=16$ nuclei, where widths are typically larger than 500 keV. In this case, the Thomas-Ehrman shift is typically of 500 keV to 1 MeV. Interestingly, the $A=17$ nuclei do not present a very large Thomas-Ehrman shift, even though their widths are of the same order of magnitude. Indeed, the largest energy difference therein is about 200 keV only. One then sees two effects competing in the generation of the Thomas-Ehrman shift. On the one hand, proton-rich nuclear states must be broad resonances for Thomas-Ehrman shifts to be the largest, as the asymptotes of the many-body wave functions of mirror nuclei are very different. On the other hand, continuum coupling strongly influences the Thomas-Ehrman shift, so that it can change in two different pairs of nuclear states, even though the nuclear excitation energies and widths of these two pairs of nuclei are similar (see Ref.\cite{PhysRevC.102.024309} for an analogous study with Li isotopes).

\begin{figure*}[!htb]
\includegraphics[width=1.30\columnwidth]{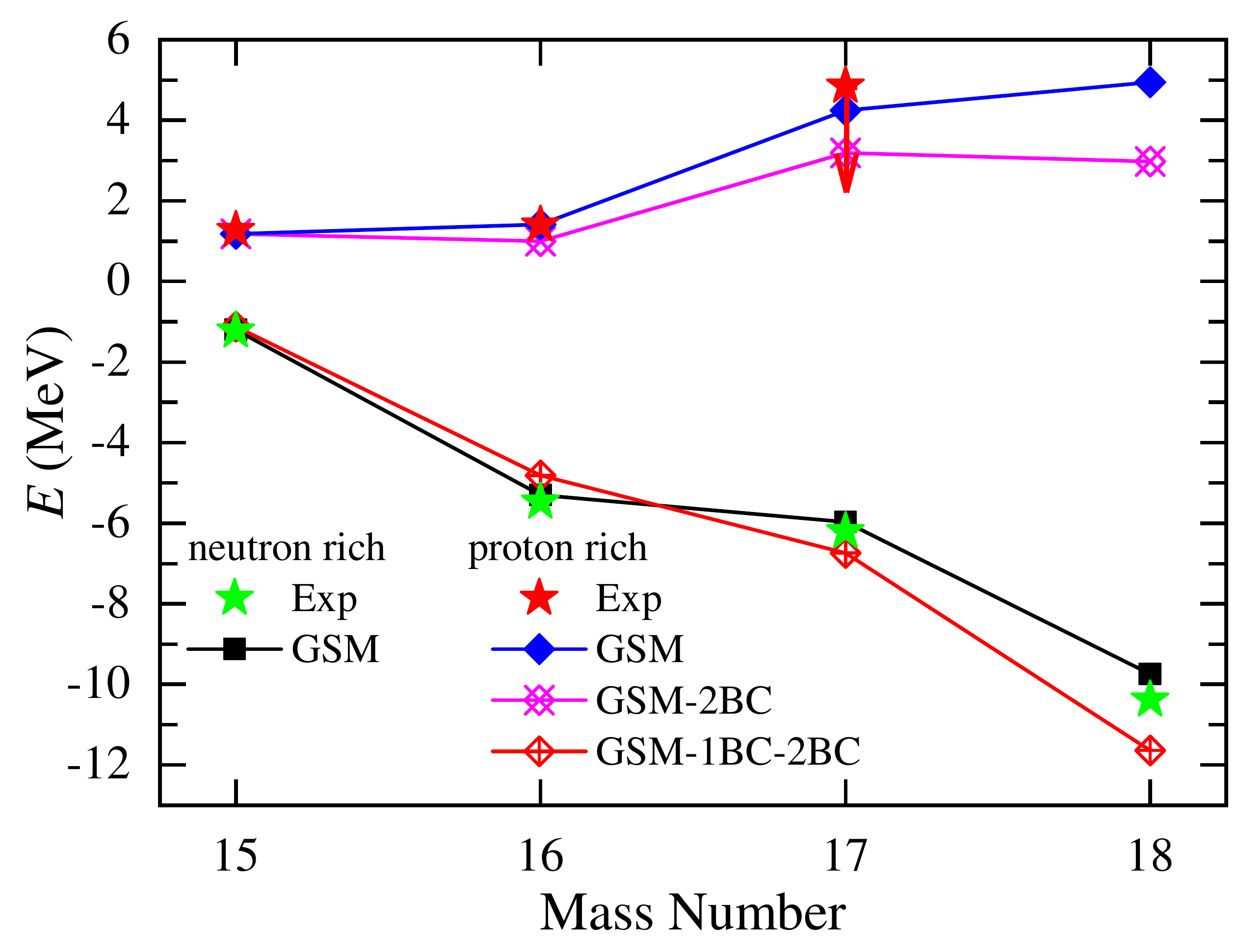}
\caption{Calculated binding energies of carbon isotopes and isotones (GSM, black squares and blue lozenges, respectively), binding energies of carbon isotones minus Coulomb two-body part contribution (GSM-2BC, purple crosses) and binding energies of carbon isotones minus one- and two-body Coulomb total contributions (GSM-1BC-2BC, red lozenges), as a function of mass number $A$. Experimental data are represented by green and red stars for carbon isotopes and isotones, respectively. All energies ($E$, in MeV) are given with respect to $^{14}$C and $^{14}$O cores for carbon isotopes and isotones, respectively. The arrow next to the ground state energy of $^{17}$Na indicates that the experimental datum is an upper limit.}{\label{Coulomb_contribution}}
\end{figure*}

Another observable measuring isospin-symmetry breaking is the Coulomb contribution to the binding energy. Indeed, as shown in Ref.\cite{PhysRevC.100.064303} in the context of proton-rich oxygen isotones, it strongly varies with many-body wave function asymptotes. Consequently, the effect of the one-body and two-body contributions of the Coulomb Hamiltonian in the binding energies of carbon isotones has been considered (see Fig. \ref{Coulomb_contribution}).
For this, one firstly suppresses the two-body Coulomb energy of carbon isotone binding energies, to see the effect of the Coulomb correlations (GSM-2BC in Fig. \ref{Coulomb_contribution}). Then, the one-body part of Coulomb energy is removed from the previous value (GSM-1BC-2BC in Fig. \ref{Coulomb_contribution}). This also provides the expectation value of the nuclear strong interaction Hamiltonian (i.e., excluding the Coulomb interaction). Hence, if the many-body wave functions of carbon isotones and isotopes of the same isospin multiplet had  the same configuration mixing, the obtained energies would be exactly that of carbon isotopes where the valence particles are only neutrons without the Coulomb interaction. The energy differences for the $A=15$ carbon mirror nuclei, which are represented by one-nucleon wave functions in the presence of a WS potential, are almost identical. Interestingly, the expectation value of the nuclear part of the Hamiltonian (i.e., excluding the Coulomb interaction) in $^{16}$Ne is slightly smaller in absolute value than that of $^{16}$C, while it is the opposite in $A=17,18$ carbon mirror nuclei (see Fig. \ref{Coulomb_contribution}). Added to that, the nuclear Hamiltonian expectation value increases in absolute value from $A=17$ to $A=18$. This behavior is, in fact, unexpected.  Indeed, one would expect energy difference to vary monotonously with proton-emission width, as isospin-symmetry breaking is the largest in this case (see Fig. \ref{Coulomb_contribution}). On the contrary, one sees almost no difference in $A=15$ carbon mirror nuclei, comparable differences of opposite signs in $A=16,17$ cases, and the largest difference in $A=18$ (see Fig. \ref{Coulomb_contribution}).

This study has then shown the subtle effects induced by isospin-symmetry breaking in mirror nuclei of bound and unbound nature. Indeed, considerations based on separation energy and width are not sufficient to assess the Coulomb contribution on proton-rich nuclei. The inter-dependence of nucleon-nucleon correlations and continuum coupling generates complex configuration mixing and thus sometimes counter-intuitive behavior of observables.

\subsection{Determination of the one-proton and two-proton decay widths of $^{16}$Ne and $^{18}$Mg in GSM} \label{1p_2p_widths}
The $^{16}$Ne and $^{18}$Mg isotopes are proton-rich unbound nuclei \cite{ensdf}. Added to that, both one-proton and two-proton separation energies are negative, so that two different particle-emission channels are open therein. Consequently, it is necessary to calculate separately one-proton and two-proton decay widths in order to delineate the detailed structure of the ground states of $^{16}$Ne and $^{18}$Mg. However, as particle-emission width is obtained from the imaginary part of the eigenenergy in GSM, one has solely access, in principle, to the total emission width.

Nevertheless, due to the special structure of $^{16}$Ne and $^{18}$Mg many-body wave functions, we have been able to indirectly determine both one-proton and two-proton decay widths therein. For this, we use the fact that one has only two open channels, on the one hand, and that their particle-emission thresholds vary differently as a function of the Hamiltonian parameters, on the other hand. 

\begin{figure*}[!htb]
\includegraphics[width=1.4\columnwidth]{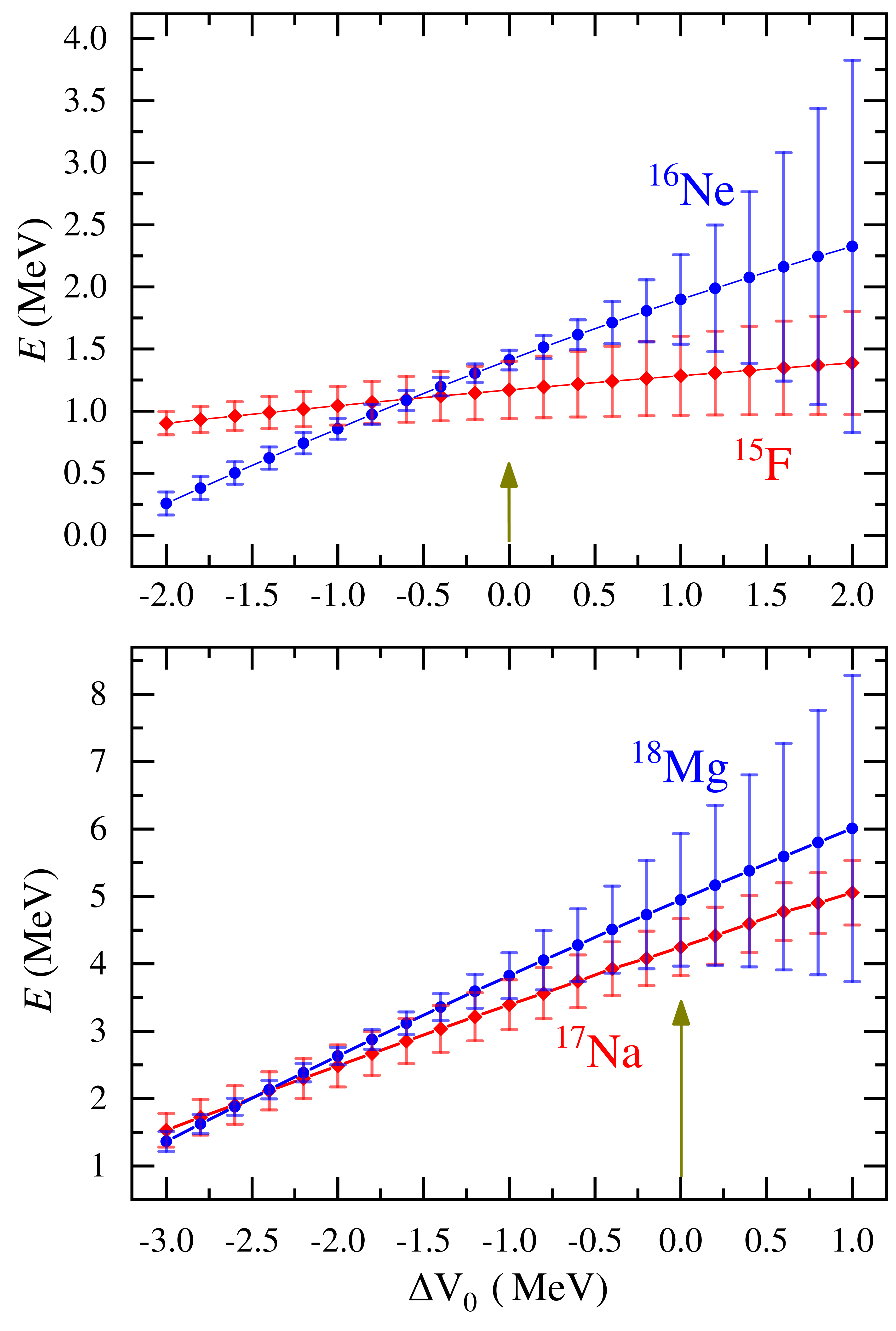}
\caption{Calculated energies and widths (in MeV) of $^{15}$F and $^{16}$Ne (upper panel) and of $^{17}$Na and $^{16}$Mg (lower panel) as a function of the difference $\Delta V_0 = V_0 - V_0^{\text{(fit)}}$ of the Woods-Saxon central potential depths (see text for definition). Energies are depicted by blue disks and red lozenges for even and odd nuclei, respectively. Widths are represented by segments centered on disks and lozenges. The widths of $^{16}$Ne and $^{18}$Mg have been multiplied by 20 for readability.
Energies are given with respect to the $^{14}$O core. The physical GSM calculation, for which $V_0 = V_0^{\text{(fit)}}$, is indicated by an arrow.}{\label{widths}}
\end{figure*}

Thus, in order to evaluate one-proton and two-proton decay widths, we changed the central potential depth $V_0$ of the WS core potential, in order for the one-proton separation energy to become positive or very negative.
The Woods-Saxon central potential depth fitted by experimental data (see Sec.(\ref{GSM_Hamiltonians})) will be denoted as $V_0^{\text{(fit)}}$ in this section, as it is used only for comparison with other calculations. The results are shown in Fig. \ref{widths}. 
As $^{16}$Ne and $^{18}$Mg possess one additional valence proton compared to $^{15}$F and $^{17}$Ne, the binding energy of the former nuclei increases faster with central potential depth than that of the latter. Consequently, it is possible to find a central potential depth for which only the two-proton decay channel is open, so that the obtained width is that of two-proton emission. 

The increase of width above one-proton emission threshold is of one-proton type only, which can be noticed from its exponential increase as a function of one-proton separation energy. The width of $A=16,18$ carbon mirror nuclei can also be compared to the widths of $A=15,17$ cases in Fig. \ref{widths}. As $A=15,17$ carbon mirror nuclei are one-proton resonances, their width increases steadily with the Hamiltonian central potential depth. On the contrary, one can see that the widths of $A=16,18$ carbon mirrors increase abruptly when the one-proton channel opens, which points out the different asymptotes of $A=16,18$ ground-state wave functions before and after one-proton emission threshold.

The two-proton decay width is almost constant with respect to the central potential depth below one-proton emission threshold and also about 500 keV to 1 MeV above (see Fig. \ref{widths}). It is thus reasonable to assume that two-proton decay width is almost independent of energy. Therefore, the value obtained in Fig. \ref{widths}, where only the two-proton channel is open, can be extrapolated to the physical case, i.e.~when $V_0 = V_0^{\text{(fit)}}$ (indicated by an arrow in Fig. \ref{widths}). This two-proton decay width is about 10-15 keV for both $^{16}$Ne and $^{18}$Mg nuclei.  As one only has two emission channels, the one-proton width is the difference between total width and the two-proton emission width of 10-15 keV. One-proton emission is negligible for $^{16}$Ne (see upper panel of Fig. \ref{widths}). Conversely, that of $^{18}$Mg is then estimated to be about 85-90 keV (see lower panel of of Fig. \ref{widths} and Fig. \ref{C_isotones_spectra}). The obtained values are in agreement with current experimental data \cite{PhysRevC.27.27,PhysRevC.17.1929,PhysRevLett.113.232501,ensdf}. 

As a consequence, despite the fact that GSM can only provide total emission widths, where all partial widths are summed, we have been able to determine both one-proton and two-proton widths of $^{16}$Ne and $^{18}$Mg. For this, one has only assumed that two-proton width can be considered as independent of energy. This is justified from two grounds. On the one hand, two-proton width is almost constant below or close to one-proton emission threshold. On the other hand, total width minus two-proton width, with the latter considered as constant, behaves as a one-proton width for separation energies above one-proton emission threshold.

\section {Summary} 

Due to its exact treatment of continuum coupling and inter-nucleon correlations, GSM is the tool of choice to study proton and neutron drip-lines. Consequently, it can help answering questions related to isotopes whose experimental study is difficult. This is the case for the $^{16}$Ne and $^{18}$Mg nuclei. Indeed, while the binding energy and decay pattern of $^{16}$Ne have been determined experimentally, the value of its two-proton emission width is only known to be smaller than 80 keV \cite{ensdf}. The situation is even more delicate for $^{18}$Mg, as it has not even been observed \cite{ensdf}. 

Consequently, in order to better understand the structure of the $^{16}$Ne and $^{18}$Mg nuclei, we performed GSM calculations of carbon isotones of $A=15\text{-}18$. These isotones are convenient to study with EFT, as they consist of a few valence protons above a $^{14}$O core. Moreover, as the EFT interaction used was fitted in another context, that of oxygen isotones, the resulted GSM calculations are predictive. 

We obtained spectra of carbon isotones reproducing known experimental data. Due to the mirror symmetry between carbon isotones and isotopes, the calculation of their spectra could be realized by using the same Hamiltonian and exchanging proton states with neutron states. The eigenstates of carbon isotopes are also well reproduced, except maybe for the $3/2^+$ eigenstate of $^{17}$Na, which is too high by 800 keV when comparing the GSM energy to that issued from other models. Consequently, except for this latter state, one can assume that the excited states of carbon isotones are close to experimental data, so that their obtained energies and widths can guide experimentalists in future experiments.

Isospin-symmetry breaking can be studied from observables associated to the mirror eigenstates of carbon isotones and isotopes. This situation is particularly interesting as one is in the presence of partial dynamical symmetry, where eigenstates possess a conserved quantum number, even though its associated operator does not commute with the Hamiltonian. The largest Thomas-Ehrman shifts were noticed when resonances of fairly large widths are present in isobaric multiplets, which can be expected. However,  Thomas-Ehrman shift is not a simple function of width, as different values occur for eigenstates of similar widths. The effect of isospin-symmetry breaking was also studied by considering the Coulomb Hamiltonian observables. For this, one suppressed the Coulomb contribution of proton-rich nuclear eigenenergies, as one would then obtain the same binding energy as those of carbon isotopes in the case of exact isospin symmetry. It was there seen that the energy differences between the nuclear Hamiltonian expectation values of carbon isotones and isotopes follow an uneven pattern, where it can be positive or negative. As a consequence, due to the presence of nucleon-nucleon correlations induced by the Coulomb and nuclear parts of the Hamiltonian, on the one hand, and of a large continuum coupling, on the other hand, isospin-symmetry breaking leads to complex effects in drip-line nuclei, whose effects cannot be predicted from simple grounds.

After studying the overall features of the isotopic and isotonic carbon chains, we proceeded to the prediction of one-proton and two-proton widths of $^{16}$Ne and $^{18}$Mg. Even though an exact separation of one-proton and two-proton widths is impossible to do in GSM, we could devise a method allowing to have an estimate of these two widths in $^{16}$Ne and $^{18}$Mg from an extrapolation method based on reasonable physical arguments. The two-proton width of $^{16}$Ne obtained with GSM is of the same order of magnitude as that arising from other models, albeit slightly larger, as it is about 10-15 keV. Added to that, in agreement with experimental data, one-proton emission is absent from $^{16}$Ne in GSM calculations. We could also provide, up to our knowledge, the first theoretical calculation of the one-proton and two-proton emission widths of $^{18}$Mg, which are respectively about 85-90 keV and 10-15 keV. This suggests that it is possible to detect $^{18}$Mg experimentally, on the one hand, and to devise an estimate of its one-proton and two-proton decay widths from experimental analysis, on the other hand.

\begin{acknowledgments}

\end{acknowledgments}
This work has been supported by the Strategic Priority Research Program of Chinese Academy of Sciences under Grant No. XDB34000000; the National Natural Science Foundation of China under Grants No. 11975282, 11835001,  11921006, and  12035001; the National Key R\&D Program of China under Grant No. 2018YFA0404401; the State Key Laboratory of Nuclear Physics and Technology, Peking University under Grant No. NPT2020KFY13; and the CUSTIPEN (China-U.S. Theory Institute for Physics with Exotic Nuclei) funded by the U.S. Department of Energy, Office of Science under Grant No. de-sc0009971. We acknowledge the High-Performance Computing Platform of Peking University for providing computational resources.

\bibliography{Ref}

\begin{thebibliography}{79}%
\makeatletter
\providecommand \@ifxundefined [1]{%
 \@ifx{#1\undefined}
}%
\providecommand \@ifnum [1]{%
 \ifnum #1\expandafter \@firstoftwo
 \else \expandafter \@secondoftwo
 \fi
}%
\providecommand \@ifx [1]{%
 \ifx #1\expandafter \@firstoftwo
 \else \expandafter \@secondoftwo
 \fi
}%
\providecommand \natexlab [1]{#1}%
\providecommand \enquote  [1]{``#1''}%
\providecommand \bibnamefont  [1]{#1}%
\providecommand \bibfnamefont [1]{#1}%
\providecommand \citenamefont [1]{#1}%
\providecommand \href@noop [0]{\@secondoftwo}%
\providecommand \href [0]{\begingroup \@sanitize@url \@href}%
\providecommand \@href[1]{\@@startlink{#1}\@@href}%
\providecommand \@@href[1]{\endgroup#1\@@endlink}%
\providecommand \@sanitize@url [0]{\catcode `\\12\catcode `\$12\catcode
  `\&12\catcode `\#12\catcode `\^12\catcode `\_12\catcode `\%12\relax}%
\providecommand \@@startlink[1]{}%
\providecommand \@@endlink[0]{}%
\providecommand \url  [0]{\begingroup\@sanitize@url \@url }%
\providecommand \@url [1]{\endgroup\@href {#1}{\urlprefix }}%
\providecommand \urlprefix  [0]{URL }%
\providecommand \Eprint [0]{\href }%
\providecommand \doibase [0]{http://dx.doi.org/}%
\providecommand \selectlanguage [0]{\@gobble}%
\providecommand \bibinfo  [0]{\@secondoftwo}%
\providecommand \bibfield  [0]{\@secondoftwo}%
\providecommand \translation [1]{[#1]}%
\providecommand \BibitemOpen [0]{}%
\providecommand \bibitemStop [0]{}%
\providecommand \bibitemNoStop [0]{.\EOS\space}%
\providecommand \EOS [0]{\spacefactor3000\relax}%
\providecommand \BibitemShut  [1]{\csname bibitem#1\endcsname}%
\let\auto@bib@innerbib\@empty
\bibitem [{\citenamefont {Al-Khalili}(2004)}]{Al-Khalili2004}%
  \BibitemOpen
  \bibfield  {author} {\bibinfo {author} {\bibfnamefont {J.}~\bibnamefont
  {Al-Khalili}},\ }\enquote {\bibinfo {title} {An introduction to halo
  nuclei},}\ in\ \href {\doibase 10.1007/978-3-540-44490-9_3} {\emph {\bibinfo
  {booktitle} {The Euroschool Lectures on Physics with Exotic Beams, Vol.
  I}}},\ \bibinfo {editor} {edited by\ \bibinfo {editor} {\bibfnamefont
  {J.}~\bibnamefont {Al-Khalili}}\ and\ \bibinfo {editor} {\bibfnamefont
  {E.}~\bibnamefont {Roeckl}}}\ (\bibinfo  {publisher} {Springer Berlin
  Heidelberg},\ \bibinfo {address} {Berlin, Heidelberg},\ \bibinfo {year}
  {2004})\ pp.\ \bibinfo {pages} {77--112}\BibitemShut {NoStop}%
\bibitem [{\citenamefont {Tanihata}\ \emph {et~al.}(1985)\citenamefont
  {Tanihata}, \citenamefont {Hamagaki}, \citenamefont {Hashimoto},
  \citenamefont {Shida}, \citenamefont {Yoshikawa}, \citenamefont {Sugimoto},
  \citenamefont {Yamakawa}, \citenamefont {Kobayashi},\ and\ \citenamefont
  {Takahashi}}]{PhysRevLett.55.2676}%
  \BibitemOpen
  \bibfield  {author} {\bibinfo {author} {\bibfnamefont {I.}~\bibnamefont
  {Tanihata}}, \bibinfo {author} {\bibfnamefont {H.}~\bibnamefont {Hamagaki}},
  \bibinfo {author} {\bibfnamefont {O.}~\bibnamefont {Hashimoto}}, \bibinfo
  {author} {\bibfnamefont {Y.}~\bibnamefont {Shida}}, \bibinfo {author}
  {\bibfnamefont {N.}~\bibnamefont {Yoshikawa}}, \bibinfo {author}
  {\bibfnamefont {K.}~\bibnamefont {Sugimoto}}, \bibinfo {author}
  {\bibfnamefont {O.}~\bibnamefont {Yamakawa}}, \bibinfo {author}
  {\bibfnamefont {T.}~\bibnamefont {Kobayashi}}, \ and\ \bibinfo {author}
  {\bibfnamefont {N.}~\bibnamefont {Takahashi}},\ }\href {\doibase
  10.1103/PhysRevLett.55.2676} {\bibfield  {journal} {\bibinfo  {journal}
  {Phys. Rev. Lett.}\ }\textbf {\bibinfo {volume} {55}},\ \bibinfo {pages}
  {2676} (\bibinfo {year} {1985})}\BibitemShut {NoStop}%
\bibitem [{\citenamefont {Mueller}\ \emph {et~al.}(2007)\citenamefont
  {Mueller}, \citenamefont {Sulai}, \citenamefont {Villari}, \citenamefont
  {Alc\'antara-N\'u\~nez}, \citenamefont {Alves-Cond\'e}, \citenamefont
  {Bailey}, \citenamefont {Drake}, \citenamefont {Dubois}, \citenamefont
  {El\'eon}, \citenamefont {Gaubert}, \citenamefont {Holt}, \citenamefont
  {Janssens}, \citenamefont {Lecesne}, \citenamefont {Lu}, \citenamefont
  {O'Connor}, \citenamefont {Saint-Laurent}, \citenamefont {Thomas},\ and\
  \citenamefont {Wang}}]{PhysRevLett.99.252501}%
  \BibitemOpen
  \bibfield  {author} {\bibinfo {author} {\bibfnamefont {P.}~\bibnamefont
  {Mueller}}, \bibinfo {author} {\bibfnamefont {I.~A.}\ \bibnamefont {Sulai}},
  \bibinfo {author} {\bibfnamefont {A.~C.~C.}\ \bibnamefont {Villari}},
  \bibinfo {author} {\bibfnamefont {J.~A.}\ \bibnamefont
  {Alc\'antara-N\'u\~nez}}, \bibinfo {author} {\bibfnamefont {R.}~\bibnamefont
  {Alves-Cond\'e}}, \bibinfo {author} {\bibfnamefont {K.}~\bibnamefont
  {Bailey}}, \bibinfo {author} {\bibfnamefont {G.~W.~F.}\ \bibnamefont
  {Drake}}, \bibinfo {author} {\bibfnamefont {M.}~\bibnamefont {Dubois}},
  \bibinfo {author} {\bibfnamefont {C.}~\bibnamefont {El\'eon}}, \bibinfo
  {author} {\bibfnamefont {G.}~\bibnamefont {Gaubert}}, \bibinfo {author}
  {\bibfnamefont {R.~J.}\ \bibnamefont {Holt}}, \bibinfo {author}
  {\bibfnamefont {R.~V.~F.}\ \bibnamefont {Janssens}}, \bibinfo {author}
  {\bibfnamefont {N.}~\bibnamefont {Lecesne}}, \bibinfo {author} {\bibfnamefont
  {Z.-T.}\ \bibnamefont {Lu}}, \bibinfo {author} {\bibfnamefont {T.~P.}\
  \bibnamefont {O'Connor}}, \bibinfo {author} {\bibfnamefont {M.-G.}\
  \bibnamefont {Saint-Laurent}}, \bibinfo {author} {\bibfnamefont {J.-C.}\
  \bibnamefont {Thomas}}, \ and\ \bibinfo {author} {\bibfnamefont {L.-B.}\
  \bibnamefont {Wang}},\ }\href {\doibase 10.1103/PhysRevLett.99.252501}
  {\bibfield  {journal} {\bibinfo  {journal} {Phys. Rev. Lett.}\ }\textbf
  {\bibinfo {volume} {99}},\ \bibinfo {pages} {252501} (\bibinfo {year}
  {2007})}\BibitemShut {NoStop}%
\bibitem [{\citenamefont {Nakamura}\ \emph {et~al.}(2014)\citenamefont
  {Nakamura}, \citenamefont {Kobayashi}, \citenamefont {Kondo}, \citenamefont
  {Satou}, \citenamefont {Tostevin}, \citenamefont {Utsuno}, \citenamefont
  {Aoi}, \citenamefont {Baba}, \citenamefont {Fukuda}, \citenamefont {Gibelin},
  \citenamefont {Inabe}, \citenamefont {Ishihara}, \citenamefont {Kameda},
  \citenamefont {Kubo}, \citenamefont {Motobayashi}, \citenamefont {Ohnishi},
  \citenamefont {Orr}, \citenamefont {Otsu}, \citenamefont {Otsuka},
  \citenamefont {Sakurai}, \citenamefont {Sumikama}, \citenamefont {Takeda},
  \citenamefont {Takeshita}, \citenamefont {Takechi}, \citenamefont {Takeuchi},
  \citenamefont {Togano},\ and\ \citenamefont
  {Yoneda}}]{PhysRevLett.112.142501}%
  \BibitemOpen
  \bibfield  {author} {\bibinfo {author} {\bibfnamefont {T.}~\bibnamefont
  {Nakamura}}, \bibinfo {author} {\bibfnamefont {N.}~\bibnamefont {Kobayashi}},
  \bibinfo {author} {\bibfnamefont {Y.}~\bibnamefont {Kondo}}, \bibinfo
  {author} {\bibfnamefont {Y.}~\bibnamefont {Satou}}, \bibinfo {author}
  {\bibfnamefont {J.~A.}\ \bibnamefont {Tostevin}}, \bibinfo {author}
  {\bibfnamefont {Y.}~\bibnamefont {Utsuno}}, \bibinfo {author} {\bibfnamefont
  {N.}~\bibnamefont {Aoi}}, \bibinfo {author} {\bibfnamefont {H.}~\bibnamefont
  {Baba}}, \bibinfo {author} {\bibfnamefont {N.}~\bibnamefont {Fukuda}},
  \bibinfo {author} {\bibfnamefont {J.}~\bibnamefont {Gibelin}}, \bibinfo
  {author} {\bibfnamefont {N.}~\bibnamefont {Inabe}}, \bibinfo {author}
  {\bibfnamefont {M.}~\bibnamefont {Ishihara}}, \bibinfo {author}
  {\bibfnamefont {D.}~\bibnamefont {Kameda}}, \bibinfo {author} {\bibfnamefont
  {T.}~\bibnamefont {Kubo}}, \bibinfo {author} {\bibfnamefont {T.}~\bibnamefont
  {Motobayashi}}, \bibinfo {author} {\bibfnamefont {T.}~\bibnamefont
  {Ohnishi}}, \bibinfo {author} {\bibfnamefont {N.~A.}\ \bibnamefont {Orr}},
  \bibinfo {author} {\bibfnamefont {H.}~\bibnamefont {Otsu}}, \bibinfo {author}
  {\bibfnamefont {T.}~\bibnamefont {Otsuka}}, \bibinfo {author} {\bibfnamefont
  {H.}~\bibnamefont {Sakurai}}, \bibinfo {author} {\bibfnamefont
  {T.}~\bibnamefont {Sumikama}}, \bibinfo {author} {\bibfnamefont
  {H.}~\bibnamefont {Takeda}}, \bibinfo {author} {\bibfnamefont
  {E.}~\bibnamefont {Takeshita}}, \bibinfo {author} {\bibfnamefont
  {M.}~\bibnamefont {Takechi}}, \bibinfo {author} {\bibfnamefont
  {S.}~\bibnamefont {Takeuchi}}, \bibinfo {author} {\bibfnamefont
  {Y.}~\bibnamefont {Togano}}, \ and\ \bibinfo {author} {\bibfnamefont
  {K.}~\bibnamefont {Yoneda}},\ }\href {\doibase
  10.1103/PhysRevLett.112.142501} {\bibfield  {journal} {\bibinfo  {journal}
  {Phys. Rev. Lett.}\ }\textbf {\bibinfo {volume} {112}},\ \bibinfo {pages}
  {142501} (\bibinfo {year} {2014})}\BibitemShut {NoStop}%
\bibitem [{\citenamefont {Kobayashi}\ \emph {et~al.}(2014)\citenamefont
  {Kobayashi}, \citenamefont {Nakamura}, \citenamefont {Kondo}, \citenamefont
  {Tostevin}, \citenamefont {Utsuno}, \citenamefont {Aoi}, \citenamefont
  {Baba}, \citenamefont {Barthelemy}, \citenamefont {Famiano}, \citenamefont
  {Fukuda}, \citenamefont {Inabe}, \citenamefont {Ishihara}, \citenamefont
  {Kanungo}, \citenamefont {Kim}, \citenamefont {Kubo}, \citenamefont {Lee},
  \citenamefont {Lee}, \citenamefont {Matsushita}, \citenamefont {Motobayashi},
  \citenamefont {Ohnishi}, \citenamefont {Orr}, \citenamefont {Otsu},
  \citenamefont {Otsuka}, \citenamefont {Sako}, \citenamefont {Sakurai},
  \citenamefont {Satou}, \citenamefont {Sumikama}, \citenamefont {Takeda},
  \citenamefont {Takeuchi}, \citenamefont {Tanaka}, \citenamefont {Togano},\
  and\ \citenamefont {Yoneda}}]{PhysRevLett.112.242501}%
  \BibitemOpen
  \bibfield  {author} {\bibinfo {author} {\bibfnamefont {N.}~\bibnamefont
  {Kobayashi}}, \bibinfo {author} {\bibfnamefont {T.}~\bibnamefont {Nakamura}},
  \bibinfo {author} {\bibfnamefont {Y.}~\bibnamefont {Kondo}}, \bibinfo
  {author} {\bibfnamefont {J.~A.}\ \bibnamefont {Tostevin}}, \bibinfo {author}
  {\bibfnamefont {Y.}~\bibnamefont {Utsuno}}, \bibinfo {author} {\bibfnamefont
  {N.}~\bibnamefont {Aoi}}, \bibinfo {author} {\bibfnamefont {H.}~\bibnamefont
  {Baba}}, \bibinfo {author} {\bibfnamefont {R.}~\bibnamefont {Barthelemy}},
  \bibinfo {author} {\bibfnamefont {M.~A.}\ \bibnamefont {Famiano}}, \bibinfo
  {author} {\bibfnamefont {N.}~\bibnamefont {Fukuda}}, \bibinfo {author}
  {\bibfnamefont {N.}~\bibnamefont {Inabe}}, \bibinfo {author} {\bibfnamefont
  {M.}~\bibnamefont {Ishihara}}, \bibinfo {author} {\bibfnamefont
  {R.}~\bibnamefont {Kanungo}}, \bibinfo {author} {\bibfnamefont
  {S.}~\bibnamefont {Kim}}, \bibinfo {author} {\bibfnamefont {T.}~\bibnamefont
  {Kubo}}, \bibinfo {author} {\bibfnamefont {G.~S.}\ \bibnamefont {Lee}},
  \bibinfo {author} {\bibfnamefont {H.~S.}\ \bibnamefont {Lee}}, \bibinfo
  {author} {\bibfnamefont {M.}~\bibnamefont {Matsushita}}, \bibinfo {author}
  {\bibfnamefont {T.}~\bibnamefont {Motobayashi}}, \bibinfo {author}
  {\bibfnamefont {T.}~\bibnamefont {Ohnishi}}, \bibinfo {author} {\bibfnamefont
  {N.~A.}\ \bibnamefont {Orr}}, \bibinfo {author} {\bibfnamefont
  {H.}~\bibnamefont {Otsu}}, \bibinfo {author} {\bibfnamefont {T.}~\bibnamefont
  {Otsuka}}, \bibinfo {author} {\bibfnamefont {T.}~\bibnamefont {Sako}},
  \bibinfo {author} {\bibfnamefont {H.}~\bibnamefont {Sakurai}}, \bibinfo
  {author} {\bibfnamefont {Y.}~\bibnamefont {Satou}}, \bibinfo {author}
  {\bibfnamefont {T.}~\bibnamefont {Sumikama}}, \bibinfo {author}
  {\bibfnamefont {H.}~\bibnamefont {Takeda}}, \bibinfo {author} {\bibfnamefont
  {S.}~\bibnamefont {Takeuchi}}, \bibinfo {author} {\bibfnamefont
  {R.}~\bibnamefont {Tanaka}}, \bibinfo {author} {\bibfnamefont
  {Y.}~\bibnamefont {Togano}}, \ and\ \bibinfo {author} {\bibfnamefont
  {K.}~\bibnamefont {Yoneda}},\ }\href {\doibase
  10.1103/PhysRevLett.112.242501} {\bibfield  {journal} {\bibinfo  {journal}
  {Phys. Rev. Lett.}\ }\textbf {\bibinfo {volume} {112}},\ \bibinfo {pages}
  {242501} (\bibinfo {year} {2014})}\BibitemShut {NoStop}%
\bibitem [{\citenamefont {Tanaka}\ \emph
  {et~al.}(2010{\natexlab{a}})\citenamefont {Tanaka}, \citenamefont
  {Yamaguchi}, \citenamefont {Suzuki}, \citenamefont {Ohtsubo}, \citenamefont
  {Fukuda}, \citenamefont {Nishimura}, \citenamefont {Takechi}, \citenamefont
  {Ogata}, \citenamefont {Ozawa}, \citenamefont {Izumikawa}, \citenamefont
  {Aiba}, \citenamefont {Aoi}, \citenamefont {Baba}, \citenamefont {Hashizume},
  \citenamefont {Inafuku}, \citenamefont {Iwasa}, \citenamefont {Kobayashi},
  \citenamefont {Komuro}, \citenamefont {Kondo}, \citenamefont {Kubo},
  \citenamefont {Kurokawa}, \citenamefont {Matsuyama}, \citenamefont
  {Michimasa}, \citenamefont {Motobayashi}, \citenamefont {Nakabayashi},
  \citenamefont {Nakajima}, \citenamefont {Nakamura}, \citenamefont {Sakurai},
  \citenamefont {Shinoda}, \citenamefont {Shinohara}, \citenamefont {Suzuki},
  \citenamefont {Takeshita}, \citenamefont {Takeuchi}, \citenamefont {Togano},
  \citenamefont {Yamada}, \citenamefont {Yasuno},\ and\ \citenamefont
  {Yoshitake}}]{PhysRevLett.104.062701}%
  \BibitemOpen
  \bibfield  {author} {\bibinfo {author} {\bibfnamefont {K.}~\bibnamefont
  {Tanaka}}, \bibinfo {author} {\bibfnamefont {T.}~\bibnamefont {Yamaguchi}},
  \bibinfo {author} {\bibfnamefont {T.}~\bibnamefont {Suzuki}}, \bibinfo
  {author} {\bibfnamefont {T.}~\bibnamefont {Ohtsubo}}, \bibinfo {author}
  {\bibfnamefont {M.}~\bibnamefont {Fukuda}}, \bibinfo {author} {\bibfnamefont
  {D.}~\bibnamefont {Nishimura}}, \bibinfo {author} {\bibfnamefont
  {M.}~\bibnamefont {Takechi}}, \bibinfo {author} {\bibfnamefont
  {K.}~\bibnamefont {Ogata}}, \bibinfo {author} {\bibfnamefont
  {A.}~\bibnamefont {Ozawa}}, \bibinfo {author} {\bibfnamefont
  {T.}~\bibnamefont {Izumikawa}}, \bibinfo {author} {\bibfnamefont
  {T.}~\bibnamefont {Aiba}}, \bibinfo {author} {\bibfnamefont {N.}~\bibnamefont
  {Aoi}}, \bibinfo {author} {\bibfnamefont {H.}~\bibnamefont {Baba}}, \bibinfo
  {author} {\bibfnamefont {Y.}~\bibnamefont {Hashizume}}, \bibinfo {author}
  {\bibfnamefont {K.}~\bibnamefont {Inafuku}}, \bibinfo {author} {\bibfnamefont
  {N.}~\bibnamefont {Iwasa}}, \bibinfo {author} {\bibfnamefont
  {K.}~\bibnamefont {Kobayashi}}, \bibinfo {author} {\bibfnamefont
  {M.}~\bibnamefont {Komuro}}, \bibinfo {author} {\bibfnamefont
  {Y.}~\bibnamefont {Kondo}}, \bibinfo {author} {\bibfnamefont
  {T.}~\bibnamefont {Kubo}}, \bibinfo {author} {\bibfnamefont {M.}~\bibnamefont
  {Kurokawa}}, \bibinfo {author} {\bibfnamefont {T.}~\bibnamefont {Matsuyama}},
  \bibinfo {author} {\bibfnamefont {S.}~\bibnamefont {Michimasa}}, \bibinfo
  {author} {\bibfnamefont {T.}~\bibnamefont {Motobayashi}}, \bibinfo {author}
  {\bibfnamefont {T.}~\bibnamefont {Nakabayashi}}, \bibinfo {author}
  {\bibfnamefont {S.}~\bibnamefont {Nakajima}}, \bibinfo {author}
  {\bibfnamefont {T.}~\bibnamefont {Nakamura}}, \bibinfo {author}
  {\bibfnamefont {H.}~\bibnamefont {Sakurai}}, \bibinfo {author} {\bibfnamefont
  {R.}~\bibnamefont {Shinoda}}, \bibinfo {author} {\bibfnamefont
  {M.}~\bibnamefont {Shinohara}}, \bibinfo {author} {\bibfnamefont
  {H.}~\bibnamefont {Suzuki}}, \bibinfo {author} {\bibfnamefont
  {E.}~\bibnamefont {Takeshita}}, \bibinfo {author} {\bibfnamefont
  {S.}~\bibnamefont {Takeuchi}}, \bibinfo {author} {\bibfnamefont
  {Y.}~\bibnamefont {Togano}}, \bibinfo {author} {\bibfnamefont
  {K.}~\bibnamefont {Yamada}}, \bibinfo {author} {\bibfnamefont
  {T.}~\bibnamefont {Yasuno}}, \ and\ \bibinfo {author} {\bibfnamefont
  {M.}~\bibnamefont {Yoshitake}},\ }\href {\doibase
  10.1103/PhysRevLett.104.062701} {\bibfield  {journal} {\bibinfo  {journal}
  {Phys. Rev. Lett.}\ }\textbf {\bibinfo {volume} {104}},\ \bibinfo {pages}
  {062701} (\bibinfo {year} {2010}{\natexlab{a}})}\BibitemShut {NoStop}%
\bibitem [{\citenamefont {Shimoura}\ \emph {et~al.}(1997)\citenamefont
  {Shimoura}, \citenamefont {Teranishi}, \citenamefont {Ando}, \citenamefont
  {Hirai}, \citenamefont {Iwasa}, \citenamefont {Kikuchi}, \citenamefont
  {Moriya}, \citenamefont {Motobayashi}, \citenamefont {Murakami},
  \citenamefont {Nakamura}, \citenamefont {Nishio}, \citenamefont {Sakurai},
  \citenamefont {Uchibori}, \citenamefont {Wabanabe}, \citenamefont
  {Yanagisawa},\ and\ \citenamefont {Ishihara}}]{Shi97}%
  \BibitemOpen
  \bibfield  {author} {\bibinfo {author} {\bibfnamefont {S.}~\bibnamefont
  {Shimoura}}, \bibinfo {author} {\bibfnamefont {T.}~\bibnamefont {Teranishi}},
  \bibinfo {author} {\bibfnamefont {Y.}~\bibnamefont {Ando}}, \bibinfo {author}
  {\bibfnamefont {M.}~\bibnamefont {Hirai}}, \bibinfo {author} {\bibfnamefont
  {N.}~\bibnamefont {Iwasa}}, \bibinfo {author} {\bibfnamefont
  {T.}~\bibnamefont {Kikuchi}}, \bibinfo {author} {\bibfnamefont
  {S.}~\bibnamefont {Moriya}}, \bibinfo {author} {\bibfnamefont
  {T.}~\bibnamefont {Motobayashi}}, \bibinfo {author} {\bibfnamefont
  {T.}~\bibnamefont {Murakami}}, \bibinfo {author} {\bibfnamefont
  {T.}~\bibnamefont {Nakamura}}, \bibinfo {author} {\bibfnamefont
  {T.}~\bibnamefont {Nishio}}, \bibinfo {author} {\bibfnamefont
  {H.}~\bibnamefont {Sakurai}}, \bibinfo {author} {\bibfnamefont
  {T.}~\bibnamefont {Uchibori}}, \bibinfo {author} {\bibfnamefont
  {Y.}~\bibnamefont {Wabanabe}}, \bibinfo {author} {\bibfnamefont
  {Y.}~\bibnamefont {Yanagisawa}}, \ and\ \bibinfo {author} {\bibfnamefont
  {M.}~\bibnamefont {Ishihara}},\ }\href@noop {} {\bibfield  {journal}
  {\bibinfo  {journal} {Nucl. Phys. A}\ }\textbf {\bibinfo {volume} {616}},\
  \bibinfo {pages} {208 } (\bibinfo {year} {1997})}\BibitemShut {NoStop}%
\bibitem [{\citenamefont {Tanihata}\ \emph {et~al.}(2013)\citenamefont
  {Tanihata}, \citenamefont {Savajols},\ and\ \citenamefont
  {Kanungo}}]{TANIHATA2013215}%
  \BibitemOpen
  \bibfield  {author} {\bibinfo {author} {\bibfnamefont {I.}~\bibnamefont
  {Tanihata}}, \bibinfo {author} {\bibfnamefont {H.}~\bibnamefont {Savajols}},
  \ and\ \bibinfo {author} {\bibfnamefont {R.}~\bibnamefont {Kanungo}},\
  }\href@noop {} {\bibfield  {journal} {\bibinfo  {journal} {Prog. Part. Nucl.
  Phys.}\ }\textbf {\bibinfo {volume} {68}},\ \bibinfo {pages} {215 } (\bibinfo
  {year} {2013})}\BibitemShut {NoStop}%
\bibitem [{\citenamefont {Riisager}(1994)}]{RevModPhys.66.1105}%
  \BibitemOpen
  \bibfield  {author} {\bibinfo {author} {\bibfnamefont {K.}~\bibnamefont
  {Riisager}},\ }\href {\doibase 10.1103/RevModPhys.66.1105} {\bibfield
  {journal} {\bibinfo  {journal} {Rev. Mod. Phys.}\ }\textbf {\bibinfo {volume}
  {66}},\ \bibinfo {pages} {1105} (\bibinfo {year} {1994})}\BibitemShut
  {NoStop}%
\bibitem [{\citenamefont {Hansen}\ \emph {et~al.}(1995)\citenamefont {Hansen},
  \citenamefont {Jensen},\ and\ \citenamefont {Jonson}}]{Hansen95}%
  \BibitemOpen
  \bibfield  {author} {\bibinfo {author} {\bibfnamefont {P.~G.}\ \bibnamefont
  {Hansen}}, \bibinfo {author} {\bibfnamefont {A.~S.}\ \bibnamefont {Jensen}},
  \ and\ \bibinfo {author} {\bibfnamefont {B.}~\bibnamefont {Jonson}},\
  }\href@noop {} {\bibfield  {journal} {\bibinfo  {journal} {Ann. Rev. Nucl.
  Part. Sci.}\ }\textbf {\bibinfo {volume} {45}},\ \bibinfo {pages} {591}
  (\bibinfo {year} {1995})}\BibitemShut {NoStop}%
\bibitem [{\citenamefont {Riisager}\ \emph {et~al.}(1992)\citenamefont
  {Riisager}, \citenamefont {Jensen},\ and\ \citenamefont
  {Møller}}]{RIISAGER1992393}%
  \BibitemOpen
  \bibfield  {author} {\bibinfo {author} {\bibfnamefont {K.}~\bibnamefont
  {Riisager}}, \bibinfo {author} {\bibfnamefont {A.}~\bibnamefont {Jensen}}, \
  and\ \bibinfo {author} {\bibfnamefont {P.}~\bibnamefont {Møller}},\ }\href
  {\doibase https://doi.org/10.1016/0375-9474(92)90691-C} {\bibfield  {journal}
  {\bibinfo  {journal} {Nucl. Phys. A}\ }\textbf {\bibinfo {volume} {548}},\
  \bibinfo {pages} {393} (\bibinfo {year} {1992})}\BibitemShut {NoStop}%
\bibitem [{\citenamefont {Morlock}\ \emph {et~al.}(1997)\citenamefont
  {Morlock}, \citenamefont {Kunz}, \citenamefont {Mayer}, \citenamefont
  {Jaeger}, \citenamefont {M\"uller}, \citenamefont {Hammer}, \citenamefont
  {Mohr}, \citenamefont {Oberhummer}, \citenamefont {Staudt},\ and\
  \citenamefont {K\"olle}}]{PhysRevLett.79.3837}%
  \BibitemOpen
  \bibfield  {author} {\bibinfo {author} {\bibfnamefont {R.}~\bibnamefont
  {Morlock}}, \bibinfo {author} {\bibfnamefont {R.}~\bibnamefont {Kunz}},
  \bibinfo {author} {\bibfnamefont {A.}~\bibnamefont {Mayer}}, \bibinfo
  {author} {\bibfnamefont {M.}~\bibnamefont {Jaeger}}, \bibinfo {author}
  {\bibfnamefont {A.}~\bibnamefont {M\"uller}}, \bibinfo {author}
  {\bibfnamefont {J.~W.}\ \bibnamefont {Hammer}}, \bibinfo {author}
  {\bibfnamefont {P.}~\bibnamefont {Mohr}}, \bibinfo {author} {\bibfnamefont
  {H.}~\bibnamefont {Oberhummer}}, \bibinfo {author} {\bibfnamefont
  {G.}~\bibnamefont {Staudt}}, \ and\ \bibinfo {author} {\bibfnamefont
  {V.}~\bibnamefont {K\"olle}},\ }\href {\doibase 10.1103/PhysRevLett.79.3837}
  {\bibfield  {journal} {\bibinfo  {journal} {Phys. Rev. Lett.}\ }\textbf
  {\bibinfo {volume} {79}},\ \bibinfo {pages} {3837} (\bibinfo {year}
  {1997})}\BibitemShut {NoStop}%
\bibitem [{\citenamefont {Yang}\ \emph {et~al.}(2021)\citenamefont {Yang},
  \citenamefont {Kubota}, \citenamefont {Corsi}, \citenamefont {Yoshida},
  \citenamefont {Sun}, \citenamefont {Li}, \citenamefont {Kimura},
  \citenamefont {Michel}, \citenamefont {Ogata}, \citenamefont {Yuan},
  \citenamefont {Yuan}, \citenamefont {Authelet}, \citenamefont {Baba},
  \citenamefont {Caesar}, \citenamefont {Calvet}, \citenamefont {Delbart},
  \citenamefont {Dozono}, \citenamefont {Feng}, \citenamefont {Flavigny},
  \citenamefont {Gheller}, \citenamefont {Gibelin}, \citenamefont {Giganon},
  \citenamefont {Gillibert}, \citenamefont {Hasegawa}, \citenamefont {Isobe},
  \citenamefont {Kanaya}, \citenamefont {Kawakami}, \citenamefont {Kim},
  \citenamefont {Kiyokawa}, \citenamefont {Kobayashi}, \citenamefont
  {Kobayashi}, \citenamefont {Kobayashi}, \citenamefont {Kondo}, \citenamefont
  {Korkulu}, \citenamefont {Koyama}, \citenamefont {Lapoux}, \citenamefont
  {Maeda}, \citenamefont {Marqu\'es}, \citenamefont {Motobayashi},
  \citenamefont {Miyazaki}, \citenamefont {Nakamura}, \citenamefont
  {Nakatsuka}, \citenamefont {Nishio}, \citenamefont {Obertelli}, \citenamefont
  {Ohkura}, \citenamefont {Orr}, \citenamefont {Ota}, \citenamefont {Otsu},
  \citenamefont {Ozaki}, \citenamefont {Panin}, \citenamefont {Paschalis},
  \citenamefont {Pollacco}, \citenamefont {Reichert}, \citenamefont {Rouss\'e},
  \citenamefont {Saito}, \citenamefont {Sakaguchi}, \citenamefont {Sako},
  \citenamefont {Santamaria}, \citenamefont {Sasano}, \citenamefont {Sato},
  \citenamefont {Shikata}, \citenamefont {Shimizu}, \citenamefont {Shindo},
  \citenamefont {Stuhl}, \citenamefont {Sumikama}, \citenamefont {Sun},
  \citenamefont {Tabata}, \citenamefont {Togano}, \citenamefont {Tsubota},
  \citenamefont {Xu}, \citenamefont {Yasuda}, \citenamefont {Yoneda},
  \citenamefont {Zenihiro}, \citenamefont {Zhou}, \citenamefont {Zuo},\ and\
  \citenamefont {Uesaka}}]{PhysRevLett.126.082501}%
  \BibitemOpen
  \bibfield  {author} {\bibinfo {author} {\bibfnamefont {Z.~H.}\ \bibnamefont
  {Yang}}, \bibinfo {author} {\bibfnamefont {Y.}~\bibnamefont {Kubota}},
  \bibinfo {author} {\bibfnamefont {A.}~\bibnamefont {Corsi}}, \bibinfo
  {author} {\bibfnamefont {K.}~\bibnamefont {Yoshida}}, \bibinfo {author}
  {\bibfnamefont {X.-X.}\ \bibnamefont {Sun}}, \bibinfo {author} {\bibfnamefont
  {J.~G.}\ \bibnamefont {Li}}, \bibinfo {author} {\bibfnamefont
  {M.}~\bibnamefont {Kimura}}, \bibinfo {author} {\bibfnamefont
  {N.}~\bibnamefont {Michel}}, \bibinfo {author} {\bibfnamefont
  {K.}~\bibnamefont {Ogata}}, \bibinfo {author} {\bibfnamefont {C.~X.}\
  \bibnamefont {Yuan}}, \bibinfo {author} {\bibfnamefont {Q.}~\bibnamefont
  {Yuan}}, \bibinfo {author} {\bibfnamefont {G.}~\bibnamefont {Authelet}},
  \bibinfo {author} {\bibfnamefont {H.}~\bibnamefont {Baba}}, \bibinfo {author}
  {\bibfnamefont {C.}~\bibnamefont {Caesar}}, \bibinfo {author} {\bibfnamefont
  {D.}~\bibnamefont {Calvet}}, \bibinfo {author} {\bibfnamefont
  {A.}~\bibnamefont {Delbart}}, \bibinfo {author} {\bibfnamefont
  {M.}~\bibnamefont {Dozono}}, \bibinfo {author} {\bibfnamefont
  {J.}~\bibnamefont {Feng}}, \bibinfo {author} {\bibfnamefont {F.}~\bibnamefont
  {Flavigny}}, \bibinfo {author} {\bibfnamefont {J.-M.}\ \bibnamefont
  {Gheller}}, \bibinfo {author} {\bibfnamefont {J.}~\bibnamefont {Gibelin}},
  \bibinfo {author} {\bibfnamefont {A.}~\bibnamefont {Giganon}}, \bibinfo
  {author} {\bibfnamefont {A.}~\bibnamefont {Gillibert}}, \bibinfo {author}
  {\bibfnamefont {K.}~\bibnamefont {Hasegawa}}, \bibinfo {author}
  {\bibfnamefont {T.}~\bibnamefont {Isobe}}, \bibinfo {author} {\bibfnamefont
  {Y.}~\bibnamefont {Kanaya}}, \bibinfo {author} {\bibfnamefont
  {S.}~\bibnamefont {Kawakami}}, \bibinfo {author} {\bibfnamefont
  {D.}~\bibnamefont {Kim}}, \bibinfo {author} {\bibfnamefont {Y.}~\bibnamefont
  {Kiyokawa}}, \bibinfo {author} {\bibfnamefont {M.}~\bibnamefont {Kobayashi}},
  \bibinfo {author} {\bibfnamefont {N.}~\bibnamefont {Kobayashi}}, \bibinfo
  {author} {\bibfnamefont {T.}~\bibnamefont {Kobayashi}}, \bibinfo {author}
  {\bibfnamefont {Y.}~\bibnamefont {Kondo}}, \bibinfo {author} {\bibfnamefont
  {Z.}~\bibnamefont {Korkulu}}, \bibinfo {author} {\bibfnamefont
  {S.}~\bibnamefont {Koyama}}, \bibinfo {author} {\bibfnamefont
  {V.}~\bibnamefont {Lapoux}}, \bibinfo {author} {\bibfnamefont
  {Y.}~\bibnamefont {Maeda}}, \bibinfo {author} {\bibfnamefont {F.~M.}\
  \bibnamefont {Marqu\'es}}, \bibinfo {author} {\bibfnamefont {T.}~\bibnamefont
  {Motobayashi}}, \bibinfo {author} {\bibfnamefont {T.}~\bibnamefont
  {Miyazaki}}, \bibinfo {author} {\bibfnamefont {T.}~\bibnamefont {Nakamura}},
  \bibinfo {author} {\bibfnamefont {N.}~\bibnamefont {Nakatsuka}}, \bibinfo
  {author} {\bibfnamefont {Y.}~\bibnamefont {Nishio}}, \bibinfo {author}
  {\bibfnamefont {A.}~\bibnamefont {Obertelli}}, \bibinfo {author}
  {\bibfnamefont {A.}~\bibnamefont {Ohkura}}, \bibinfo {author} {\bibfnamefont
  {N.~A.}\ \bibnamefont {Orr}}, \bibinfo {author} {\bibfnamefont
  {S.}~\bibnamefont {Ota}}, \bibinfo {author} {\bibfnamefont {H.}~\bibnamefont
  {Otsu}}, \bibinfo {author} {\bibfnamefont {T.}~\bibnamefont {Ozaki}},
  \bibinfo {author} {\bibfnamefont {V.}~\bibnamefont {Panin}}, \bibinfo
  {author} {\bibfnamefont {S.}~\bibnamefont {Paschalis}}, \bibinfo {author}
  {\bibfnamefont {E.~C.}\ \bibnamefont {Pollacco}}, \bibinfo {author}
  {\bibfnamefont {S.}~\bibnamefont {Reichert}}, \bibinfo {author}
  {\bibfnamefont {J.-Y.}\ \bibnamefont {Rouss\'e}}, \bibinfo {author}
  {\bibfnamefont {A.~T.}\ \bibnamefont {Saito}}, \bibinfo {author}
  {\bibfnamefont {S.}~\bibnamefont {Sakaguchi}}, \bibinfo {author}
  {\bibfnamefont {M.}~\bibnamefont {Sako}}, \bibinfo {author} {\bibfnamefont
  {C.}~\bibnamefont {Santamaria}}, \bibinfo {author} {\bibfnamefont
  {M.}~\bibnamefont {Sasano}}, \bibinfo {author} {\bibfnamefont
  {H.}~\bibnamefont {Sato}}, \bibinfo {author} {\bibfnamefont {M.}~\bibnamefont
  {Shikata}}, \bibinfo {author} {\bibfnamefont {Y.}~\bibnamefont {Shimizu}},
  \bibinfo {author} {\bibfnamefont {Y.}~\bibnamefont {Shindo}}, \bibinfo
  {author} {\bibfnamefont {L.}~\bibnamefont {Stuhl}}, \bibinfo {author}
  {\bibfnamefont {T.}~\bibnamefont {Sumikama}}, \bibinfo {author}
  {\bibfnamefont {Y.~L.}\ \bibnamefont {Sun}}, \bibinfo {author} {\bibfnamefont
  {M.}~\bibnamefont {Tabata}}, \bibinfo {author} {\bibfnamefont
  {Y.}~\bibnamefont {Togano}}, \bibinfo {author} {\bibfnamefont
  {J.}~\bibnamefont {Tsubota}}, \bibinfo {author} {\bibfnamefont {F.~R.}\
  \bibnamefont {Xu}}, \bibinfo {author} {\bibfnamefont {J.}~\bibnamefont
  {Yasuda}}, \bibinfo {author} {\bibfnamefont {K.}~\bibnamefont {Yoneda}},
  \bibinfo {author} {\bibfnamefont {J.}~\bibnamefont {Zenihiro}}, \bibinfo
  {author} {\bibfnamefont {S.-G.}\ \bibnamefont {Zhou}}, \bibinfo {author}
  {\bibfnamefont {W.}~\bibnamefont {Zuo}}, \ and\ \bibinfo {author}
  {\bibfnamefont {T.}~\bibnamefont {Uesaka}},\ }\href {\doibase
  10.1103/PhysRevLett.126.082501} {\bibfield  {journal} {\bibinfo  {journal}
  {Phys. Rev. Lett.}\ }\textbf {\bibinfo {volume} {126}},\ \bibinfo {pages}
  {082501} (\bibinfo {year} {2021})}\BibitemShut {NoStop}%
\bibitem [{\citenamefont {Wuosmaa}\ \emph {et~al.}(2008)\citenamefont
  {Wuosmaa}, \citenamefont {Schiffer}, \citenamefont {Rehm}, \citenamefont
  {Greene}, \citenamefont {Henderson}, \citenamefont {Janssens}, \citenamefont
  {Jiang}, \citenamefont {Jisonna}, \citenamefont {Lighthall}, \citenamefont
  {Marley}, \citenamefont {Moore}, \citenamefont {Pardo}, \citenamefont
  {Patel}, \citenamefont {Paul}, \citenamefont {Peterson}, \citenamefont
  {Pieper}, \citenamefont {Savard}, \citenamefont {Segel}, \citenamefont
  {Siemssen}, \citenamefont {Tang},\ and\ \citenamefont
  {Wiringa}}]{PhysRevC.78.041302}%
  \BibitemOpen
  \bibfield  {author} {\bibinfo {author} {\bibfnamefont {A.~H.}\ \bibnamefont
  {Wuosmaa}}, \bibinfo {author} {\bibfnamefont {J.~P.}\ \bibnamefont
  {Schiffer}}, \bibinfo {author} {\bibfnamefont {K.~E.}\ \bibnamefont {Rehm}},
  \bibinfo {author} {\bibfnamefont {J.~P.}\ \bibnamefont {Greene}}, \bibinfo
  {author} {\bibfnamefont {D.~J.}\ \bibnamefont {Henderson}}, \bibinfo {author}
  {\bibfnamefont {R.~V.~F.}\ \bibnamefont {Janssens}}, \bibinfo {author}
  {\bibfnamefont {C.~L.}\ \bibnamefont {Jiang}}, \bibinfo {author}
  {\bibfnamefont {L.}~\bibnamefont {Jisonna}}, \bibinfo {author} {\bibfnamefont
  {J.~C.}\ \bibnamefont {Lighthall}}, \bibinfo {author} {\bibfnamefont {S.~T.}\
  \bibnamefont {Marley}}, \bibinfo {author} {\bibfnamefont {E.~F.}\
  \bibnamefont {Moore}}, \bibinfo {author} {\bibfnamefont {R.~C.}\ \bibnamefont
  {Pardo}}, \bibinfo {author} {\bibfnamefont {N.}~\bibnamefont {Patel}},
  \bibinfo {author} {\bibfnamefont {M.}~\bibnamefont {Paul}}, \bibinfo {author}
  {\bibfnamefont {D.}~\bibnamefont {Peterson}}, \bibinfo {author}
  {\bibfnamefont {S.~C.}\ \bibnamefont {Pieper}}, \bibinfo {author}
  {\bibfnamefont {G.}~\bibnamefont {Savard}}, \bibinfo {author} {\bibfnamefont
  {R.~E.}\ \bibnamefont {Segel}}, \bibinfo {author} {\bibfnamefont {R.~H.}\
  \bibnamefont {Siemssen}}, \bibinfo {author} {\bibfnamefont {X.~D.}\
  \bibnamefont {Tang}}, \ and\ \bibinfo {author} {\bibfnamefont {R.~B.}\
  \bibnamefont {Wiringa}},\ }\href {\doibase 10.1103/PhysRevC.78.041302}
  {\bibfield  {journal} {\bibinfo  {journal} {Phys. Rev. C}\ }\textbf {\bibinfo
  {volume} {78}},\ \bibinfo {pages} {041302(R)} (\bibinfo {year}
  {2008})}\BibitemShut {NoStop}%
\bibitem [{\citenamefont {Christian}\ \emph {et~al.}(2012)\citenamefont
  {Christian}, \citenamefont {Frank}, \citenamefont {Ash}, \citenamefont
  {Baumann}, \citenamefont {DeYoung}, \citenamefont {Finck}, \citenamefont
  {Gade}, \citenamefont {Grinyer}, \citenamefont {Luther}, \citenamefont
  {Mosby}, \citenamefont {Mosby}, \citenamefont {Smith}, \citenamefont
  {Snyder}, \citenamefont {Spyrou}, \citenamefont {Strongman}, \citenamefont
  {Thoennessen}, \citenamefont {Warren}, \citenamefont {Weisshaar},\ and\
  \citenamefont {Wersal}}]{PhysRevC.85.034327}%
  \BibitemOpen
  \bibfield  {author} {\bibinfo {author} {\bibfnamefont {G.}~\bibnamefont
  {Christian}}, \bibinfo {author} {\bibfnamefont {N.}~\bibnamefont {Frank}},
  \bibinfo {author} {\bibfnamefont {S.}~\bibnamefont {Ash}}, \bibinfo {author}
  {\bibfnamefont {T.}~\bibnamefont {Baumann}}, \bibinfo {author} {\bibfnamefont
  {P.~A.}\ \bibnamefont {DeYoung}}, \bibinfo {author} {\bibfnamefont {J.~E.}\
  \bibnamefont {Finck}}, \bibinfo {author} {\bibfnamefont {A.}~\bibnamefont
  {Gade}}, \bibinfo {author} {\bibfnamefont {G.~F.}\ \bibnamefont {Grinyer}},
  \bibinfo {author} {\bibfnamefont {B.}~\bibnamefont {Luther}}, \bibinfo
  {author} {\bibfnamefont {M.}~\bibnamefont {Mosby}}, \bibinfo {author}
  {\bibfnamefont {S.}~\bibnamefont {Mosby}}, \bibinfo {author} {\bibfnamefont
  {J.~K.}\ \bibnamefont {Smith}}, \bibinfo {author} {\bibfnamefont
  {J.}~\bibnamefont {Snyder}}, \bibinfo {author} {\bibfnamefont
  {A.}~\bibnamefont {Spyrou}}, \bibinfo {author} {\bibfnamefont {M.~J.}\
  \bibnamefont {Strongman}}, \bibinfo {author} {\bibfnamefont {M.}~\bibnamefont
  {Thoennessen}}, \bibinfo {author} {\bibfnamefont {M.}~\bibnamefont {Warren}},
  \bibinfo {author} {\bibfnamefont {D.}~\bibnamefont {Weisshaar}}, \ and\
  \bibinfo {author} {\bibfnamefont {A.}~\bibnamefont {Wersal}},\ }\href
  {\doibase 10.1103/PhysRevC.85.034327} {\bibfield  {journal} {\bibinfo
  {journal} {Phys. Rev. C}\ }\textbf {\bibinfo {volume} {85}},\ \bibinfo
  {pages} {034327} (\bibinfo {year} {2012})}\BibitemShut {NoStop}%
\bibitem [{\citenamefont {Watanabe}\ \emph {et~al.}(2014)\citenamefont
  {Watanabe}, \citenamefont {Minomo}, \citenamefont {Shimada}, \citenamefont
  {Tagami}, \citenamefont {Kimura}, \citenamefont {Takechi}, \citenamefont
  {Fukuda}, \citenamefont {Nishimura}, \citenamefont {Suzuki}, \citenamefont
  {Matsumoto}, \citenamefont {Shimizu},\ and\ \citenamefont
  {Yahiro}}]{PhysRevC.89.044610}%
  \BibitemOpen
  \bibfield  {author} {\bibinfo {author} {\bibfnamefont {S.}~\bibnamefont
  {Watanabe}}, \bibinfo {author} {\bibfnamefont {K.}~\bibnamefont {Minomo}},
  \bibinfo {author} {\bibfnamefont {M.}~\bibnamefont {Shimada}}, \bibinfo
  {author} {\bibfnamefont {S.}~\bibnamefont {Tagami}}, \bibinfo {author}
  {\bibfnamefont {M.}~\bibnamefont {Kimura}}, \bibinfo {author} {\bibfnamefont
  {M.}~\bibnamefont {Takechi}}, \bibinfo {author} {\bibfnamefont
  {M.}~\bibnamefont {Fukuda}}, \bibinfo {author} {\bibfnamefont
  {D.}~\bibnamefont {Nishimura}}, \bibinfo {author} {\bibfnamefont
  {T.}~\bibnamefont {Suzuki}}, \bibinfo {author} {\bibfnamefont
  {T.}~\bibnamefont {Matsumoto}}, \bibinfo {author} {\bibfnamefont {Y.~R.}\
  \bibnamefont {Shimizu}}, \ and\ \bibinfo {author} {\bibfnamefont
  {M.}~\bibnamefont {Yahiro}},\ }\href {\doibase 10.1103/PhysRevC.89.044610}
  {\bibfield  {journal} {\bibinfo  {journal} {Phys. Rev. C}\ }\textbf {\bibinfo
  {volume} {89}},\ \bibinfo {pages} {044610} (\bibinfo {year}
  {2014})}\BibitemShut {NoStop}%
\bibitem [{\citenamefont {Charity}\ \emph {et~al.}(2020)\citenamefont
  {Charity}, \citenamefont {Sobotka},\ and\ \citenamefont
  {Tostevin}}]{PhysRevC.102.044614}%
  \BibitemOpen
  \bibfield  {author} {\bibinfo {author} {\bibfnamefont {R.~J.}\ \bibnamefont
  {Charity}}, \bibinfo {author} {\bibfnamefont {L.~G.}\ \bibnamefont
  {Sobotka}}, \ and\ \bibinfo {author} {\bibfnamefont {J.~A.}\ \bibnamefont
  {Tostevin}},\ }\href {\doibase 10.1103/PhysRevC.102.044614} {\bibfield
  {journal} {\bibinfo  {journal} {Phys. Rev. C}\ }\textbf {\bibinfo {volume}
  {102}},\ \bibinfo {pages} {044614} (\bibinfo {year} {2020})}\BibitemShut
  {NoStop}%
\bibitem [{\citenamefont {Lunderberg}\ \emph {et~al.}(2012)\citenamefont
  {Lunderberg}, \citenamefont {DeYoung}, \citenamefont {Kohley}, \citenamefont
  {Attanayake}, \citenamefont {Baumann}, \citenamefont {Bazin}, \citenamefont
  {Christian}, \citenamefont {Divaratne}, \citenamefont {Grimes}, \citenamefont
  {Haagsma}, \citenamefont {Finck}, \citenamefont {Frank}, \citenamefont
  {Luther}, \citenamefont {Mosby}, \citenamefont {Nagi}, \citenamefont
  {Peaslee}, \citenamefont {Schiller}, \citenamefont {Snyder}, \citenamefont
  {Spyrou}, \citenamefont {Strongman},\ and\ \citenamefont
  {Thoennessen}}]{PhysRevLett.108.142503}%
  \BibitemOpen
  \bibfield  {author} {\bibinfo {author} {\bibfnamefont {E.}~\bibnamefont
  {Lunderberg}}, \bibinfo {author} {\bibfnamefont {P.~A.}\ \bibnamefont
  {DeYoung}}, \bibinfo {author} {\bibfnamefont {Z.}~\bibnamefont {Kohley}},
  \bibinfo {author} {\bibfnamefont {H.}~\bibnamefont {Attanayake}}, \bibinfo
  {author} {\bibfnamefont {T.}~\bibnamefont {Baumann}}, \bibinfo {author}
  {\bibfnamefont {D.}~\bibnamefont {Bazin}}, \bibinfo {author} {\bibfnamefont
  {G.}~\bibnamefont {Christian}}, \bibinfo {author} {\bibfnamefont
  {D.}~\bibnamefont {Divaratne}}, \bibinfo {author} {\bibfnamefont {S.~M.}\
  \bibnamefont {Grimes}}, \bibinfo {author} {\bibfnamefont {A.}~\bibnamefont
  {Haagsma}}, \bibinfo {author} {\bibfnamefont {J.~E.}\ \bibnamefont {Finck}},
  \bibinfo {author} {\bibfnamefont {N.}~\bibnamefont {Frank}}, \bibinfo
  {author} {\bibfnamefont {B.}~\bibnamefont {Luther}}, \bibinfo {author}
  {\bibfnamefont {S.}~\bibnamefont {Mosby}}, \bibinfo {author} {\bibfnamefont
  {T.}~\bibnamefont {Nagi}}, \bibinfo {author} {\bibfnamefont {G.~F.}\
  \bibnamefont {Peaslee}}, \bibinfo {author} {\bibfnamefont {A.}~\bibnamefont
  {Schiller}}, \bibinfo {author} {\bibfnamefont {J.}~\bibnamefont {Snyder}},
  \bibinfo {author} {\bibfnamefont {A.}~\bibnamefont {Spyrou}}, \bibinfo
  {author} {\bibfnamefont {M.~J.}\ \bibnamefont {Strongman}}, \ and\ \bibinfo
  {author} {\bibfnamefont {M.}~\bibnamefont {Thoennessen}},\ }\href {\doibase
  10.1103/PhysRevLett.108.142503} {\bibfield  {journal} {\bibinfo  {journal}
  {Phys. Rev. Lett.}\ }\textbf {\bibinfo {volume} {108}},\ \bibinfo {pages}
  {142503} (\bibinfo {year} {2012})}\BibitemShut {NoStop}%
\bibitem [{\citenamefont {Baumann}\ \emph {et~al.}(2007)\citenamefont
  {Baumann}, \citenamefont {Amthor}, \citenamefont {Bazin}, \citenamefont
  {Brown}, \citenamefont {Folden~III}, \citenamefont {Gade}, \citenamefont
  {Ginter}, \citenamefont {Hausmann}, \citenamefont {Matoš}, \citenamefont
  {Morrissey}, \citenamefont {Portillo}, \citenamefont {Schiller},
  \citenamefont {Sherrill}, \citenamefont {Stolz}, \citenamefont {Tarasov},\
  and\ \citenamefont {Thoennessen}}]{Baumann20071022}%
  \BibitemOpen
  \bibfield  {author} {\bibinfo {author} {\bibfnamefont {T.}~\bibnamefont
  {Baumann}}, \bibinfo {author} {\bibfnamefont {A.}~\bibnamefont {Amthor}},
  \bibinfo {author} {\bibfnamefont {D.}~\bibnamefont {Bazin}}, \bibinfo
  {author} {\bibfnamefont {B.}~\bibnamefont {Brown}}, \bibinfo {author}
  {\bibfnamefont {C.}~\bibnamefont {Folden~III}}, \bibinfo {author}
  {\bibfnamefont {A.}~\bibnamefont {Gade}}, \bibinfo {author} {\bibfnamefont
  {T.}~\bibnamefont {Ginter}}, \bibinfo {author} {\bibfnamefont
  {M.}~\bibnamefont {Hausmann}}, \bibinfo {author} {\bibfnamefont
  {M.}~\bibnamefont {Matoš}}, \bibinfo {author} {\bibfnamefont
  {D.}~\bibnamefont {Morrissey}}, \bibinfo {author} {\bibfnamefont
  {M.}~\bibnamefont {Portillo}}, \bibinfo {author} {\bibfnamefont
  {A.}~\bibnamefont {Schiller}}, \bibinfo {author} {\bibfnamefont
  {B.}~\bibnamefont {Sherrill}}, \bibinfo {author} {\bibfnamefont
  {A.}~\bibnamefont {Stolz}}, \bibinfo {author} {\bibfnamefont
  {O.}~\bibnamefont {Tarasov}}, \ and\ \bibinfo {author} {\bibfnamefont
  {M.}~\bibnamefont {Thoennessen}},\ }\href@noop {} {\bibfield  {journal}
  {\bibinfo  {journal} {Nature}\ }\textbf {\bibinfo {volume} {449}},\ \bibinfo
  {pages} {1022} (\bibinfo {year} {2007})}\BibitemShut {NoStop}%
\bibitem [{\citenamefont {Freer}\ \emph {et~al.}(2018)\citenamefont {Freer},
  \citenamefont {Horiuchi}, \citenamefont {Kanada-En'yo}, \citenamefont {Lee},\
  and\ \citenamefont {Mei\ss{}ner}}]{RevModPhys.90.035004}%
  \BibitemOpen
  \bibfield  {author} {\bibinfo {author} {\bibfnamefont {M.}~\bibnamefont
  {Freer}}, \bibinfo {author} {\bibfnamefont {H.}~\bibnamefont {Horiuchi}},
  \bibinfo {author} {\bibfnamefont {Y.}~\bibnamefont {Kanada-En'yo}}, \bibinfo
  {author} {\bibfnamefont {D.}~\bibnamefont {Lee}}, \ and\ \bibinfo {author}
  {\bibfnamefont {U.-G.}\ \bibnamefont {Mei\ss{}ner}},\ }\href {\doibase
  10.1103/RevModPhys.90.035004} {\bibfield  {journal} {\bibinfo  {journal}
  {Rev. Mod. Phys.}\ }\textbf {\bibinfo {volume} {90}},\ \bibinfo {pages}
  {035004} (\bibinfo {year} {2018})}\BibitemShut {NoStop}%
\bibitem [{\citenamefont {Yang}\ \emph {et~al.}(2014)\citenamefont {Yang},
  \citenamefont {Ye}, \citenamefont {Li}, \citenamefont {Lou}, \citenamefont
  {Wang}, \citenamefont {Jiang}, \citenamefont {Ge}, \citenamefont {Li},
  \citenamefont {Hua}, \citenamefont {Li}, \citenamefont {Xu}, \citenamefont
  {Pei}, \citenamefont {Qiao}, \citenamefont {You}, \citenamefont {Wang},
  \citenamefont {Tian}, \citenamefont {Li}, \citenamefont {Sun}, \citenamefont
  {Liu}, \citenamefont {Chen}, \citenamefont {Wu}, \citenamefont {Li},
  \citenamefont {Jiang}, \citenamefont {Wen}, \citenamefont {Yang},
  \citenamefont {Yang}, \citenamefont {Ma}, \citenamefont {Ma}, \citenamefont
  {Jin}, \citenamefont {Han},\ and\ \citenamefont
  {Lee}}]{PhysRevLett.112.162501}%
  \BibitemOpen
  \bibfield  {author} {\bibinfo {author} {\bibfnamefont {Z.~H.}\ \bibnamefont
  {Yang}}, \bibinfo {author} {\bibfnamefont {Y.~L.}\ \bibnamefont {Ye}},
  \bibinfo {author} {\bibfnamefont {Z.~H.}\ \bibnamefont {Li}}, \bibinfo
  {author} {\bibfnamefont {J.~L.}\ \bibnamefont {Lou}}, \bibinfo {author}
  {\bibfnamefont {J.~S.}\ \bibnamefont {Wang}}, \bibinfo {author}
  {\bibfnamefont {D.~X.}\ \bibnamefont {Jiang}}, \bibinfo {author}
  {\bibfnamefont {Y.~C.}\ \bibnamefont {Ge}}, \bibinfo {author} {\bibfnamefont
  {Q.~T.}\ \bibnamefont {Li}}, \bibinfo {author} {\bibfnamefont
  {H.}~\bibnamefont {Hua}}, \bibinfo {author} {\bibfnamefont {X.~Q.}\
  \bibnamefont {Li}}, \bibinfo {author} {\bibfnamefont {F.~R.}\ \bibnamefont
  {Xu}}, \bibinfo {author} {\bibfnamefont {J.~C.}\ \bibnamefont {Pei}},
  \bibinfo {author} {\bibfnamefont {R.}~\bibnamefont {Qiao}}, \bibinfo {author}
  {\bibfnamefont {H.~B.}\ \bibnamefont {You}}, \bibinfo {author} {\bibfnamefont
  {H.}~\bibnamefont {Wang}}, \bibinfo {author} {\bibfnamefont {Z.~Y.}\
  \bibnamefont {Tian}}, \bibinfo {author} {\bibfnamefont {K.~A.}\ \bibnamefont
  {Li}}, \bibinfo {author} {\bibfnamefont {Y.~L.}\ \bibnamefont {Sun}},
  \bibinfo {author} {\bibfnamefont {H.~N.}\ \bibnamefont {Liu}}, \bibinfo
  {author} {\bibfnamefont {J.}~\bibnamefont {Chen}}, \bibinfo {author}
  {\bibfnamefont {J.}~\bibnamefont {Wu}}, \bibinfo {author} {\bibfnamefont
  {J.}~\bibnamefont {Li}}, \bibinfo {author} {\bibfnamefont {W.}~\bibnamefont
  {Jiang}}, \bibinfo {author} {\bibfnamefont {C.}~\bibnamefont {Wen}}, \bibinfo
  {author} {\bibfnamefont {B.}~\bibnamefont {Yang}}, \bibinfo {author}
  {\bibfnamefont {Y.~Y.}\ \bibnamefont {Yang}}, \bibinfo {author}
  {\bibfnamefont {P.}~\bibnamefont {Ma}}, \bibinfo {author} {\bibfnamefont
  {J.~B.}\ \bibnamefont {Ma}}, \bibinfo {author} {\bibfnamefont {S.~L.}\
  \bibnamefont {Jin}}, \bibinfo {author} {\bibfnamefont {J.~L.}\ \bibnamefont
  {Han}}, \ and\ \bibinfo {author} {\bibfnamefont {J.}~\bibnamefont {Lee}},\
  }\href {\doibase 10.1103/PhysRevLett.112.162501} {\bibfield  {journal}
  {\bibinfo  {journal} {Phys. Rev. Lett.}\ }\textbf {\bibinfo {volume} {112}},\
  \bibinfo {pages} {162501} (\bibinfo {year} {2014})}\BibitemShut {NoStop}%
\bibitem [{\citenamefont {Horiuchi}\ \emph {et~al.}(2012)\citenamefont
  {Horiuchi}, \citenamefont {Ikeda},\ and\ \citenamefont
  {Kat{\=o}}}]{PTP_clusters}%
  \BibitemOpen
  \bibfield  {author} {\bibinfo {author} {\bibfnamefont {H.}~\bibnamefont
  {Horiuchi}}, \bibinfo {author} {\bibfnamefont {K.}~\bibnamefont {Ikeda}}, \
  and\ \bibinfo {author} {\bibnamefont {Kat{\=o}}},\ }\href@noop {} {\bibfield
  {journal} {\bibinfo  {journal} {Prog. Theor. Phys. Suppl., Extra Number}\
  }\textbf {\bibinfo {volume} {192}},\ \bibinfo {pages} {1} (\bibinfo {year}
  {2012})}\BibitemShut {NoStop}%
\bibitem [{\citenamefont {Dobaczewski}\ \emph {et~al.}(2007)\citenamefont
  {Dobaczewski}, \citenamefont {Michel}, \citenamefont {Nazarewicz},
  \citenamefont {P{\l}oszajczak},\ and\ \citenamefont
  {Rotureau}}]{DOBACZEWSKI2007432}%
  \BibitemOpen
  \bibfield  {author} {\bibinfo {author} {\bibfnamefont {J.}~\bibnamefont
  {Dobaczewski}}, \bibinfo {author} {\bibfnamefont {N.}~\bibnamefont {Michel}},
  \bibinfo {author} {\bibfnamefont {W.}~\bibnamefont {Nazarewicz}}, \bibinfo
  {author} {\bibfnamefont {M.}~\bibnamefont {P{\l}oszajczak}}, \ and\ \bibinfo
  {author} {\bibfnamefont {J.}~\bibnamefont {Rotureau}},\ }\href@noop {}
  {\bibfield  {journal} {\bibinfo  {journal} {Prog. Part. Nucl. Phys.}\
  }\textbf {\bibinfo {volume} {59}},\ \bibinfo {pages} {432 } (\bibinfo {year}
  {2007})}\BibitemShut {NoStop}%
\bibitem [{\citenamefont {Utsuno}\ \emph {et~al.}(2001)\citenamefont {Utsuno},
  \citenamefont {Otsuka}, \citenamefont {Mizusaki},\ and\ \citenamefont
  {Honma}}]{PhysRevC.64.011301}%
  \BibitemOpen
  \bibfield  {author} {\bibinfo {author} {\bibfnamefont {Y.}~\bibnamefont
  {Utsuno}}, \bibinfo {author} {\bibfnamefont {T.}~\bibnamefont {Otsuka}},
  \bibinfo {author} {\bibfnamefont {T.}~\bibnamefont {Mizusaki}}, \ and\
  \bibinfo {author} {\bibfnamefont {M.}~\bibnamefont {Honma}},\ }\href
  {\doibase 10.1103/PhysRevC.64.011301} {\bibfield  {journal} {\bibinfo
  {journal} {Phys. Rev. C}\ }\textbf {\bibinfo {volume} {64}},\ \bibinfo
  {pages} {011301(R)} (\bibinfo {year} {2001})}\BibitemShut {NoStop}%
\bibitem [{\citenamefont {Michel}\ \emph
  {et~al.}(2020{\natexlab{a}})\citenamefont {Michel}, \citenamefont {Li},
  \citenamefont {Xu},\ and\ \citenamefont {Zuo}}]{PhysRevC.101.031301}%
  \BibitemOpen
  \bibfield  {author} {\bibinfo {author} {\bibfnamefont {N.}~\bibnamefont
  {Michel}}, \bibinfo {author} {\bibfnamefont {J.~G.}\ \bibnamefont {Li}},
  \bibinfo {author} {\bibfnamefont {F.~R.}\ \bibnamefont {Xu}}, \ and\ \bibinfo
  {author} {\bibfnamefont {W.}~\bibnamefont {Zuo}},\ }\href {\doibase
  10.1103/PhysRevC.101.031301} {\bibfield  {journal} {\bibinfo  {journal}
  {Phys. Rev. C}\ }\textbf {\bibinfo {volume} {101}},\ \bibinfo {pages}
  {031301(R)} (\bibinfo {year} {2020}{\natexlab{a}})}\BibitemShut {NoStop}%
\bibitem [{\citenamefont {Michel}\ \emph {et~al.}(2003)\citenamefont {Michel},
  \citenamefont {Nazarewicz}, \citenamefont {P\l{}oszajczak},\ and\
  \citenamefont {Oko\l{}owicz}}]{PhysRevC.67.054311}%
  \BibitemOpen
  \bibfield  {author} {\bibinfo {author} {\bibfnamefont {N.}~\bibnamefont
  {Michel}}, \bibinfo {author} {\bibfnamefont {W.}~\bibnamefont {Nazarewicz}},
  \bibinfo {author} {\bibfnamefont {M.}~\bibnamefont {P\l{}oszajczak}}, \ and\
  \bibinfo {author} {\bibfnamefont {J.}~\bibnamefont {Oko\l{}owicz}},\
  }\href@noop {} {\bibfield  {journal} {\bibinfo  {journal} {Phys. Rev. C}\
  }\textbf {\bibinfo {volume} {67}},\ \bibinfo {pages} {054311} (\bibinfo
  {year} {2003})}\BibitemShut {NoStop}%
\bibitem [{ens()}]{ensdf}%
  \BibitemOpen
  \href@noop {} {}\bibinfo {howpublished}
  {\url{https://www.nndc.bnl.gov/ensdf/}}\BibitemShut {NoStop}%
\bibitem [{\citenamefont {Sorlin}\ and\ \citenamefont
  {Porquet}(2008)}]{SORLIN2008602}%
  \BibitemOpen
  \bibfield  {author} {\bibinfo {author} {\bibfnamefont {O.}~\bibnamefont
  {Sorlin}}\ and\ \bibinfo {author} {\bibfnamefont {M.-G.}\ \bibnamefont
  {Porquet}},\ }\href {\doibase https://doi.org/10.1016/j.ppnp.2008.05.001}
  {\bibfield  {journal} {\bibinfo  {journal} {Prog. Part. Nucl. Phys.}\
  }\textbf {\bibinfo {volume} {61}},\ \bibinfo {pages} {602 } (\bibinfo {year}
  {2008})}\BibitemShut {NoStop}%
\bibitem [{\citenamefont {Forss\'en}\ \emph {et~al.}(2013)\citenamefont
  {Forss\'en}, \citenamefont {Hagen}, \citenamefont {Hjorth-Jensen},
  \citenamefont {Nazarewicz},\ and\ \citenamefont
  {Rotureau}}]{1402-4896-2013-T152-014022}%
  \BibitemOpen
  \bibfield  {author} {\bibinfo {author} {\bibfnamefont {C.}~\bibnamefont
  {Forss\'en}}, \bibinfo {author} {\bibfnamefont {G.}~\bibnamefont {Hagen}},
  \bibinfo {author} {\bibfnamefont {M.}~\bibnamefont {Hjorth-Jensen}}, \bibinfo
  {author} {\bibfnamefont {W.}~\bibnamefont {Nazarewicz}}, \ and\ \bibinfo
  {author} {\bibfnamefont {J.}~\bibnamefont {Rotureau}},\ }\href
  {http://stacks.iop.org/1402-4896/2013/i=T152/a=014022} {\bibfield  {journal}
  {\bibinfo  {journal} {Phys. Scri}\ }\textbf {\bibinfo {volume} {2013}},\
  \bibinfo {pages} {014022} (\bibinfo {year} {2013})}\BibitemShut {NoStop}%
\bibitem [{\citenamefont {Michel}\ \emph {et~al.}(2009)\citenamefont {Michel},
  \citenamefont {Nazarewicz}, \citenamefont {Płoszajczak},\ and\ \citenamefont
  {Vertse}}]{0954-3899-36-1-013101}%
  \BibitemOpen
  \bibfield  {author} {\bibinfo {author} {\bibfnamefont {N.}~\bibnamefont
  {Michel}}, \bibinfo {author} {\bibfnamefont {W.}~\bibnamefont {Nazarewicz}},
  \bibinfo {author} {\bibfnamefont {M.}~\bibnamefont {Płoszajczak}}, \ and\
  \bibinfo {author} {\bibfnamefont {T.}~\bibnamefont {Vertse}},\ }\href
  {http://stacks.iop.org/0954-3899/36/i=1/a=013101} {\bibfield  {journal}
  {\bibinfo  {journal} {Jour. Phys. G. Nucl. Part. Phys.}\ }\textbf {\bibinfo
  {volume} {36}},\ \bibinfo {pages} {013101} (\bibinfo {year}
  {2009})}\BibitemShut {NoStop}%
\bibitem [{\citenamefont {Navr{\'{a}}til}\ \emph {et~al.}(2016)\citenamefont
  {Navr{\'{a}}til}, \citenamefont {Quaglioni}, \citenamefont {Hupin},
  \citenamefont {Romero-Redondo},\ and\ \citenamefont {Calci}}]{Navratil_2016}%
  \BibitemOpen
  \bibfield  {author} {\bibinfo {author} {\bibfnamefont {P.}~\bibnamefont
  {Navr{\'{a}}til}}, \bibinfo {author} {\bibfnamefont {S.}~\bibnamefont
  {Quaglioni}}, \bibinfo {author} {\bibfnamefont {G.}~\bibnamefont {Hupin}},
  \bibinfo {author} {\bibfnamefont {C.}~\bibnamefont {Romero-Redondo}}, \ and\
  \bibinfo {author} {\bibfnamefont {A.}~\bibnamefont {Calci}},\ }\href@noop {}
  {\bibfield  {journal} {\bibinfo  {journal} {Phys. Scr.}\ }\textbf {\bibinfo
  {volume} {91}},\ \bibinfo {pages} {053002} (\bibinfo {year}
  {2016})}\BibitemShut {NoStop}%
\bibitem [{\citenamefont {Pf\"utzner}\ \emph {et~al.}(2012)\citenamefont
  {Pf\"utzner}, \citenamefont {Karny}, \citenamefont {Grigorenko},\ and\
  \citenamefont {Riisager}}]{RevModPhys.84.567}%
  \BibitemOpen
  \bibfield  {author} {\bibinfo {author} {\bibfnamefont {M.}~\bibnamefont
  {Pf\"utzner}}, \bibinfo {author} {\bibfnamefont {M.}~\bibnamefont {Karny}},
  \bibinfo {author} {\bibfnamefont {L.~V.}\ \bibnamefont {Grigorenko}}, \ and\
  \bibinfo {author} {\bibfnamefont {K.}~\bibnamefont {Riisager}},\ }\href
  {\doibase 10.1103/RevModPhys.84.567} {\bibfield  {journal} {\bibinfo
  {journal} {Rev. Mod. Phys.}\ }\textbf {\bibinfo {volume} {84}},\ \bibinfo
  {pages} {567} (\bibinfo {year} {2012})}\BibitemShut {NoStop}%
\bibitem [{\citenamefont {Tanaka}\ \emph
  {et~al.}(2010{\natexlab{b}})\citenamefont {Tanaka}, \citenamefont {Fukuda},
  \citenamefont {Mihara}, \citenamefont {Takechi}, \citenamefont {Nishimura},
  \citenamefont {Chinda}, \citenamefont {Sumikama}, \citenamefont {Kudo},
  \citenamefont {Matsuta}, \citenamefont {Minamisono}, \citenamefont {Suzuki},
  \citenamefont {Ohtsubo}, \citenamefont {Izumikawa}, \citenamefont {Momota},
  \citenamefont {Yamaguchi}, \citenamefont {Onishi}, \citenamefont {Ozawa},
  \citenamefont {Tanihata},\ and\ \citenamefont {Zheng}}]{PhysRevC.82.044309}%
  \BibitemOpen
  \bibfield  {author} {\bibinfo {author} {\bibfnamefont {K.}~\bibnamefont
  {Tanaka}}, \bibinfo {author} {\bibfnamefont {M.}~\bibnamefont {Fukuda}},
  \bibinfo {author} {\bibfnamefont {M.}~\bibnamefont {Mihara}}, \bibinfo
  {author} {\bibfnamefont {M.}~\bibnamefont {Takechi}}, \bibinfo {author}
  {\bibfnamefont {D.}~\bibnamefont {Nishimura}}, \bibinfo {author}
  {\bibfnamefont {T.}~\bibnamefont {Chinda}}, \bibinfo {author} {\bibfnamefont
  {T.}~\bibnamefont {Sumikama}}, \bibinfo {author} {\bibfnamefont
  {S.}~\bibnamefont {Kudo}}, \bibinfo {author} {\bibfnamefont {K.}~\bibnamefont
  {Matsuta}}, \bibinfo {author} {\bibfnamefont {T.}~\bibnamefont {Minamisono}},
  \bibinfo {author} {\bibfnamefont {T.}~\bibnamefont {Suzuki}}, \bibinfo
  {author} {\bibfnamefont {T.}~\bibnamefont {Ohtsubo}}, \bibinfo {author}
  {\bibfnamefont {T.}~\bibnamefont {Izumikawa}}, \bibinfo {author}
  {\bibfnamefont {S.}~\bibnamefont {Momota}}, \bibinfo {author} {\bibfnamefont
  {T.}~\bibnamefont {Yamaguchi}}, \bibinfo {author} {\bibfnamefont
  {T.}~\bibnamefont {Onishi}}, \bibinfo {author} {\bibfnamefont
  {A.}~\bibnamefont {Ozawa}}, \bibinfo {author} {\bibfnamefont
  {I.}~\bibnamefont {Tanihata}}, \ and\ \bibinfo {author} {\bibfnamefont
  {T.}~\bibnamefont {Zheng}},\ }\href {\doibase 10.1103/PhysRevC.82.044309}
  {\bibfield  {journal} {\bibinfo  {journal} {Phys. Rev. C}\ }\textbf {\bibinfo
  {volume} {82}},\ \bibinfo {pages} {044309} (\bibinfo {year}
  {2010}{\natexlab{b}})}\BibitemShut {NoStop}%
\bibitem [{\citenamefont {Lehr}\ \emph {et~al.}(2021)\citenamefont {Lehr},
  \citenamefont {Wamers}, \citenamefont {Aksouh}, \citenamefont {Aksyutina},
  \citenamefont {Alvarez-Pol}, \citenamefont {Atar}, \citenamefont {Aumann},
  \citenamefont {Beceiro-Novo}, \citenamefont {Bertulani}, \citenamefont
  {Boretzky}, \citenamefont {Borge}, \citenamefont {Caesar}, \citenamefont
  {Chartier}, \citenamefont {Chatillon}, \citenamefont {Chulkov}, \citenamefont
  {Cortina-Gil}, \citenamefont {Fernandez}, \citenamefont {Emling},
  \citenamefont {Ershova}, \citenamefont {Fraile}, \citenamefont {Fynbo},
  \citenamefont {Galaviz}, \citenamefont {Geissel}, \citenamefont {Heil},
  \citenamefont {Heine}, \citenamefont {Hoffmann}, \citenamefont {Holl},
  \citenamefont {Johansson}, \citenamefont {Jonson}, \citenamefont
  {Karagiannis}, \citenamefont {Kiselev}, \citenamefont {Kratz}, \citenamefont
  {Kulessa}, \citenamefont {Kurz}, \citenamefont {Langer}, \citenamefont
  {Lantz}, \citenamefont {Bleis}, \citenamefont {Lemmon}, \citenamefont
  {Litvinov}, \citenamefont {Loeher}, \citenamefont {Mahata}, \citenamefont
  {Marganiec-Galazka}, \citenamefont {Muentz}, \citenamefont {Nilsson},
  \citenamefont {Nociforo}, \citenamefont {Ott}, \citenamefont {Panin},
  \citenamefont {Paschalis}, \citenamefont {Perea}, \citenamefont {Plag},
  \citenamefont {Reifarth}, \citenamefont {Richter}, \citenamefont {Riisager},
  \citenamefont {Rodriguez-Tajes}, \citenamefont {Rossi}, \citenamefont
  {Savran}, \citenamefont {Scheit}, \citenamefont {Schrieder}, \citenamefont
  {Schrock}, \citenamefont {Simon}, \citenamefont {Stroth}, \citenamefont
  {Suemmerer}, \citenamefont {Tengblad}, \citenamefont {Weick},\ and\
  \citenamefont {Wimmer}}]{lehr2021unveiling}%
  \BibitemOpen
  \bibfield  {author} {\bibinfo {author} {\bibfnamefont {C.}~\bibnamefont
  {Lehr}}, \bibinfo {author} {\bibfnamefont {F.}~\bibnamefont {Wamers}},
  \bibinfo {author} {\bibfnamefont {F.}~\bibnamefont {Aksouh}}, \bibinfo
  {author} {\bibfnamefont {Y.}~\bibnamefont {Aksyutina}}, \bibinfo {author}
  {\bibfnamefont {H.}~\bibnamefont {Alvarez-Pol}}, \bibinfo {author}
  {\bibfnamefont {L.}~\bibnamefont {Atar}}, \bibinfo {author} {\bibfnamefont
  {T.}~\bibnamefont {Aumann}}, \bibinfo {author} {\bibfnamefont
  {S.}~\bibnamefont {Beceiro-Novo}}, \bibinfo {author} {\bibfnamefont {C.~A.}\
  \bibnamefont {Bertulani}}, \bibinfo {author} {\bibfnamefont {K.}~\bibnamefont
  {Boretzky}}, \bibinfo {author} {\bibfnamefont {M.~J.~G.}\ \bibnamefont
  {Borge}}, \bibinfo {author} {\bibfnamefont {C.}~\bibnamefont {Caesar}},
  \bibinfo {author} {\bibfnamefont {M.}~\bibnamefont {Chartier}}, \bibinfo
  {author} {\bibfnamefont {A.}~\bibnamefont {Chatillon}}, \bibinfo {author}
  {\bibfnamefont {L.~V.}\ \bibnamefont {Chulkov}}, \bibinfo {author}
  {\bibfnamefont {D.}~\bibnamefont {Cortina-Gil}}, \bibinfo {author}
  {\bibfnamefont {P.~D.}\ \bibnamefont {Fernandez}}, \bibinfo {author}
  {\bibfnamefont {H.}~\bibnamefont {Emling}}, \bibinfo {author} {\bibfnamefont
  {O.}~\bibnamefont {Ershova}}, \bibinfo {author} {\bibfnamefont {L.~M.}\
  \bibnamefont {Fraile}}, \bibinfo {author} {\bibfnamefont {H.~O.~U.}\
  \bibnamefont {Fynbo}}, \bibinfo {author} {\bibfnamefont {D.}~\bibnamefont
  {Galaviz}}, \bibinfo {author} {\bibfnamefont {H.}~\bibnamefont {Geissel}},
  \bibinfo {author} {\bibfnamefont {M.}~\bibnamefont {Heil}}, \bibinfo {author}
  {\bibfnamefont {M.}~\bibnamefont {Heine}}, \bibinfo {author} {\bibfnamefont
  {D.~H.~H.}\ \bibnamefont {Hoffmann}}, \bibinfo {author} {\bibfnamefont
  {M.}~\bibnamefont {Holl}}, \bibinfo {author} {\bibfnamefont {H.~T.}\
  \bibnamefont {Johansson}}, \bibinfo {author} {\bibfnamefont {B.}~\bibnamefont
  {Jonson}}, \bibinfo {author} {\bibfnamefont {C.}~\bibnamefont {Karagiannis}},
  \bibinfo {author} {\bibfnamefont {O.~A.}\ \bibnamefont {Kiselev}}, \bibinfo
  {author} {\bibfnamefont {J.~V.}\ \bibnamefont {Kratz}}, \bibinfo {author}
  {\bibfnamefont {R.}~\bibnamefont {Kulessa}}, \bibinfo {author} {\bibfnamefont
  {N.}~\bibnamefont {Kurz}}, \bibinfo {author} {\bibfnamefont {C.}~\bibnamefont
  {Langer}}, \bibinfo {author} {\bibfnamefont {M.}~\bibnamefont {Lantz}},
  \bibinfo {author} {\bibfnamefont {T.~L.}\ \bibnamefont {Bleis}}, \bibinfo
  {author} {\bibfnamefont {R.}~\bibnamefont {Lemmon}}, \bibinfo {author}
  {\bibfnamefont {Y.~A.}\ \bibnamefont {Litvinov}}, \bibinfo {author}
  {\bibfnamefont {B.}~\bibnamefont {Loeher}}, \bibinfo {author} {\bibfnamefont
  {K.}~\bibnamefont {Mahata}}, \bibinfo {author} {\bibfnamefont
  {J.}~\bibnamefont {Marganiec-Galazka}}, \bibinfo {author} {\bibfnamefont
  {C.}~\bibnamefont {Muentz}}, \bibinfo {author} {\bibfnamefont
  {T.}~\bibnamefont {Nilsson}}, \bibinfo {author} {\bibfnamefont
  {C.}~\bibnamefont {Nociforo}}, \bibinfo {author} {\bibfnamefont
  {W.}~\bibnamefont {Ott}}, \bibinfo {author} {\bibfnamefont {V.}~\bibnamefont
  {Panin}}, \bibinfo {author} {\bibfnamefont {S.}~\bibnamefont {Paschalis}},
  \bibinfo {author} {\bibfnamefont {A.}~\bibnamefont {Perea}}, \bibinfo
  {author} {\bibfnamefont {R.}~\bibnamefont {Plag}}, \bibinfo {author}
  {\bibfnamefont {R.}~\bibnamefont {Reifarth}}, \bibinfo {author}
  {\bibfnamefont {A.}~\bibnamefont {Richter}}, \bibinfo {author} {\bibfnamefont
  {K.}~\bibnamefont {Riisager}}, \bibinfo {author} {\bibfnamefont
  {C.}~\bibnamefont {Rodriguez-Tajes}}, \bibinfo {author} {\bibfnamefont
  {D.}~\bibnamefont {Rossi}}, \bibinfo {author} {\bibfnamefont
  {D.}~\bibnamefont {Savran}}, \bibinfo {author} {\bibfnamefont
  {H.}~\bibnamefont {Scheit}}, \bibinfo {author} {\bibfnamefont
  {G.}~\bibnamefont {Schrieder}}, \bibinfo {author} {\bibfnamefont
  {P.}~\bibnamefont {Schrock}}, \bibinfo {author} {\bibfnamefont
  {H.}~\bibnamefont {Simon}}, \bibinfo {author} {\bibfnamefont
  {J.}~\bibnamefont {Stroth}}, \bibinfo {author} {\bibfnamefont
  {K.}~\bibnamefont {Suemmerer}}, \bibinfo {author} {\bibfnamefont
  {O.}~\bibnamefont {Tengblad}}, \bibinfo {author} {\bibfnamefont
  {H.}~\bibnamefont {Weick}}, \ and\ \bibinfo {author} {\bibfnamefont
  {C.}~\bibnamefont {Wimmer}},\ }\href@noop {} {\bibfield  {journal} {\bibinfo
  {journal} {arXiv:2101.11474 [nucl-ex]}\ } (\bibinfo {year}
  {2021})}\BibitemShut {NoStop}%
\bibitem [{\citenamefont {Ma}\ \emph {et~al.}(2020)\citenamefont {Ma},
  \citenamefont {Xu}, \citenamefont {Michel}, \citenamefont {Zhang},
  \citenamefont {Li}, \citenamefont {Hu}, \citenamefont {Coraggio},
  \citenamefont {Itaco},\ and\ \citenamefont {Gargano}}]{MA2020135673}%
  \BibitemOpen
  \bibfield  {author} {\bibinfo {author} {\bibfnamefont {Y.~Z.}\ \bibnamefont
  {Ma}}, \bibinfo {author} {\bibfnamefont {F.~R.}\ \bibnamefont {Xu}}, \bibinfo
  {author} {\bibfnamefont {N.}~\bibnamefont {Michel}}, \bibinfo {author}
  {\bibfnamefont {S.}~\bibnamefont {Zhang}}, \bibinfo {author} {\bibfnamefont
  {J.~G.}\ \bibnamefont {Li}}, \bibinfo {author} {\bibfnamefont {B.~S.}\
  \bibnamefont {Hu}}, \bibinfo {author} {\bibfnamefont {L.}~\bibnamefont
  {Coraggio}}, \bibinfo {author} {\bibfnamefont {N.}~\bibnamefont {Itaco}}, \
  and\ \bibinfo {author} {\bibfnamefont {A.}~\bibnamefont {Gargano}},\ }\href
  {\doibase https://doi.org/10.1016/j.physletb.2020.135673} {\bibfield
  {journal} {\bibinfo  {journal} {Phys. Lett. B}\ }\textbf {\bibinfo {volume}
  {808}},\ \bibinfo {pages} {135673} (\bibinfo {year} {2020})}\BibitemShut
  {NoStop}%
\bibitem [{\citenamefont {Michel}\ \emph {et~al.}(2019)\citenamefont {Michel},
  \citenamefont {Li}, \citenamefont {Xu},\ and\ \citenamefont
  {Zuo}}]{PhysRevC.100.064303}%
  \BibitemOpen
  \bibfield  {author} {\bibinfo {author} {\bibfnamefont {N.}~\bibnamefont
  {Michel}}, \bibinfo {author} {\bibfnamefont {J.~G.}\ \bibnamefont {Li}},
  \bibinfo {author} {\bibfnamefont {F.~R.}\ \bibnamefont {Xu}}, \ and\ \bibinfo
  {author} {\bibfnamefont {W.}~\bibnamefont {Zuo}},\ }\href {\doibase
  10.1103/PhysRevC.100.064303} {\bibfield  {journal} {\bibinfo  {journal}
  {Phys. Rev. C}\ }\textbf {\bibinfo {volume} {100}},\ \bibinfo {pages}
  {064303} (\bibinfo {year} {2019})}\BibitemShut {NoStop}%
\bibitem [{\citenamefont {Ormand}(1997)}]{PhysRevC.55.2407}%
  \BibitemOpen
  \bibfield  {author} {\bibinfo {author} {\bibfnamefont {W.~E.}\ \bibnamefont
  {Ormand}},\ }\href {\doibase 10.1103/PhysRevC.55.2407} {\bibfield  {journal}
  {\bibinfo  {journal} {Phys. Rev. C}\ }\textbf {\bibinfo {volume} {55}},\
  \bibinfo {pages} {2407} (\bibinfo {year} {1997})}\BibitemShut {NoStop}%
\bibitem [{\citenamefont {Britz}\ \emph {et~al.}(1998)\citenamefont {Britz},
  \citenamefont {Pape},\ and\ \citenamefont {Antony}}]{IMME}%
  \BibitemOpen
  \bibfield  {author} {\bibinfo {author} {\bibfnamefont {J.}~\bibnamefont
  {Britz}}, \bibinfo {author} {\bibfnamefont {A.}~\bibnamefont {Pape}}, \ and\
  \bibinfo {author} {\bibfnamefont {M.~S.}\ \bibnamefont {Antony}},\
  }\href@noop {} {\bibfield  {journal} {\bibinfo  {journal} {At. Data Nucl.
  Data Tab.}\ }\textbf {\bibinfo {volume} {69}},\ \bibinfo {pages} {125}
  (\bibinfo {year} {1998})}\BibitemShut {NoStop}%
\bibitem [{\citenamefont {Brown}\ and\ \citenamefont
  {Richter}(2006)}]{PhysRevC.74.034315}%
  \BibitemOpen
  \bibfield  {author} {\bibinfo {author} {\bibfnamefont {B.~A.}\ \bibnamefont
  {Brown}}\ and\ \bibinfo {author} {\bibfnamefont {W.~A.}\ \bibnamefont
  {Richter}},\ }\href {\doibase 10.1103/PhysRevC.74.034315} {\bibfield
  {journal} {\bibinfo  {journal} {Phys. Rev. C}\ }\textbf {\bibinfo {volume}
  {74}},\ \bibinfo {pages} {034315} (\bibinfo {year} {2006})}\BibitemShut
  {NoStop}%
\bibitem [{\citenamefont {De~Grancey}\ \emph {et~al.}(2016)\citenamefont
  {De~Grancey} \emph {et~al.}}]{DeGrancey:2016bez}%
  \BibitemOpen
  \bibfield  {author} {\bibinfo {author} {\bibfnamefont {F.}~\bibnamefont
  {De~Grancey}} \emph {et~al.},\ }\href {\doibase
  10.1016/j.physletb.2016.04.051} {\bibfield  {journal} {\bibinfo  {journal}
  {Phys. Lett.}\ }\textbf {\bibinfo {volume} {B758}},\ \bibinfo {pages} {26}
  (\bibinfo {year} {2016})}\BibitemShut {NoStop}%
\bibitem [{\citenamefont {Goldanskii}(1960)}]{goldanskii60}%
  \BibitemOpen
  \bibfield  {author} {\bibinfo {author} {\bibfnamefont {V.~I.}\ \bibnamefont
  {Goldanskii}},\ }\href@noop {} {\bibfield  {journal} {\bibinfo  {journal}
  {Nucl. Phys.}\ }\textbf {\bibinfo {volume} {19}},\ \bibinfo {pages} {482}
  (\bibinfo {year} {1960})}\BibitemShut {NoStop}%
\bibitem [{\citenamefont {Pomorski}\ \emph {et~al.}(2011)\citenamefont
  {Pomorski}, \citenamefont {Pf\"utzner}, \citenamefont {Dominik},
  \citenamefont {Grzywacz}, \citenamefont {Baumann}, \citenamefont {Berryman},
  \citenamefont {Czyrkowski}, \citenamefont {D\c{a}browski}, \citenamefont
  {Ginter}, \citenamefont {Johnson}, \citenamefont {Kami\ifmmode~\acute{n}\else
  \'{n}\fi{}ski}, \citenamefont {Ku\ifmmode~\acute{z}\else \'{z}\fi{}niak},
  \citenamefont {Larson}, \citenamefont {Liddick}, \citenamefont {Madurga},
  \citenamefont {Mazzocchi}, \citenamefont {Mianowski}, \citenamefont
  {Miernik}, \citenamefont {Miller}, \citenamefont {Paulauskas}, \citenamefont
  {Pereira}, \citenamefont {Rykaczewski}, \citenamefont {Stolz},\ and\
  \citenamefont {Suchyta}}]{PhysRevC.83.061303}%
  \BibitemOpen
  \bibfield  {author} {\bibinfo {author} {\bibfnamefont {M.}~\bibnamefont
  {Pomorski}}, \bibinfo {author} {\bibfnamefont {M.}~\bibnamefont
  {Pf\"utzner}}, \bibinfo {author} {\bibfnamefont {W.}~\bibnamefont {Dominik}},
  \bibinfo {author} {\bibfnamefont {R.}~\bibnamefont {Grzywacz}}, \bibinfo
  {author} {\bibfnamefont {T.}~\bibnamefont {Baumann}}, \bibinfo {author}
  {\bibfnamefont {J.~S.}\ \bibnamefont {Berryman}}, \bibinfo {author}
  {\bibfnamefont {H.}~\bibnamefont {Czyrkowski}}, \bibinfo {author}
  {\bibfnamefont {R.}~\bibnamefont {D\c{a}browski}}, \bibinfo {author}
  {\bibfnamefont {T.}~\bibnamefont {Ginter}}, \bibinfo {author} {\bibfnamefont
  {J.}~\bibnamefont {Johnson}}, \bibinfo {author} {\bibfnamefont
  {G.}~\bibnamefont {Kami\ifmmode~\acute{n}\else \'{n}\fi{}ski}}, \bibinfo
  {author} {\bibfnamefont {A.}~\bibnamefont {Ku\ifmmode~\acute{z}\else
  \'{z}\fi{}niak}}, \bibinfo {author} {\bibfnamefont {N.}~\bibnamefont
  {Larson}}, \bibinfo {author} {\bibfnamefont {S.~N.}\ \bibnamefont {Liddick}},
  \bibinfo {author} {\bibfnamefont {M.}~\bibnamefont {Madurga}}, \bibinfo
  {author} {\bibfnamefont {C.}~\bibnamefont {Mazzocchi}}, \bibinfo {author}
  {\bibfnamefont {S.}~\bibnamefont {Mianowski}}, \bibinfo {author}
  {\bibfnamefont {K.}~\bibnamefont {Miernik}}, \bibinfo {author} {\bibfnamefont
  {D.}~\bibnamefont {Miller}}, \bibinfo {author} {\bibfnamefont
  {S.}~\bibnamefont {Paulauskas}}, \bibinfo {author} {\bibfnamefont
  {J.}~\bibnamefont {Pereira}}, \bibinfo {author} {\bibfnamefont {K.~P.}\
  \bibnamefont {Rykaczewski}}, \bibinfo {author} {\bibfnamefont
  {A.}~\bibnamefont {Stolz}}, \ and\ \bibinfo {author} {\bibfnamefont
  {S.}~\bibnamefont {Suchyta}},\ }\href {\doibase 10.1103/PhysRevC.83.061303}
  {\bibfield  {journal} {\bibinfo  {journal} {Phys. Rev. C}\ }\textbf {\bibinfo
  {volume} {83}},\ \bibinfo {pages} {061303(R)} (\bibinfo {year}
  {2011})}\BibitemShut {NoStop}%
\bibitem [{\citenamefont {Miernik}\ \emph {et~al.}(2007)\citenamefont
  {Miernik}, \citenamefont {Dominik}, \citenamefont {Janas}, \citenamefont
  {Pf\"utzner}, \citenamefont {Grigorenko}, \citenamefont {Bingham},
  \citenamefont {Czyrkowski}, \citenamefont {\ifmmode~\acute{C}\else
  \'{C}\fi{}wiok}, \citenamefont {Darby}, \citenamefont {D\c{a}browski},
  \citenamefont {Ginter}, \citenamefont {Grzywacz}, \citenamefont {Karny},
  \citenamefont {Korgul}, \citenamefont {Ku\ifmmode~\acute{s}\else
  \'{s}\fi{}mierz}, \citenamefont {Liddick}, \citenamefont {Rajabali},
  \citenamefont {Rykaczewski},\ and\ \citenamefont
  {Stolz}}]{PhysRevLett.99.192501}%
  \BibitemOpen
  \bibfield  {author} {\bibinfo {author} {\bibfnamefont {K.}~\bibnamefont
  {Miernik}}, \bibinfo {author} {\bibfnamefont {W.}~\bibnamefont {Dominik}},
  \bibinfo {author} {\bibfnamefont {Z.}~\bibnamefont {Janas}}, \bibinfo
  {author} {\bibfnamefont {M.}~\bibnamefont {Pf\"utzner}}, \bibinfo {author}
  {\bibfnamefont {L.}~\bibnamefont {Grigorenko}}, \bibinfo {author}
  {\bibfnamefont {C.~R.}\ \bibnamefont {Bingham}}, \bibinfo {author}
  {\bibfnamefont {H.}~\bibnamefont {Czyrkowski}}, \bibinfo {author}
  {\bibfnamefont {M.}~\bibnamefont {\ifmmode~\acute{C}\else \'{C}\fi{}wiok}},
  \bibinfo {author} {\bibfnamefont {I.~G.}\ \bibnamefont {Darby}}, \bibinfo
  {author} {\bibfnamefont {R.}~\bibnamefont {D\c{a}browski}}, \bibinfo {author}
  {\bibfnamefont {T.}~\bibnamefont {Ginter}}, \bibinfo {author} {\bibfnamefont
  {R.}~\bibnamefont {Grzywacz}}, \bibinfo {author} {\bibfnamefont
  {M.}~\bibnamefont {Karny}}, \bibinfo {author} {\bibfnamefont
  {A.}~\bibnamefont {Korgul}}, \bibinfo {author} {\bibfnamefont
  {W.}~\bibnamefont {Ku\ifmmode~\acute{s}\else \'{s}\fi{}mierz}}, \bibinfo
  {author} {\bibfnamefont {S.~N.}\ \bibnamefont {Liddick}}, \bibinfo {author}
  {\bibfnamefont {M.}~\bibnamefont {Rajabali}}, \bibinfo {author}
  {\bibfnamefont {K.}~\bibnamefont {Rykaczewski}}, \ and\ \bibinfo {author}
  {\bibfnamefont {A.}~\bibnamefont {Stolz}},\ }\href {\doibase
  10.1103/PhysRevLett.99.192501} {\bibfield  {journal} {\bibinfo  {journal}
  {Phys. Rev. Lett.}\ }\textbf {\bibinfo {volume} {99}},\ \bibinfo {pages}
  {192501} (\bibinfo {year} {2007})}\BibitemShut {NoStop}%
\bibitem [{\citenamefont {Blank}\ \emph {et~al.}(2005)\citenamefont {Blank},
  \citenamefont {Bey}, \citenamefont {Canchel}, \citenamefont {Dossat},
  \citenamefont {Fleury}, \citenamefont {Giovinazzo}, \citenamefont {Matea},
  \citenamefont {Adimi}, \citenamefont {De~Oliveira}, \citenamefont {Stefan},
  \citenamefont {Georgiev}, \citenamefont {Gr\'evy}, \citenamefont {Thomas},
  \citenamefont {Borcea}, \citenamefont {Cortina}, \citenamefont {Caamano},
  \citenamefont {Stanoiu}, \citenamefont {Aksouh}, \citenamefont {Brown},
  \citenamefont {Barker},\ and\ \citenamefont
  {Richter}}]{PhysRevLett.94.232501}%
  \BibitemOpen
  \bibfield  {author} {\bibinfo {author} {\bibfnamefont {B.}~\bibnamefont
  {Blank}}, \bibinfo {author} {\bibfnamefont {A.}~\bibnamefont {Bey}}, \bibinfo
  {author} {\bibfnamefont {G.}~\bibnamefont {Canchel}}, \bibinfo {author}
  {\bibfnamefont {C.}~\bibnamefont {Dossat}}, \bibinfo {author} {\bibfnamefont
  {A.}~\bibnamefont {Fleury}}, \bibinfo {author} {\bibfnamefont
  {J.}~\bibnamefont {Giovinazzo}}, \bibinfo {author} {\bibfnamefont
  {I.}~\bibnamefont {Matea}}, \bibinfo {author} {\bibfnamefont
  {N.}~\bibnamefont {Adimi}}, \bibinfo {author} {\bibfnamefont
  {F.}~\bibnamefont {De~Oliveira}}, \bibinfo {author} {\bibfnamefont
  {I.}~\bibnamefont {Stefan}}, \bibinfo {author} {\bibfnamefont
  {G.}~\bibnamefont {Georgiev}}, \bibinfo {author} {\bibfnamefont
  {S.}~\bibnamefont {Gr\'evy}}, \bibinfo {author} {\bibfnamefont {J.~C.}\
  \bibnamefont {Thomas}}, \bibinfo {author} {\bibfnamefont {C.}~\bibnamefont
  {Borcea}}, \bibinfo {author} {\bibfnamefont {D.}~\bibnamefont {Cortina}},
  \bibinfo {author} {\bibfnamefont {M.}~\bibnamefont {Caamano}}, \bibinfo
  {author} {\bibfnamefont {M.}~\bibnamefont {Stanoiu}}, \bibinfo {author}
  {\bibfnamefont {F.}~\bibnamefont {Aksouh}}, \bibinfo {author} {\bibfnamefont
  {B.~A.}\ \bibnamefont {Brown}}, \bibinfo {author} {\bibfnamefont {F.~C.}\
  \bibnamefont {Barker}}, \ and\ \bibinfo {author} {\bibfnamefont {W.~A.}\
  \bibnamefont {Richter}},\ }\href {\doibase 10.1103/PhysRevLett.94.232501}
  {\bibfield  {journal} {\bibinfo  {journal} {Phys. Rev. Lett.}\ }\textbf
  {\bibinfo {volume} {94}},\ \bibinfo {pages} {232501} (\bibinfo {year}
  {2005})}\BibitemShut {NoStop}%
\bibitem [{\citenamefont {Goigoux}\ \emph {et~al.}(2016)\citenamefont
  {Goigoux}, \citenamefont {Ascher}, \citenamefont {Blank}, \citenamefont
  {Gerbaux}, \citenamefont {Giovinazzo}, \citenamefont {Gr\'evy}, \citenamefont
  {Kurtukian~Nieto}, \citenamefont {Magron}, \citenamefont {Doornenbal},
  \citenamefont {Kiss}, \citenamefont {Nishimura}, \citenamefont
  {S\"oderstr\"om}, \citenamefont {Phong}, \citenamefont {Wu}, \citenamefont
  {Ahn}, \citenamefont {Fukuda}, \citenamefont {Inabe}, \citenamefont {Kubo},
  \citenamefont {Kubono}, \citenamefont {Sakurai}, \citenamefont {Shimizu},
  \citenamefont {Sumikama}, \citenamefont {Suzuki}, \citenamefont {Takeda},
  \citenamefont {Agramunt}, \citenamefont {Algora}, \citenamefont {Guadilla},
  \citenamefont {Montaner-Piza}, \citenamefont {Morales}, \citenamefont
  {Orrigo}, \citenamefont {Rubio}, \citenamefont {Fujita}, \citenamefont
  {Tanaka}, \citenamefont {Gelletly}, \citenamefont {Aguilera}, \citenamefont
  {Molina}, \citenamefont {Diel}, \citenamefont {Lubos}, \citenamefont
  {de~Angelis}, \citenamefont {Napoli}, \citenamefont {Borcea}, \citenamefont
  {Boso}, \citenamefont {Cakirli}, \citenamefont {Ganioglu}, \citenamefont
  {Chiba}, \citenamefont {Nishimura}, \citenamefont {Oikawa}, \citenamefont
  {Takei}, \citenamefont {Yagi}, \citenamefont {Wimmer}, \citenamefont
  {de~France}, \citenamefont {Go},\ and\ \citenamefont
  {Brown}}]{PhysRevLett.117.162501}%
  \BibitemOpen
  \bibfield  {author} {\bibinfo {author} {\bibfnamefont {T.}~\bibnamefont
  {Goigoux}}, \bibinfo {author} {\bibfnamefont {P.}~\bibnamefont {Ascher}},
  \bibinfo {author} {\bibfnamefont {B.}~\bibnamefont {Blank}}, \bibinfo
  {author} {\bibfnamefont {M.}~\bibnamefont {Gerbaux}}, \bibinfo {author}
  {\bibfnamefont {J.}~\bibnamefont {Giovinazzo}}, \bibinfo {author}
  {\bibfnamefont {S.}~\bibnamefont {Gr\'evy}}, \bibinfo {author} {\bibfnamefont
  {T.}~\bibnamefont {Kurtukian~Nieto}}, \bibinfo {author} {\bibfnamefont
  {C.}~\bibnamefont {Magron}}, \bibinfo {author} {\bibfnamefont
  {P.}~\bibnamefont {Doornenbal}}, \bibinfo {author} {\bibfnamefont {G.~G.}\
  \bibnamefont {Kiss}}, \bibinfo {author} {\bibfnamefont {S.}~\bibnamefont
  {Nishimura}}, \bibinfo {author} {\bibfnamefont {P.-A.}\ \bibnamefont
  {S\"oderstr\"om}}, \bibinfo {author} {\bibfnamefont {V.~H.}\ \bibnamefont
  {Phong}}, \bibinfo {author} {\bibfnamefont {J.}~\bibnamefont {Wu}}, \bibinfo
  {author} {\bibfnamefont {D.~S.}\ \bibnamefont {Ahn}}, \bibinfo {author}
  {\bibfnamefont {N.}~\bibnamefont {Fukuda}}, \bibinfo {author} {\bibfnamefont
  {N.}~\bibnamefont {Inabe}}, \bibinfo {author} {\bibfnamefont
  {T.}~\bibnamefont {Kubo}}, \bibinfo {author} {\bibfnamefont {S.}~\bibnamefont
  {Kubono}}, \bibinfo {author} {\bibfnamefont {H.}~\bibnamefont {Sakurai}},
  \bibinfo {author} {\bibfnamefont {Y.}~\bibnamefont {Shimizu}}, \bibinfo
  {author} {\bibfnamefont {T.}~\bibnamefont {Sumikama}}, \bibinfo {author}
  {\bibfnamefont {H.}~\bibnamefont {Suzuki}}, \bibinfo {author} {\bibfnamefont
  {H.}~\bibnamefont {Takeda}}, \bibinfo {author} {\bibfnamefont
  {J.}~\bibnamefont {Agramunt}}, \bibinfo {author} {\bibfnamefont
  {A.}~\bibnamefont {Algora}}, \bibinfo {author} {\bibfnamefont
  {V.}~\bibnamefont {Guadilla}}, \bibinfo {author} {\bibfnamefont
  {A.}~\bibnamefont {Montaner-Piza}}, \bibinfo {author} {\bibfnamefont {A.~I.}\
  \bibnamefont {Morales}}, \bibinfo {author} {\bibfnamefont {S.~E.~A.}\
  \bibnamefont {Orrigo}}, \bibinfo {author} {\bibfnamefont {B.}~\bibnamefont
  {Rubio}}, \bibinfo {author} {\bibfnamefont {Y.}~\bibnamefont {Fujita}},
  \bibinfo {author} {\bibfnamefont {M.}~\bibnamefont {Tanaka}}, \bibinfo
  {author} {\bibfnamefont {W.}~\bibnamefont {Gelletly}}, \bibinfo {author}
  {\bibfnamefont {P.}~\bibnamefont {Aguilera}}, \bibinfo {author}
  {\bibfnamefont {F.}~\bibnamefont {Molina}}, \bibinfo {author} {\bibfnamefont
  {F.}~\bibnamefont {Diel}}, \bibinfo {author} {\bibfnamefont {D.}~\bibnamefont
  {Lubos}}, \bibinfo {author} {\bibfnamefont {G.}~\bibnamefont {de~Angelis}},
  \bibinfo {author} {\bibfnamefont {D.}~\bibnamefont {Napoli}}, \bibinfo
  {author} {\bibfnamefont {C.}~\bibnamefont {Borcea}}, \bibinfo {author}
  {\bibfnamefont {A.}~\bibnamefont {Boso}}, \bibinfo {author} {\bibfnamefont
  {R.~B.}\ \bibnamefont {Cakirli}}, \bibinfo {author} {\bibfnamefont
  {E.}~\bibnamefont {Ganioglu}}, \bibinfo {author} {\bibfnamefont
  {J.}~\bibnamefont {Chiba}}, \bibinfo {author} {\bibfnamefont
  {D.}~\bibnamefont {Nishimura}}, \bibinfo {author} {\bibfnamefont
  {H.}~\bibnamefont {Oikawa}}, \bibinfo {author} {\bibfnamefont
  {Y.}~\bibnamefont {Takei}}, \bibinfo {author} {\bibfnamefont
  {S.}~\bibnamefont {Yagi}}, \bibinfo {author} {\bibfnamefont {K.}~\bibnamefont
  {Wimmer}}, \bibinfo {author} {\bibfnamefont {G.}~\bibnamefont {de~France}},
  \bibinfo {author} {\bibfnamefont {S.}~\bibnamefont {Go}}, \ and\ \bibinfo
  {author} {\bibfnamefont {B.~A.}\ \bibnamefont {Brown}},\ }\href {\doibase
  10.1103/PhysRevLett.117.162501} {\bibfield  {journal} {\bibinfo  {journal}
  {Phys. Rev. Lett.}\ }\textbf {\bibinfo {volume} {117}},\ \bibinfo {pages}
  {162501} (\bibinfo {year} {2016})}\BibitemShut {NoStop}%
\bibitem [{\citenamefont {Wang}\ and\ \citenamefont
  {Nazarewicz}(2018)}]{PhysRevLett.120.212502}%
  \BibitemOpen
  \bibfield  {author} {\bibinfo {author} {\bibfnamefont {S.~M.}\ \bibnamefont
  {Wang}}\ and\ \bibinfo {author} {\bibfnamefont {W.}~\bibnamefont
  {Nazarewicz}},\ }\href {\doibase 10.1103/PhysRevLett.120.212502} {\bibfield
  {journal} {\bibinfo  {journal} {Phys. Rev. Lett.}\ }\textbf {\bibinfo
  {volume} {120}},\ \bibinfo {pages} {212502} (\bibinfo {year}
  {2018})}\BibitemShut {NoStop}%
\bibitem [{\citenamefont {Kryger}\ \emph {et~al.}(1995)\citenamefont {Kryger},
  \citenamefont {Azhari}, \citenamefont {Hellstr\"om}, \citenamefont {Kelley},
  \citenamefont {Kubo}, \citenamefont {Pfaff}, \citenamefont {Ramakrishnan},
  \citenamefont {Sherrill}, \citenamefont {Thoennessen}, \citenamefont
  {Yokoyama}, \citenamefont {Charity}, \citenamefont {Dempsey}, \citenamefont
  {Kirov}, \citenamefont {Robertson}, \citenamefont {Sarantites}, \citenamefont
  {Sobotka},\ and\ \citenamefont {Winger}}]{PhysRevLett.74.860}%
  \BibitemOpen
  \bibfield  {author} {\bibinfo {author} {\bibfnamefont {R.~A.}\ \bibnamefont
  {Kryger}}, \bibinfo {author} {\bibfnamefont {A.}~\bibnamefont {Azhari}},
  \bibinfo {author} {\bibfnamefont {M.}~\bibnamefont {Hellstr\"om}}, \bibinfo
  {author} {\bibfnamefont {J.~H.}\ \bibnamefont {Kelley}}, \bibinfo {author}
  {\bibfnamefont {T.}~\bibnamefont {Kubo}}, \bibinfo {author} {\bibfnamefont
  {R.}~\bibnamefont {Pfaff}}, \bibinfo {author} {\bibfnamefont
  {E.}~\bibnamefont {Ramakrishnan}}, \bibinfo {author} {\bibfnamefont {B.~M.}\
  \bibnamefont {Sherrill}}, \bibinfo {author} {\bibfnamefont {M.}~\bibnamefont
  {Thoennessen}}, \bibinfo {author} {\bibfnamefont {S.}~\bibnamefont
  {Yokoyama}}, \bibinfo {author} {\bibfnamefont {R.~J.}\ \bibnamefont
  {Charity}}, \bibinfo {author} {\bibfnamefont {J.}~\bibnamefont {Dempsey}},
  \bibinfo {author} {\bibfnamefont {A.}~\bibnamefont {Kirov}}, \bibinfo
  {author} {\bibfnamefont {N.}~\bibnamefont {Robertson}}, \bibinfo {author}
  {\bibfnamefont {D.~G.}\ \bibnamefont {Sarantites}}, \bibinfo {author}
  {\bibfnamefont {L.~G.}\ \bibnamefont {Sobotka}}, \ and\ \bibinfo {author}
  {\bibfnamefont {J.~A.}\ \bibnamefont {Winger}},\ }\href {\doibase
  10.1103/PhysRevLett.74.860} {\bibfield  {journal} {\bibinfo  {journal} {Phys.
  Rev. Lett.}\ }\textbf {\bibinfo {volume} {74}},\ \bibinfo {pages} {860}
  (\bibinfo {year} {1995})}\BibitemShut {NoStop}%
\bibitem [{\citenamefont {Mukha}\ \emph {et~al.}(2010)\citenamefont {Mukha},
  \citenamefont {S\"ummerer}, \citenamefont {Acosta}, \citenamefont {Alvarez},
  \citenamefont {Casarejos}, \citenamefont {Chatillon}, \citenamefont
  {Cortina-Gil}, \citenamefont {Egorova}, \citenamefont {Espino}, \citenamefont
  {Fomichev}, \citenamefont {Garc\'{\i}a-Ramos}, \citenamefont {Geissel},
  \citenamefont {G\'omez-Camacho}, \citenamefont {Grigorenko}, \citenamefont
  {Hofmann}, \citenamefont {Kiselev}, \citenamefont {Korsheninnikov},
  \citenamefont {Kurz}, \citenamefont {Litvinov}, \citenamefont {Litvinova},
  \citenamefont {Martel}, \citenamefont {Nociforo}, \citenamefont {Ott},
  \citenamefont {Pf\"utzner}, \citenamefont {Rodr\'{\i}guez-Tajes},
  \citenamefont {Roeckl}, \citenamefont {Stanoiu}, \citenamefont {Timofeyuk},
  \citenamefont {Weick},\ and\ \citenamefont {Woods}}]{PhysRevC.82.054315}%
  \BibitemOpen
  \bibfield  {author} {\bibinfo {author} {\bibfnamefont {I.}~\bibnamefont
  {Mukha}}, \bibinfo {author} {\bibfnamefont {K.}~\bibnamefont {S\"ummerer}},
  \bibinfo {author} {\bibfnamefont {L.}~\bibnamefont {Acosta}}, \bibinfo
  {author} {\bibfnamefont {M.~A.~G.}\ \bibnamefont {Alvarez}}, \bibinfo
  {author} {\bibfnamefont {E.}~\bibnamefont {Casarejos}}, \bibinfo {author}
  {\bibfnamefont {A.}~\bibnamefont {Chatillon}}, \bibinfo {author}
  {\bibfnamefont {D.}~\bibnamefont {Cortina-Gil}}, \bibinfo {author}
  {\bibfnamefont {I.~A.}\ \bibnamefont {Egorova}}, \bibinfo {author}
  {\bibfnamefont {J.~M.}\ \bibnamefont {Espino}}, \bibinfo {author}
  {\bibfnamefont {A.}~\bibnamefont {Fomichev}}, \bibinfo {author}
  {\bibfnamefont {J.~E.}\ \bibnamefont {Garc\'{\i}a-Ramos}}, \bibinfo {author}
  {\bibfnamefont {H.}~\bibnamefont {Geissel}}, \bibinfo {author} {\bibfnamefont
  {J.}~\bibnamefont {G\'omez-Camacho}}, \bibinfo {author} {\bibfnamefont
  {L.}~\bibnamefont {Grigorenko}}, \bibinfo {author} {\bibfnamefont
  {J.}~\bibnamefont {Hofmann}}, \bibinfo {author} {\bibfnamefont
  {O.}~\bibnamefont {Kiselev}}, \bibinfo {author} {\bibfnamefont
  {A.}~\bibnamefont {Korsheninnikov}}, \bibinfo {author} {\bibfnamefont
  {N.}~\bibnamefont {Kurz}}, \bibinfo {author} {\bibfnamefont {Y.~A.}\
  \bibnamefont {Litvinov}}, \bibinfo {author} {\bibfnamefont {E.}~\bibnamefont
  {Litvinova}}, \bibinfo {author} {\bibfnamefont {I.}~\bibnamefont {Martel}},
  \bibinfo {author} {\bibfnamefont {C.}~\bibnamefont {Nociforo}}, \bibinfo
  {author} {\bibfnamefont {W.}~\bibnamefont {Ott}}, \bibinfo {author}
  {\bibfnamefont {M.}~\bibnamefont {Pf\"utzner}}, \bibinfo {author}
  {\bibfnamefont {C.}~\bibnamefont {Rodr\'{\i}guez-Tajes}}, \bibinfo {author}
  {\bibfnamefont {E.}~\bibnamefont {Roeckl}}, \bibinfo {author} {\bibfnamefont
  {M.}~\bibnamefont {Stanoiu}}, \bibinfo {author} {\bibfnamefont {N.~K.}\
  \bibnamefont {Timofeyuk}}, \bibinfo {author} {\bibfnamefont {H.}~\bibnamefont
  {Weick}}, \ and\ \bibinfo {author} {\bibfnamefont {P.~J.}\ \bibnamefont
  {Woods}},\ }\href {\doibase 10.1103/PhysRevC.82.054315} {\bibfield  {journal}
  {\bibinfo  {journal} {Phys. Rev. C}\ }\textbf {\bibinfo {volume} {82}},\
  \bibinfo {pages} {054315} (\bibinfo {year} {2010})}\BibitemShut {NoStop}%
\bibitem [{\citenamefont {Mukha}\ \emph {et~al.}(2008)\citenamefont {Mukha},
  \citenamefont {Grigorenko}, \citenamefont {S\"ummerer}, \citenamefont
  {Acosta}, \citenamefont {Alvarez}, \citenamefont {Casarejos}, \citenamefont
  {Chatillon}, \citenamefont {Cortina-Gil}, \citenamefont {Espino},
  \citenamefont {Fomichev}, \citenamefont {Garc\'{\i}a-Ramos}, \citenamefont
  {Geissel}, \citenamefont {G\'omez-Camacho}, \citenamefont {Hofmann},
  \citenamefont {Kiselev}, \citenamefont {Korsheninnikov}, \citenamefont
  {Kurz}, \citenamefont {Litvinov}, \citenamefont {Martel}, \citenamefont
  {Nociforo}, \citenamefont {Ott}, \citenamefont {Pf\"utzner}, \citenamefont
  {Rodr\'{\i}guez-Tajes}, \citenamefont {Roeckl}, \citenamefont {Stanoiu},
  \citenamefont {Weick},\ and\ \citenamefont {Woods}}]{PhysRevC.77.061303}%
  \BibitemOpen
  \bibfield  {author} {\bibinfo {author} {\bibfnamefont {I.}~\bibnamefont
  {Mukha}}, \bibinfo {author} {\bibfnamefont {L.}~\bibnamefont {Grigorenko}},
  \bibinfo {author} {\bibfnamefont {K.}~\bibnamefont {S\"ummerer}}, \bibinfo
  {author} {\bibfnamefont {L.}~\bibnamefont {Acosta}}, \bibinfo {author}
  {\bibfnamefont {M.~A.~G.}\ \bibnamefont {Alvarez}}, \bibinfo {author}
  {\bibfnamefont {E.}~\bibnamefont {Casarejos}}, \bibinfo {author}
  {\bibfnamefont {A.}~\bibnamefont {Chatillon}}, \bibinfo {author}
  {\bibfnamefont {D.}~\bibnamefont {Cortina-Gil}}, \bibinfo {author}
  {\bibfnamefont {J.~M.}\ \bibnamefont {Espino}}, \bibinfo {author}
  {\bibfnamefont {A.}~\bibnamefont {Fomichev}}, \bibinfo {author}
  {\bibfnamefont {J.~E.}\ \bibnamefont {Garc\'{\i}a-Ramos}}, \bibinfo {author}
  {\bibfnamefont {H.}~\bibnamefont {Geissel}}, \bibinfo {author} {\bibfnamefont
  {J.}~\bibnamefont {G\'omez-Camacho}}, \bibinfo {author} {\bibfnamefont
  {J.}~\bibnamefont {Hofmann}}, \bibinfo {author} {\bibfnamefont
  {O.}~\bibnamefont {Kiselev}}, \bibinfo {author} {\bibfnamefont
  {A.}~\bibnamefont {Korsheninnikov}}, \bibinfo {author} {\bibfnamefont
  {N.}~\bibnamefont {Kurz}}, \bibinfo {author} {\bibfnamefont {Y.}~\bibnamefont
  {Litvinov}}, \bibinfo {author} {\bibfnamefont {I.}~\bibnamefont {Martel}},
  \bibinfo {author} {\bibfnamefont {C.}~\bibnamefont {Nociforo}}, \bibinfo
  {author} {\bibfnamefont {W.}~\bibnamefont {Ott}}, \bibinfo {author}
  {\bibfnamefont {M.}~\bibnamefont {Pf\"utzner}}, \bibinfo {author}
  {\bibfnamefont {C.}~\bibnamefont {Rodr\'{\i}guez-Tajes}}, \bibinfo {author}
  {\bibfnamefont {E.}~\bibnamefont {Roeckl}}, \bibinfo {author} {\bibfnamefont
  {M.}~\bibnamefont {Stanoiu}}, \bibinfo {author} {\bibfnamefont
  {H.}~\bibnamefont {Weick}}, \ and\ \bibinfo {author} {\bibfnamefont {P.~J.}\
  \bibnamefont {Woods}},\ }\href {\doibase 10.1103/PhysRevC.77.061303}
  {\bibfield  {journal} {\bibinfo  {journal} {Phys. Rev. C}\ }\textbf {\bibinfo
  {volume} {77}},\ \bibinfo {pages} {061303(R)} (\bibinfo {year}
  {2008})}\BibitemShut {NoStop}%
\bibitem [{\citenamefont {Tian}\ \emph {et~al.}(2013)\citenamefont {Tian},
  \citenamefont {Wang}, \citenamefont {Li},\ and\ \citenamefont
  {Li}}]{PhysRevC.87.014313}%
  \BibitemOpen
  \bibfield  {author} {\bibinfo {author} {\bibfnamefont {J.}~\bibnamefont
  {Tian}}, \bibinfo {author} {\bibfnamefont {N.}~\bibnamefont {Wang}}, \bibinfo
  {author} {\bibfnamefont {C.}~\bibnamefont {Li}}, \ and\ \bibinfo {author}
  {\bibfnamefont {J.}~\bibnamefont {Li}},\ }\href {\doibase
  10.1103/PhysRevC.87.014313} {\bibfield  {journal} {\bibinfo  {journal} {Phys.
  Rev. C}\ }\textbf {\bibinfo {volume} {87}},\ \bibinfo {pages} {014313}
  (\bibinfo {year} {2013})}\BibitemShut {NoStop}%
\bibitem [{\citenamefont {Rotureau}\ \emph {et~al.}(2005)\citenamefont
  {Rotureau}, \citenamefont {Oko\l{}owicz},\ and\ \citenamefont
  {P\l{}oszajczak}}]{PhysRevLett.95.042503}%
  \BibitemOpen
  \bibfield  {author} {\bibinfo {author} {\bibfnamefont {J.}~\bibnamefont
  {Rotureau}}, \bibinfo {author} {\bibfnamefont {J.}~\bibnamefont
  {Oko\l{}owicz}}, \ and\ \bibinfo {author} {\bibfnamefont {M.}~\bibnamefont
  {P\l{}oszajczak}},\ }\href {\doibase 10.1103/PhysRevLett.95.042503}
  {\bibfield  {journal} {\bibinfo  {journal} {Phys. Rev. Lett.}\ }\textbf
  {\bibinfo {volume} {95}},\ \bibinfo {pages} {042503} (\bibinfo {year}
  {2005})}\BibitemShut {NoStop}%
\bibitem [{\citenamefont {Rotureau}\ \emph {et~al.}(2006)\citenamefont
  {Rotureau}, \citenamefont {Okołowicz},\ and\ \citenamefont
  {Płoszajczak}}]{ROTUREAU200613}%
  \BibitemOpen
  \bibfield  {author} {\bibinfo {author} {\bibfnamefont {J.}~\bibnamefont
  {Rotureau}}, \bibinfo {author} {\bibfnamefont {J.}~\bibnamefont
  {Okołowicz}}, \ and\ \bibinfo {author} {\bibfnamefont {M.}~\bibnamefont
  {Płoszajczak}},\ }\href {\doibase
  https://doi.org/10.1016/j.nuclphysa.2005.12.005} {\bibfield  {journal}
  {\bibinfo  {journal} {Nuclear Physics A}\ }\textbf {\bibinfo {volume}
  {767}},\ \bibinfo {pages} {13} (\bibinfo {year} {2006})}\BibitemShut
  {NoStop}%
\bibitem [{\citenamefont {Blank}\ and\ \citenamefont
  {P{\l}oszajczak}(2008)}]{Blank_2008}%
  \BibitemOpen
  \bibfield  {author} {\bibinfo {author} {\bibfnamefont {B.}~\bibnamefont
  {Blank}}\ and\ \bibinfo {author} {\bibfnamefont {M.}~\bibnamefont
  {P{\l}oszajczak}},\ }\href {\doibase 10.1088/0034-4885/71/4/046301}
  {\bibfield  {journal} {\bibinfo  {journal} {Rep. Prog. Phys.}\ }\textbf
  {\bibinfo {volume} {71}},\ \bibinfo {pages} {046301} (\bibinfo {year}
  {2008})}\BibitemShut {NoStop}%
\bibitem [{\citenamefont {Brown}\ \emph {et~al.}(2015)\citenamefont {Brown},
  \citenamefont {Charity}, \citenamefont {Sobotka}, \citenamefont {Grigorenko},
  \citenamefont {Golubkova}, \citenamefont {Bedoor}, \citenamefont {Buhro},
  \citenamefont {Chajecki}, \citenamefont {Elson}, \citenamefont {Lynch},
  \citenamefont {Manfredi}, \citenamefont {McNeel}, \citenamefont {Reviol},
  \citenamefont {Shane}, \citenamefont {Showalter}, \citenamefont {Tsang},
  \citenamefont {Winkelbauer},\ and\ \citenamefont
  {Wuosmaa}}]{PhysRevC.92.034329}%
  \BibitemOpen
  \bibfield  {author} {\bibinfo {author} {\bibfnamefont {K.~W.}\ \bibnamefont
  {Brown}}, \bibinfo {author} {\bibfnamefont {R.~J.}\ \bibnamefont {Charity}},
  \bibinfo {author} {\bibfnamefont {L.~G.}\ \bibnamefont {Sobotka}}, \bibinfo
  {author} {\bibfnamefont {L.~V.}\ \bibnamefont {Grigorenko}}, \bibinfo
  {author} {\bibfnamefont {T.~A.}\ \bibnamefont {Golubkova}}, \bibinfo {author}
  {\bibfnamefont {S.}~\bibnamefont {Bedoor}}, \bibinfo {author} {\bibfnamefont
  {W.~W.}\ \bibnamefont {Buhro}}, \bibinfo {author} {\bibfnamefont
  {Z.}~\bibnamefont {Chajecki}}, \bibinfo {author} {\bibfnamefont {J.~M.}\
  \bibnamefont {Elson}}, \bibinfo {author} {\bibfnamefont {W.~G.}\ \bibnamefont
  {Lynch}}, \bibinfo {author} {\bibfnamefont {J.}~\bibnamefont {Manfredi}},
  \bibinfo {author} {\bibfnamefont {D.~G.}\ \bibnamefont {McNeel}}, \bibinfo
  {author} {\bibfnamefont {W.}~\bibnamefont {Reviol}}, \bibinfo {author}
  {\bibfnamefont {R.}~\bibnamefont {Shane}}, \bibinfo {author} {\bibfnamefont
  {R.~H.}\ \bibnamefont {Showalter}}, \bibinfo {author} {\bibfnamefont {M.~B.}\
  \bibnamefont {Tsang}}, \bibinfo {author} {\bibfnamefont {J.~R.}\ \bibnamefont
  {Winkelbauer}}, \ and\ \bibinfo {author} {\bibfnamefont {A.~H.}\ \bibnamefont
  {Wuosmaa}},\ }\href {\doibase 10.1103/PhysRevC.92.034329} {\bibfield
  {journal} {\bibinfo  {journal} {Phys. Rev. C}\ }\textbf {\bibinfo {volume}
  {92}},\ \bibinfo {pages} {034329} (\bibinfo {year} {2015})}\BibitemShut
  {NoStop}%
\bibitem [{\citenamefont {Michel}\ \emph {et~al.}(2002)\citenamefont {Michel},
  \citenamefont {Nazarewicz}, \citenamefont {P\l{}oszajczak},\ and\
  \citenamefont {Bennaceur}}]{PhysRevLett.89.042502}%
  \BibitemOpen
  \bibfield  {author} {\bibinfo {author} {\bibfnamefont {N.}~\bibnamefont
  {Michel}}, \bibinfo {author} {\bibfnamefont {W.}~\bibnamefont {Nazarewicz}},
  \bibinfo {author} {\bibfnamefont {M.}~\bibnamefont {P\l{}oszajczak}}, \ and\
  \bibinfo {author} {\bibfnamefont {K.}~\bibnamefont {Bennaceur}},\ }\href
  {\doibase 10.1103/PhysRevLett.89.042502} {\bibfield  {journal} {\bibinfo
  {journal} {Phys. Rev. Lett.}\ }\textbf {\bibinfo {volume} {89}},\ \bibinfo
  {pages} {042502} (\bibinfo {year} {2002})}\BibitemShut {NoStop}%
\bibitem [{\citenamefont {Fossez}\ \emph {et~al.}(2017)\citenamefont {Fossez},
  \citenamefont {Rotureau}, \citenamefont {Michel},\ and\ \citenamefont
  {Nazarewicz}}]{PhysRevC.96.024308}%
  \BibitemOpen
  \bibfield  {author} {\bibinfo {author} {\bibfnamefont {K.}~\bibnamefont
  {Fossez}}, \bibinfo {author} {\bibfnamefont {J.}~\bibnamefont {Rotureau}},
  \bibinfo {author} {\bibfnamefont {N.}~\bibnamefont {Michel}}, \ and\ \bibinfo
  {author} {\bibfnamefont {W.}~\bibnamefont {Nazarewicz}},\ }\href {\doibase
  10.1103/PhysRevC.96.024308} {\bibfield  {journal} {\bibinfo  {journal} {Phys.
  Rev. C}\ }\textbf {\bibinfo {volume} {96}},\ \bibinfo {pages} {024308}
  (\bibinfo {year} {2017})}\BibitemShut {NoStop}%
\bibitem [{\citenamefont {Li}\ \emph {et~al.}(2021)\citenamefont {Li},
  \citenamefont {Michel}, \citenamefont {Zuo},\ and\ \citenamefont
  {Xu}}]{PhysRevC.103.034305}%
  \BibitemOpen
  \bibfield  {author} {\bibinfo {author} {\bibfnamefont {J.~G.}\ \bibnamefont
  {Li}}, \bibinfo {author} {\bibfnamefont {N.}~\bibnamefont {Michel}}, \bibinfo
  {author} {\bibfnamefont {W.}~\bibnamefont {Zuo}}, \ and\ \bibinfo {author}
  {\bibfnamefont {F.~R.}\ \bibnamefont {Xu}},\ }\href {\doibase
  10.1103/PhysRevC.103.034305} {\bibfield  {journal} {\bibinfo  {journal}
  {Phys. Rev. C}\ }\textbf {\bibinfo {volume} {103}},\ \bibinfo {pages}
  {034305} (\bibinfo {year} {2021})}\BibitemShut {NoStop}%
\bibitem [{\citenamefont {Leviatan}(2011)}]{LEVIATAN201193}%
  \BibitemOpen
  \bibfield  {author} {\bibinfo {author} {\bibfnamefont {A.}~\bibnamefont
  {Leviatan}},\ }\href {\doibase https://doi.org/10.1016/j.ppnp.2010.08.001}
  {\bibfield  {journal} {\bibinfo  {journal} {Prog. Part. Nucl. Phys.}\
  }\textbf {\bibinfo {volume} {66}},\ \bibinfo {pages} {93} (\bibinfo {year}
  {2011})}\BibitemShut {NoStop}%
\bibitem [{\citenamefont {Leviatan}\ and\ \citenamefont
  {Van~Isacker}(2002)}]{PhysRevLett.89.222501}%
  \BibitemOpen
  \bibfield  {author} {\bibinfo {author} {\bibfnamefont {A.}~\bibnamefont
  {Leviatan}}\ and\ \bibinfo {author} {\bibfnamefont {P.}~\bibnamefont
  {Van~Isacker}},\ }\href {\doibase 10.1103/PhysRevLett.89.222501} {\bibfield
  {journal} {\bibinfo  {journal} {Phys. Rev. Lett.}\ }\textbf {\bibinfo
  {volume} {89}},\ \bibinfo {pages} {222501} (\bibinfo {year}
  {2002})}\BibitemShut {NoStop}%
\bibitem [{\citenamefont {Berggren}(1968)}]{BERGGREN1968265}%
  \BibitemOpen
  \bibfield  {author} {\bibinfo {author} {\bibfnamefont {T.}~\bibnamefont
  {Berggren}},\ }\href {\doibase https://doi.org/10.1016/0375-9474(68)90593-9}
  {\bibfield  {journal} {\bibinfo  {journal} {Nucl. Phys. A}\ }\textbf
  {\bibinfo {volume} {109}},\ \bibinfo {pages} {265 } (\bibinfo {year}
  {1968})}\BibitemShut {NoStop}%
\bibitem [{\citenamefont {Newton}(2013)}]{newton1966scattering}%
  \BibitemOpen
  \bibfield  {author} {\bibinfo {author} {\bibfnamefont {R.}~\bibnamefont
  {Newton}},\ }\href@noop {} {\emph {\bibinfo {title} {Scattering Theory of
  Waves and Particles}}}\ (\bibinfo  {publisher} {Dover Publications; Second
  Edition; New York},\ \bibinfo {year} {2013})\BibitemShut {NoStop}%
\bibitem [{\citenamefont {Michel}(2011)}]{PhysRevC.83.034325}%
  \BibitemOpen
  \bibfield  {author} {\bibinfo {author} {\bibfnamefont {N.}~\bibnamefont
  {Michel}},\ }\href {\doibase 10.1103/PhysRevC.83.034325} {\bibfield
  {journal} {\bibinfo  {journal} {Phys. Rev. C}\ }\textbf {\bibinfo {volume}
  {83}},\ \bibinfo {pages} {034325} (\bibinfo {year} {2011})}\BibitemShut
  {NoStop}%
\bibitem [{\citenamefont {Sleijpen}\ and\ \citenamefont {van~der
  Vorst}(1996)}]{Jacobi_Davidson}%
  \BibitemOpen
  \bibfield  {author} {\bibinfo {author} {\bibfnamefont {G.~J.~G.}\
  \bibnamefont {Sleijpen}}\ and\ \bibinfo {author} {\bibfnamefont {H.~A.}\
  \bibnamefont {van~der Vorst}},\ }\href@noop {} {\bibfield  {journal}
  {\bibinfo  {journal} {SIAM J. Matrix Anal. Appl.}\ }\textbf {\bibinfo
  {volume} {17}},\ \bibinfo {pages} {401} (\bibinfo {year} {1996})}\BibitemShut
  {NoStop}%
\bibitem [{\citenamefont {Michel}\ \emph
  {et~al.}(2020{\natexlab{b}})\citenamefont {Michel}, \citenamefont {Aktulga},\
  and\ \citenamefont {Jaganathen}}]{MICHEL2020106978}%
  \BibitemOpen
  \bibfield  {author} {\bibinfo {author} {\bibfnamefont {N.}~\bibnamefont
  {Michel}}, \bibinfo {author} {\bibfnamefont {H.}~\bibnamefont {Aktulga}}, \
  and\ \bibinfo {author} {\bibfnamefont {Y.}~\bibnamefont {Jaganathen}},\
  }\href@noop {} {\bibfield  {journal} {\bibinfo  {journal} {Comput. Phys.
  Comm.}\ }\textbf {\bibinfo {volume} {247}},\ \bibinfo {pages} {106978}
  (\bibinfo {year} {2020}{\natexlab{b}})}\BibitemShut {NoStop}%
\bibitem [{\citenamefont {Machleidt}\ and\ \citenamefont
  {Entem}(2011)}]{MACHLEIDT20111}%
  \BibitemOpen
  \bibfield  {author} {\bibinfo {author} {\bibfnamefont {R.}~\bibnamefont
  {Machleidt}}\ and\ \bibinfo {author} {\bibfnamefont {D.}~\bibnamefont
  {Entem}},\ }\href {\doibase https://doi.org/10.1016/j.physrep.2011.02.001}
  {\bibfield  {journal} {\bibinfo  {journal} {Phys. Rep}\ }\textbf {\bibinfo
  {volume} {503}},\ \bibinfo {pages} {1 } (\bibinfo {year} {2011})}\BibitemShut
  {NoStop}%
\bibitem [{\citenamefont {Brown}\ \emph {et~al.}(2014)\citenamefont {Brown},
  \citenamefont {Charity}, \citenamefont {Sobotka}, \citenamefont {Chajecki},
  \citenamefont {Grigorenko}, \citenamefont {Egorova}, \citenamefont
  {Parfenova}, \citenamefont {Zhukov}, \citenamefont {Bedoor}, \citenamefont
  {Buhro}, \citenamefont {Elson}, \citenamefont {Lynch}, \citenamefont
  {Manfredi}, \citenamefont {McNeel}, \citenamefont {Reviol}, \citenamefont
  {Shane}, \citenamefont {Showalter}, \citenamefont {Tsang}, \citenamefont
  {Winkelbauer},\ and\ \citenamefont {Wuosmaa}}]{PhysRevLett.113.232501}%
  \BibitemOpen
  \bibfield  {author} {\bibinfo {author} {\bibfnamefont {K.~W.}\ \bibnamefont
  {Brown}}, \bibinfo {author} {\bibfnamefont {R.~J.}\ \bibnamefont {Charity}},
  \bibinfo {author} {\bibfnamefont {L.~G.}\ \bibnamefont {Sobotka}}, \bibinfo
  {author} {\bibfnamefont {Z.}~\bibnamefont {Chajecki}}, \bibinfo {author}
  {\bibfnamefont {L.~V.}\ \bibnamefont {Grigorenko}}, \bibinfo {author}
  {\bibfnamefont {I.~A.}\ \bibnamefont {Egorova}}, \bibinfo {author}
  {\bibfnamefont {Y.~L.}\ \bibnamefont {Parfenova}}, \bibinfo {author}
  {\bibfnamefont {M.~V.}\ \bibnamefont {Zhukov}}, \bibinfo {author}
  {\bibfnamefont {S.}~\bibnamefont {Bedoor}}, \bibinfo {author} {\bibfnamefont
  {W.~W.}\ \bibnamefont {Buhro}}, \bibinfo {author} {\bibfnamefont {J.~M.}\
  \bibnamefont {Elson}}, \bibinfo {author} {\bibfnamefont {W.~G.}\ \bibnamefont
  {Lynch}}, \bibinfo {author} {\bibfnamefont {J.}~\bibnamefont {Manfredi}},
  \bibinfo {author} {\bibfnamefont {D.~G.}\ \bibnamefont {McNeel}}, \bibinfo
  {author} {\bibfnamefont {W.}~\bibnamefont {Reviol}}, \bibinfo {author}
  {\bibfnamefont {R.}~\bibnamefont {Shane}}, \bibinfo {author} {\bibfnamefont
  {R.~H.}\ \bibnamefont {Showalter}}, \bibinfo {author} {\bibfnamefont {M.~B.}\
  \bibnamefont {Tsang}}, \bibinfo {author} {\bibfnamefont {J.~R.}\ \bibnamefont
  {Winkelbauer}}, \ and\ \bibinfo {author} {\bibfnamefont {A.~H.}\ \bibnamefont
  {Wuosmaa}},\ }\href {\doibase 10.1103/PhysRevLett.113.232501} {\bibfield
  {journal} {\bibinfo  {journal} {Phys. Rev. Lett.}\ }\textbf {\bibinfo
  {volume} {113}},\ \bibinfo {pages} {232501} (\bibinfo {year}
  {2014})}\BibitemShut {NoStop}%
\bibitem [{\citenamefont {Woodward}\ \emph {et~al.}(1983)\citenamefont
  {Woodward}, \citenamefont {Tribble},\ and\ \citenamefont
  {Tanner}}]{PhysRevC.27.27}%
  \BibitemOpen
  \bibfield  {author} {\bibinfo {author} {\bibfnamefont {C.~J.}\ \bibnamefont
  {Woodward}}, \bibinfo {author} {\bibfnamefont {R.~E.}\ \bibnamefont
  {Tribble}}, \ and\ \bibinfo {author} {\bibfnamefont {D.~M.}\ \bibnamefont
  {Tanner}},\ }\href {\doibase 10.1103/PhysRevC.27.27} {\bibfield  {journal}
  {\bibinfo  {journal} {Phys. Rev. C}\ }\textbf {\bibinfo {volume} {27}},\
  \bibinfo {pages} {27} (\bibinfo {year} {1983})}\BibitemShut {NoStop}%
\bibitem [{\citenamefont {KeKelis}\ \emph {et~al.}(1978)\citenamefont
  {KeKelis}, \citenamefont {Zisman}, \citenamefont {Scott}, \citenamefont
  {Jahn}, \citenamefont {Vieira}, \citenamefont {Cerny},\ and\ \citenamefont
  {Ajzenberg-Selove}}]{PhysRevC.17.1929}%
  \BibitemOpen
  \bibfield  {author} {\bibinfo {author} {\bibfnamefont {G.~J.}\ \bibnamefont
  {KeKelis}}, \bibinfo {author} {\bibfnamefont {M.~S.}\ \bibnamefont {Zisman}},
  \bibinfo {author} {\bibfnamefont {D.~K.}\ \bibnamefont {Scott}}, \bibinfo
  {author} {\bibfnamefont {R.}~\bibnamefont {Jahn}}, \bibinfo {author}
  {\bibfnamefont {D.~J.}\ \bibnamefont {Vieira}}, \bibinfo {author}
  {\bibfnamefont {J.}~\bibnamefont {Cerny}}, \ and\ \bibinfo {author}
  {\bibfnamefont {F.}~\bibnamefont {Ajzenberg-Selove}},\ }\href {\doibase
  10.1103/PhysRevC.17.1929} {\bibfield  {journal} {\bibinfo  {journal} {Phys.
  Rev. C}\ }\textbf {\bibinfo {volume} {17}},\ \bibinfo {pages} {1929}
  (\bibinfo {year} {1978})}\BibitemShut {NoStop}%
\bibitem [{\citenamefont {Grigorenko}\ \emph {et~al.}(2002)\citenamefont
  {Grigorenko}, \citenamefont {Mukha}, \citenamefont {Thompson},\ and\
  \citenamefont {Zhukov}}]{PhysRevLett.88.042502}%
  \BibitemOpen
  \bibfield  {author} {\bibinfo {author} {\bibfnamefont {L.~V.}\ \bibnamefont
  {Grigorenko}}, \bibinfo {author} {\bibfnamefont {I.~G.}\ \bibnamefont
  {Mukha}}, \bibinfo {author} {\bibfnamefont {I.~J.}\ \bibnamefont {Thompson}},
  \ and\ \bibinfo {author} {\bibfnamefont {M.~V.}\ \bibnamefont {Zhukov}},\
  }\href {\doibase 10.1103/PhysRevLett.88.042502} {\bibfield  {journal}
  {\bibinfo  {journal} {Phys. Rev. Lett.}\ }\textbf {\bibinfo {volume} {88}},\
  \bibinfo {pages} {042502} (\bibinfo {year} {2002})}\BibitemShut {NoStop}%
\bibitem [{\citenamefont {Wamers}\ \emph {et~al.}(2014)\citenamefont {Wamers},
  \citenamefont {Marganiec}, \citenamefont {Aksouh}, \citenamefont {Aksyutina},
  \citenamefont {\'Alvarez-Pol}, \citenamefont {Aumann}, \citenamefont
  {Beceiro-Novo}, \citenamefont {Boretzky}, \citenamefont {Borge},
  \citenamefont {Chartier}, \citenamefont {Chatillon}, \citenamefont {Chulkov},
  \citenamefont {Cortina-Gil}, \citenamefont {Emling}, \citenamefont {Ershova},
  \citenamefont {Fraile}, \citenamefont {Fynbo}, \citenamefont {Galaviz},
  \citenamefont {Geissel}, \citenamefont {Heil}, \citenamefont {Hoffmann},
  \citenamefont {Johansson}, \citenamefont {Jonson}, \citenamefont
  {Karagiannis}, \citenamefont {Kiselev}, \citenamefont {Kratz}, \citenamefont
  {Kulessa}, \citenamefont {Kurz}, \citenamefont {Langer}, \citenamefont
  {Lantz}, \citenamefont {Le~Bleis}, \citenamefont {Lemmon}, \citenamefont
  {Litvinov}, \citenamefont {Mahata}, \citenamefont {M\"untz}, \citenamefont
  {Nilsson}, \citenamefont {Nociforo}, \citenamefont {Nyman}, \citenamefont
  {Ott}, \citenamefont {Panin}, \citenamefont {Paschalis}, \citenamefont
  {Perea}, \citenamefont {Plag}, \citenamefont {Reifarth}, \citenamefont
  {Richter}, \citenamefont {Rodriguez-Tajes}, \citenamefont {Rossi},
  \citenamefont {Riisager}, \citenamefont {Savran}, \citenamefont {Schrieder},
  \citenamefont {Simon}, \citenamefont {Stroth}, \citenamefont {S\"ummerer},
  \citenamefont {Tengblad}, \citenamefont {Weick}, \citenamefont {Wimmer},\
  and\ \citenamefont {Zhukov}}]{PhysRevLett.112.132502}%
  \BibitemOpen
  \bibfield  {author} {\bibinfo {author} {\bibfnamefont {F.}~\bibnamefont
  {Wamers}}, \bibinfo {author} {\bibfnamefont {J.}~\bibnamefont {Marganiec}},
  \bibinfo {author} {\bibfnamefont {F.}~\bibnamefont {Aksouh}}, \bibinfo
  {author} {\bibfnamefont {Y.}~\bibnamefont {Aksyutina}}, \bibinfo {author}
  {\bibfnamefont {H.}~\bibnamefont {\'Alvarez-Pol}}, \bibinfo {author}
  {\bibfnamefont {T.}~\bibnamefont {Aumann}}, \bibinfo {author} {\bibfnamefont
  {S.}~\bibnamefont {Beceiro-Novo}}, \bibinfo {author} {\bibfnamefont
  {K.}~\bibnamefont {Boretzky}}, \bibinfo {author} {\bibfnamefont {M.~J.~G.}\
  \bibnamefont {Borge}}, \bibinfo {author} {\bibfnamefont {M.}~\bibnamefont
  {Chartier}}, \bibinfo {author} {\bibfnamefont {A.}~\bibnamefont {Chatillon}},
  \bibinfo {author} {\bibfnamefont {L.~V.}\ \bibnamefont {Chulkov}}, \bibinfo
  {author} {\bibfnamefont {D.}~\bibnamefont {Cortina-Gil}}, \bibinfo {author}
  {\bibfnamefont {H.}~\bibnamefont {Emling}}, \bibinfo {author} {\bibfnamefont
  {O.}~\bibnamefont {Ershova}}, \bibinfo {author} {\bibfnamefont {L.~M.}\
  \bibnamefont {Fraile}}, \bibinfo {author} {\bibfnamefont {H.~O.~U.}\
  \bibnamefont {Fynbo}}, \bibinfo {author} {\bibfnamefont {D.}~\bibnamefont
  {Galaviz}}, \bibinfo {author} {\bibfnamefont {H.}~\bibnamefont {Geissel}},
  \bibinfo {author} {\bibfnamefont {M.}~\bibnamefont {Heil}}, \bibinfo {author}
  {\bibfnamefont {D.~H.~H.}\ \bibnamefont {Hoffmann}}, \bibinfo {author}
  {\bibfnamefont {H.~T.}\ \bibnamefont {Johansson}}, \bibinfo {author}
  {\bibfnamefont {B.}~\bibnamefont {Jonson}}, \bibinfo {author} {\bibfnamefont
  {C.}~\bibnamefont {Karagiannis}}, \bibinfo {author} {\bibfnamefont {O.~A.}\
  \bibnamefont {Kiselev}}, \bibinfo {author} {\bibfnamefont {J.~V.}\
  \bibnamefont {Kratz}}, \bibinfo {author} {\bibfnamefont {R.}~\bibnamefont
  {Kulessa}}, \bibinfo {author} {\bibfnamefont {N.}~\bibnamefont {Kurz}},
  \bibinfo {author} {\bibfnamefont {C.}~\bibnamefont {Langer}}, \bibinfo
  {author} {\bibfnamefont {M.}~\bibnamefont {Lantz}}, \bibinfo {author}
  {\bibfnamefont {T.}~\bibnamefont {Le~Bleis}}, \bibinfo {author}
  {\bibfnamefont {R.}~\bibnamefont {Lemmon}}, \bibinfo {author} {\bibfnamefont
  {Y.~A.}\ \bibnamefont {Litvinov}}, \bibinfo {author} {\bibfnamefont
  {K.}~\bibnamefont {Mahata}}, \bibinfo {author} {\bibfnamefont
  {C.}~\bibnamefont {M\"untz}}, \bibinfo {author} {\bibfnamefont
  {T.}~\bibnamefont {Nilsson}}, \bibinfo {author} {\bibfnamefont
  {C.}~\bibnamefont {Nociforo}}, \bibinfo {author} {\bibfnamefont
  {G.}~\bibnamefont {Nyman}}, \bibinfo {author} {\bibfnamefont
  {W.}~\bibnamefont {Ott}}, \bibinfo {author} {\bibfnamefont {V.}~\bibnamefont
  {Panin}}, \bibinfo {author} {\bibfnamefont {S.}~\bibnamefont {Paschalis}},
  \bibinfo {author} {\bibfnamefont {A.}~\bibnamefont {Perea}}, \bibinfo
  {author} {\bibfnamefont {R.}~\bibnamefont {Plag}}, \bibinfo {author}
  {\bibfnamefont {R.}~\bibnamefont {Reifarth}}, \bibinfo {author}
  {\bibfnamefont {A.}~\bibnamefont {Richter}}, \bibinfo {author} {\bibfnamefont
  {C.}~\bibnamefont {Rodriguez-Tajes}}, \bibinfo {author} {\bibfnamefont
  {D.}~\bibnamefont {Rossi}}, \bibinfo {author} {\bibfnamefont
  {K.}~\bibnamefont {Riisager}}, \bibinfo {author} {\bibfnamefont
  {D.}~\bibnamefont {Savran}}, \bibinfo {author} {\bibfnamefont
  {G.}~\bibnamefont {Schrieder}}, \bibinfo {author} {\bibfnamefont
  {H.}~\bibnamefont {Simon}}, \bibinfo {author} {\bibfnamefont
  {J.}~\bibnamefont {Stroth}}, \bibinfo {author} {\bibfnamefont
  {K.}~\bibnamefont {S\"ummerer}}, \bibinfo {author} {\bibfnamefont
  {O.}~\bibnamefont {Tengblad}}, \bibinfo {author} {\bibfnamefont
  {H.}~\bibnamefont {Weick}}, \bibinfo {author} {\bibfnamefont
  {C.}~\bibnamefont {Wimmer}}, \ and\ \bibinfo {author} {\bibfnamefont {M.~V.}\
  \bibnamefont {Zhukov}},\ }\href {\doibase 10.1103/PhysRevLett.112.132502}
  {\bibfield  {journal} {\bibinfo  {journal} {Phys. Rev. Lett.}\ }\textbf
  {\bibinfo {volume} {112}},\ \bibinfo {pages} {132502} (\bibinfo {year}
  {2014})}\BibitemShut {NoStop}%
\bibitem [{\citenamefont {Fortune}(2017)}]{PhysRevC.96.064313}%
  \BibitemOpen
  \bibfield  {author} {\bibinfo {author} {\bibfnamefont {H.~T.}\ \bibnamefont
  {Fortune}},\ }\href {\doibase 10.1103/PhysRevC.96.064313} {\bibfield
  {journal} {\bibinfo  {journal} {Phys. Rev. C}\ }\textbf {\bibinfo {volume}
  {96}},\ \bibinfo {pages} {064313} (\bibinfo {year} {2017})}\BibitemShut
  {NoStop}%
\bibitem [{\citenamefont {Fortune}\ and\ \citenamefont
  {Sherr}(2013)}]{PhysRevC.87.044315}%
  \BibitemOpen
  \bibfield  {author} {\bibinfo {author} {\bibfnamefont {H.~T.}\ \bibnamefont
  {Fortune}}\ and\ \bibinfo {author} {\bibfnamefont {R.}~\bibnamefont
  {Sherr}},\ }\href {\doibase 10.1103/PhysRevC.87.044315} {\bibfield  {journal}
  {\bibinfo  {journal} {Phys. Rev. C}\ }\textbf {\bibinfo {volume} {87}},\
  \bibinfo {pages} {044315} (\bibinfo {year} {2013})}\BibitemShut {NoStop}%
\bibitem [{\citenamefont {Fortune}\ and\ \citenamefont
  {Sherr}(2010)}]{PhysRevC.82.027310}%
  \BibitemOpen
  \bibfield  {author} {\bibinfo {author} {\bibfnamefont {H.~T.}\ \bibnamefont
  {Fortune}}\ and\ \bibinfo {author} {\bibfnamefont {R.}~\bibnamefont
  {Sherr}},\ }\href {\doibase 10.1103/PhysRevC.82.027310} {\bibfield  {journal}
  {\bibinfo  {journal} {Phys. Rev. C}\ }\textbf {\bibinfo {volume} {82}},\
  \bibinfo {pages} {027310} (\bibinfo {year} {2010})}\BibitemShut {NoStop}%
\bibitem [{\citenamefont {Fortune}(2016)}]{PhysRevC.94.044305}%
  \BibitemOpen
  \bibfield  {author} {\bibinfo {author} {\bibfnamefont {H.~T.}\ \bibnamefont
  {Fortune}},\ }\href {\doibase 10.1103/PhysRevC.94.044305} {\bibfield
  {journal} {\bibinfo  {journal} {Phys. Rev. C}\ }\textbf {\bibinfo {volume}
  {94}},\ \bibinfo {pages} {044305} (\bibinfo {year} {2016})}\BibitemShut
  {NoStop}%
\bibitem [{\citenamefont {Jansen}\ \emph {et~al.}(2014)\citenamefont {Jansen},
  \citenamefont {Engel}, \citenamefont {Hagen}, \citenamefont {Navratil},\ and\
  \citenamefont {Signoracci}}]{PhysRevLett.113.142502}%
  \BibitemOpen
  \bibfield  {author} {\bibinfo {author} {\bibfnamefont {G.~R.}\ \bibnamefont
  {Jansen}}, \bibinfo {author} {\bibfnamefont {J.}~\bibnamefont {Engel}},
  \bibinfo {author} {\bibfnamefont {G.}~\bibnamefont {Hagen}}, \bibinfo
  {author} {\bibfnamefont {P.}~\bibnamefont {Navratil}}, \ and\ \bibinfo
  {author} {\bibfnamefont {A.}~\bibnamefont {Signoracci}},\ }\href {\doibase
  10.1103/PhysRevLett.113.142502} {\bibfield  {journal} {\bibinfo  {journal}
  {Phys. Rev. Lett.}\ }\textbf {\bibinfo {volume} {113}},\ \bibinfo {pages}
  {142502} (\bibinfo {year} {2014})}\BibitemShut {NoStop}%
\bibitem [{\citenamefont {Michel}\ \emph {et~al.}(2010)\citenamefont {Michel},
  \citenamefont {Nazarewicz},\ and\ \citenamefont
  {P\l{}oszajczak}}]{PhysRevC.82.044315}%
  \BibitemOpen
  \bibfield  {author} {\bibinfo {author} {\bibfnamefont {N.}~\bibnamefont
  {Michel}}, \bibinfo {author} {\bibfnamefont {W.}~\bibnamefont {Nazarewicz}},
  \ and\ \bibinfo {author} {\bibfnamefont {M.}~\bibnamefont {P\l{}oszajczak}},\
  }\href {\doibase 10.1103/PhysRevC.82.044315} {\bibfield  {journal} {\bibinfo
  {journal} {Phys. Rev. C}\ }\textbf {\bibinfo {volume} {82}},\ \bibinfo
  {pages} {044315} (\bibinfo {year} {2010})}\BibitemShut {NoStop}%
\bibitem [{\citenamefont {Ehrman}(1951)}]{PhysRev.81.412}%
  \BibitemOpen
  \bibfield  {author} {\bibinfo {author} {\bibfnamefont {J.~B.}\ \bibnamefont
  {Ehrman}},\ }\href {\doibase 10.1103/PhysRev.81.412} {\bibfield  {journal}
  {\bibinfo  {journal} {Phys. Rev.}\ }\textbf {\bibinfo {volume} {81}},\
  \bibinfo {pages} {412} (\bibinfo {year} {1951})}\BibitemShut {NoStop}%
\bibitem [{\citenamefont {Thomas}(1952)}]{PhysRev.88.1109}%
  \BibitemOpen
  \bibfield  {author} {\bibinfo {author} {\bibfnamefont {R.~G.}\ \bibnamefont
  {Thomas}},\ }\href {\doibase 10.1103/PhysRev.88.1109} {\bibfield  {journal}
  {\bibinfo  {journal} {Phys. Rev.}\ }\textbf {\bibinfo {volume} {88}},\
  \bibinfo {pages} {1109} (\bibinfo {year} {1952})}\BibitemShut {NoStop}%
\bibitem [{\citenamefont {Mao}\ \emph {et~al.}(2020)\citenamefont {Mao},
  \citenamefont {Rotureau}, \citenamefont {Nazarewicz}, \citenamefont {Michel},
  \citenamefont {Id~Betan},\ and\ \citenamefont
  {Jaganathen}}]{PhysRevC.102.024309}%
  \BibitemOpen
  \bibfield  {author} {\bibinfo {author} {\bibfnamefont {X.}~\bibnamefont
  {Mao}}, \bibinfo {author} {\bibfnamefont {J.}~\bibnamefont {Rotureau}},
  \bibinfo {author} {\bibfnamefont {W.}~\bibnamefont {Nazarewicz}}, \bibinfo
  {author} {\bibfnamefont {N.}~\bibnamefont {Michel}}, \bibinfo {author}
  {\bibfnamefont {R.~M.}\ \bibnamefont {Id~Betan}}, \ and\ \bibinfo {author}
  {\bibfnamefont {Y.}~\bibnamefont {Jaganathen}},\ }\href {\doibase
  10.1103/PhysRevC.102.024309} {\bibfield  {journal} {\bibinfo  {journal}
  {Phys. Rev. C}\ }\textbf {\bibinfo {volume} {102}},\ \bibinfo {pages}
  {024309} (\bibinfo {year} {2020})}\BibitemShut {NoStop}%
\end{thebibliography}%
\end{document}